\documentclass[a4paper,11pt]{article}
\pdfoutput=1
\usepackage{jinstpub} 
\usepackage{booktabs}

\title{A New Cryogenic Apparatus to Search for the Neutron Electric Dipole Moment}

\author[a,b,c]{M. W. Ahmed,}
\author[d]{R. Alarcon,}
\author[e]{A. Aleksandrova,}
\author[f,g]{S. Bae\ss{}ler,}
\author[h]{L. Barron-Palos,}
\author[i]{L. M. Bartoszek,}
\author[j]{D. H. Beck,}
\author[e]{M. Behzadipour,}
\author[c]{I. Berkutov,}
\author[k]{J. Bessuille,}
\author[l]{M. Blatnik,}
\author[e]{M. Broering,}
\author[g]{L. J. Broussard,}
\author[b]{M. Busch,}
\author[l]{R. Carr,}
\author[g]{V. Cianciolo,}
\author[m]{S. M. Clayton,}
\author[m]{M. D. Cooper,}
\author[e]{C. Crawford,}
\author[m]{S. A. Currie,}
\author[i]{C. Daurer,}
\author[d]{R. Dipert,}
\author[k]{K. Dow,}
\author[n]{D. Dutta,}
\author[g,o]{Y. Efremenko,}
\author[j]{C. B. Erickson,}
\author[l,1]{B. W. Filippone\note{Corresponding author},}
\author[o]{N. Fomin,}
\author[b]{H. Gao,}
\author[p,c]{R. Golub,}
\author[q]{C. R. Gould,}
\author[o,g]{G. Greene,}
\author[p,c]{D. G. Haase,}
\author[k]{D. Hasell,}
\author[p]{A. I. Hawari,}
\author[r] {M.E. Hayden,}
\author[s]{A. Holley,}
\author[l]{R. J. Holt,}
\author[p,g,c]{P. R. Huffman,}
\author[k]{E. Ihloff,}
\author[o]{S. K. Imam,}
\author[m]{T. M. Ito,}
\author[t]{M. Karcz,}
\author[k]{J. Kelsey,}
\author[b,c]{D. P. Kendellen,}
\author[k]{Y. J. Kim,}
\author[p,q,c]{E. Korobkina,}
\author[e]{W. Korsch,}
\author[u]{S. K. Lamoreaux,}
\author[n]{E. Leggett,}
\author[p,c]{K. K. H. Leung,}
\author[p]{A. Lipman,}
\author[t]{C. Y. Liu,}
\author[t]{J. Long,}
\author[m]{S. W. T. MacDonald,}
\author[m]{M. Makela,}
\author[m]{A. Matlashov,}
\author[k,2]{J. D. Maxwell\note{Present Address: Thomas Jefferson National Accelerator Facility},}
\author[l,3]{M. Mendenhall,\note{Present Address: Lawrence Livermore National Laboratory}}
\author[t]{H. O. Meyer,}
\author[k]{R. G. Milner,}
\author[g]{P. E. Mueller,}
\author[e]{N. Nouri,}
\author[m]{C. M. O'Shaughnessy,}
\author[l]{C. Osthelder,}
\author[j]{J. C. Peng,}
\author[g]{S. I. Penttila,}
\author[m]{N. S. Phan,}
\author[e]{B. Plaster,}
\author[m,g]{J. C. Ramsey,}
\author[j,4]{T. M. Rao\note{Present Address: North Carolina State University},}
\author[k]{R. P. Redwine,}
\author[p,c,5]{A. Reid\note{Present Address: Indiana University},}
\author[e]{A. Saftah,}
\author[v]{G. M. Seidel,}
\author[w]{I. Silvera,}
\author[l]{S. Slutsky,}
\author[m]{E. Smith,}
\author[t]{W. M. Snow,}
\author[m]{W. Sondheim,}
\author[p,c]{S. Sosothikul,}
\author[x]{T. D. S. Stanislaus,}
\author[l]{X. Sun,}
\author[l]{C. M. Swank,}
\author[m]{Z. Tang,}
\author[r,6]{R. Tavakoli Dinani\note{Present Address: KU Leuven},}
\author[k]{E. Tsentalovich,}
\author[k]{C. Vidal,}
\author[h,i]{W. Wei,}
\author[p,c]{C. R. White,}
\author[j]{S. E. Williamson,}
\author[j]{L. Yang,}
\author[g]{W. Yao,}
\author[p]{and A. R. Young}
\affiliation[a]{Department of Mathematics and Physics, North Carolina Central University, \\  Durham, NC 27707, U.S.A.}
\affiliation[b]{Department of Physics, Duke University, \\ Durham, NC 27708, U.S.A.}
\affiliation[c]{Triangle Universities Nuclear Laboratory, \\ Durham, North Carolina 27708, U.S.A.}
\affiliation[d]{Department of Physics, Arizona State University,\\ Tempe, Arizona 85287-1504, U.S.A.}
\affiliation[e]{Department of Physics and Astronomy, University of Kentucky,\\ Lexington, KY, 40506, U.S.A.}
\affiliation[f]{Physics Department, University of Virginia, \\ 382 McCormick Road, Charlottesville, Virginia 22904, U.S.A.}
\affiliation[g]{Physics Division, Oak Ridge National Laboratory\\  Oak Ridge, TN 37831, U.S.A.}
\affiliation[h]{Instituto de F\'isica, Universidad Nacional Aut\'onoma de M\'exico, \\  Apartado Postal 20-364, 01000, Mexico.}
\affiliation[i]{Bartoszek Engineering, 818 W. Downer Place, \\ Aurora, IL 60506-4904, U.S.A.}
\affiliation[j]{Department of Physics, University of Illinois at Urbana-Champaign,\\ 1110 W. Green St., Urbana, IL, 61801-3090, U.S.A.}
\affiliation[k]{Bates Laboratory, Massachusetts Institute of Technology,\\ Cambridge, Massachusetts 02139, U.S.A.}
\affiliation[l]{Kellogg Radiation Laboratory, California Institute of Technology, \\ Pasadena, CA 91125, U.S.A.}
\affiliation[m]{Physics Division, Los Alamos National Laboratory\\ Los Alamos, New Mexico, 87545, U.S.A.}
\affiliation[n]{Department of Physics, Mississippi State University, // Mississippi State 39762, USA}
\affiliation[o]{Department of Physics and Astronomy, University of Tennessee,\\  Knoxville, TN , 37996, U.S.A.}
\affiliation[p]{Department of Physics, North Carolina State University, \\ Raleigh, North Carolina 27695, U.S.A.}
\affiliation[q]{Department of Nuclear Engineering, North Carolina State University, \\ Raleigh, NC 27695, U.S.A.}
\affiliation[r]{Department of Physics, Simon Fraser University,  \\ 8888 University Drive, Burnaby BC, V5A 1S6, Canada.}
\affiliation[s]{Department of Physics, Tennessee Technological University, \\ Cooteville, Tennessee 38501, U.S.A.}
\affiliation[t]{Department of Physics, Indiana University, \\ Bloomington, Indiana 47408, U.S.A.}
\affiliation[u]{Department of Physics, Yale University, \\ P.O. Box 208120, New Haven, Connecticut 06520-8120, USA}
\affiliation[v]{Department of Physics, Brown University, \\  Providence, Rhode Island 02912, USA}
\affiliation[w]{Lyman Laboratory of Physics, Harvard University, \\ Cambridge, Massachusetts 02138, U.S.A.}
\affiliation[x]{Department of Physics, Valparaiso University, \\ Valparaiso, Indiana 46383}

\emailAdd{bradf@caltech.edu}

\abstract{A cryogenic apparatus is described that enables a new experiment, nEDM@SNS, with a major improvement in sensitivity compared to the existing limit in the search for a neutron Electric Dipole Moment (EDM). This apparatus uses superfluid $^4$He to produce a high density of Ultra-Cold Neutrons (UCN) which are contained in a suitably coated pair of measurement cells. The experiment, to be operated at the Spallation Neutron Source at Oak Ridge National Laboratory, uses polarized $^3$He from an Atomic Beam Source injected into the superfluid $^4$He and transported to the measurement cells where it serves as a co-magnetometer. The superfluid $^4$He is also used as an insulating medium allowing significantly higher electric fields, compared to previous experiments, to be maintained across the measurement cells. These features provide an ultimate statistical uncertainty for the EDM of $2-3\times 10^{-28}$ e-cm, with anticipated systematic uncertainties below this level.}

\begin{document}
\maketitle
\flushbottom

\section{Overview}

\subsection{Introduction}

Because of powerful theoretical motivation, experimental searches for the neutron's electric dipole moment have been underway for more than sixty years, beginning with the pioneering studies of Purcell and Ramsey~\cite{PR50,SPR57}. A recent review of both the experimental and theoretical status of general Electric Dipole Moment (EDM) searches is given in Ref.~\cite{CFRS19}. 

Here, a cryogenic apparatus is described that enables a new experiment, nEDM@SNS, which can provide a significant improvement in sensitivity compared to the existing limit in the search for a neutron Electric Dipole Moment (EDM). This experiment, based on the a concept discussed in Ref.~\cite{GL94}, uses superfluid $^4$He to produce a high density of Ultra-Cold Neutrons (UCN) that are then trapped inside a pair of measurement cells. The experiment, to be operated at the Spallation Neutron Source (SNS) at Oak Ridge National Laboratory, uses polarized $^3$He from an Atomic Beam Source injected into the superfluid $^4$He and transported to the measurement cells as a co-magnetometer. The superfluid $^4$He is also used as an insulating medium allowing significantly higher electric fields compared to previous experiments to be maintained across the measurement cells. These features provide an ultimate statistical uncertainty for the EDM of $2-3\times 10^{-28}$ e-cm, with anticipated systematic uncertainties below this level. The overall method, the performance requirements needed to achieve the sensitivity goals, and the technical details of each component of the apparatus are presented below.  
\subsection{Theoretical Prelude}
The search for a neutron electric dipole moment, $d_n$, has, as its primary goal, the discovery of new physics in the CP violating sector of particle interactions. A focus on CP violation is suggested by the critical importance which this symmetry has assumed in constructing theories of modern particle physics. More broadly, it acknowledges the importance of CP violation in shaping our understanding of the origins and evolution of the universe. In particular, explaining the origin of the baryonic matter of the universe via high sensitivity searches for the neutron EDM (nEDM) and other EDMs is an important goal for nuclear physics as discussed in the recent Long Range Plan for Nuclear Science \cite{LRP15}. While the CP violation present in the Standard Model (SM) suffices to explain what has been observed in the kaon and B meson systems, it is not sufficient to explain the significant excess of baryons over antibaryons in the present universe. The new measurement discussed here, with its substantially greater sensitivity to new CP violation, provides a powerful tool in this quest.

The role of symmetry, including the observed breaking of the discrete symmetries of parity P and CP, has been particularly significant for the construction of the SM. Parity violation, which has been measured in many systems, is well represented in the SM through a definitive chiral V-A coupling of fermions to gauge bosons. The information available on CP violation, while much more limited, has had a profound impact. Indeed, the decay of neutral kaons anticipated the three-generation structure of the SM as we now know it.

Although the deeper reasons for the P and CP violation of the SM have yet to be understood, CP violation is arguably the more mysterious of the two. It occurs in two places within the model: as a complex phase in the Cabibbo-Kobayashi-Maskawa (CKM) matrix that characterizes charge changing weak interactions of quarks, and as an allowed term in the QCD Lagrangian that does not vanish (as in QED) due to the non-abelian nature of the theory. The CP violation observed in the neutral kaon system and in the decays of B mesons is consistent with the presence of the phase factor. On the other hand, present limits on $d_n$ \cite{NEDM15} and the $^{199}$Hg EDM \cite{HG16} imply that the coefficient of the CP violating term in the strong Lagrangian is exceedingly small. In neither case is the strength of the associated CP violation sufficient to explain the observed abundance of baryonic matter over antimatter. Thus, searching for new sources of CP violation has become an attractive focus in the quest for new physics.

\subsection{Basics of EDM Searches}
\label{basics}
A non-zero permanent electric dipole moment (EDM), characterized by $\vec{d} = e\vec{x},$
($e$ is the elementary charge and $\vec x$ is a characteristic displacement) is seen by means of the interaction Hamiltonian%
\begin{equation}
H=-e\vec{x}\cdot\vec{E}\, ,
\end{equation}
where $\vec{E}$ is the electric field.
For a system with fixed spin $\vec{s}$ the matrix elements of any vector operator are
proportional to the spin, $\vec{d} = d\vec{s}/s$, giving
\begin{equation}
H=-\frac{2ex}{\hbar}\left(  \vec{s}\cdot\vec{E}\right)\, ,
\end{equation}
for a system with spin $s=\hbar/2$. In this form, the interaction is seen to violate both parity $\left(
\vec{x}\rightarrow-\vec{x}\right)  $ and time reversal
$\left(  t\rightarrow-t\right)  $ invariance. In addition particles have a
magnetic dipole moment so the total interaction is
\begin{equation}
H=-\frac{2ex}{\hbar}\left(  \vec{s}\cdot\vec{E}\right)
-\gamma\left(  \vec{s}\cdot\vec{B}\right)
\end{equation}
where the magnetic moment is given by $\vec{\mu}=
\gamma\vec{s}$ with $\gamma$ being the gyromagnetic ratio. This
interaction results in the spin precessing with an angular Larmor frequency given by%
\begin{align}
\omega_{\uparrow\uparrow} & =\frac{2ex}{\hbar}|\vec E|+\gamma |\vec B|\label{1}%
\end{align}
where the $\uparrow\uparrow$ indicates that $\vec{E}$ and $\vec{B}$ are parallel.

The most sensitive searches for electric dipole moments involve
looking for a change in Larmor precession frequency, or energy splittings, when an electric
field is reversed relative to a parallel magnetic field, in which case the
change in angular frequency is $\Delta\omega=\omega_{\downarrow\uparrow}-\omega_{\uparrow\uparrow}=\frac{4ex}{\hbar}|\vec E|=\frac{4d}{\hbar}|\vec E|.$ From (\ref{1}), the statistical
uncertainty for a measured EDM is
\begin{equation}
\sigma_d =\frac{\hbar}{4|\vec E|}\sigma_{\Delta\omega}
\end{equation}
where the statistical uncertainty in the measurement of an angular precession frequency  $\sigma_\omega=1/(\tau\sqrt{N})$, with $\tau$ the observation time and $N$ the number of observations. 
This gives
\begin{equation}
\sigma_d=\frac{\sqrt{2}\hbar}{4|\vec E|\tau\sqrt{N}}\label{1a}%
\end{equation}
assuming that $\sigma_{\omega_{\downarrow\uparrow}}=\sigma_{\omega_{\uparrow\uparrow}}$. Thus $G = |\vec E|\tau\sqrt{N}$
$\ $is a figure of merit that can be used to compare different experiments.

\subsection{Concept of the nEDM@SNS Experiment}
\label{sec:NEDM_Concept}
The first step in improving the sensitivity of an EDM search is to maximize the factors contributing to the 
figure of merit, $G,$ above. In order to maximize the observation time $\tau$, most modern neutron EDM experiments use Ultra-Cold Neutrons (UCN). This refers to neutrons
with energies so low, $\lesssim 300$~neV that they are reflected from selected
solid materials for all angles of incidence (total reflection). One way of
producing high densities of UCN is by exposing a volume of liquid $^{4}$He to
an incident neutron beam. Interaction of the cold neutron beam with superfluid He results in `superthermal' production~\cite{GOL-PEN77}, capable of producing very high densities of UCN. At sufficiently low temperatures, $T < 0.7$K, and with the use of appropriate materials for storage cell walls, UCN can be stored in
liquid $^{4}$He for long periods of time~\cite{Brome05} allowing maximization of $N$ and
$\tau$ in equation (\ref{1a}). Liquid $^4$He is also a very good
electrical insulator (see e.g. Ref.~\cite{ITO16} and references therein) so that the E field can also be maximized.

The neutron magnetic moment has a scale of order $\mu_n \approx e c
\lambda_{c}$ where $\lambda_{c}$ is the
Compton wavelength of the neutron.  This length scale of order $\sim 10^{-13}$ cm is to be
compared with the nEDM@SNS sensitivity of some $10^{-28}$ cm. The fact that
the electric field used in previous experiments is effectively $10^{4}$ times
larger than the magnetic field does not come close to compensating for this
discrepancy. The fact that the magnetic interaction is about $10^{11}$ times larger than the sought for electric interaction at the proposed nEDM@SNS sensitivity requires control of
magnetic field fluctuations to an extremely small level. Thus all experiments generally
rely on an extensive system of different magnetometers. These magnetometers should provide the average value of the magnetic field over the volume where the
neutrons are precessing. Early experiments had difficulties using
magnetometers external to the storage region because of fields induced by
leakage currents associated with the high electric field and other sources of
uncontrollable field gradients. It was suggested that it would be better to
use a `co-magnetometer', that is a magnetometer occupying the same volume as
the UCN. This was first used in Ref.~\cite{HARRIS99} where a gas of polarized Hg atoms was
introduced into the UCN storage chamber. To show how subtle these experiments
are, the small displacement $\left(  \sim 1\mbox{~mm}\right)  $ between the centers of
gravity of the Hg atoms and the UCN played an important part in the
experiment as discussed in Ref.~\cite{PEND04}. 

The experiment described here will use polarized $^3$He atoms as a co-magnetometer. Importantly, at low concentrations, these atoms also remain in solution in the liquid $^4$He and retain their polarization for long periods of time. $^{3}$He is a
strong absorber of neutrons, with a very large unpolarized cross section, $\sim 5000$ barns for thermal neutrons, increasing as $\left(  \sim 1/v\right)$ as the velocity decreases. With this large cross section, the fractional concentration of $^{3}$He compatible with a long $\tau$, i.e. comparable to the neutron lifetime, is quite small $\left(  \sim 10^{-10}\right)$. The absorption reaction is%
\begin{equation}
n + {^3He} \rightarrow p+{^3H}+765~\mbox{keV.}\label{1aa}%
\end{equation} 

The energy released from neutron capture is transferred to the liquid helium and produces scintillation light in the extreme
ultra-violet region (80~nm). However, the absorption cross section is strongly spin-dependent, with a much larger capture rate when the neutron and $^3$He spins are anti-parallel. Thus, if both species are precessing together in a magnetic field $B_0$, the scintillation rate varies as $\left( 1-\cos\theta_{n3}\right)$, where $\theta_{n3}$ is the angle between the neutron and $^{3}$He spins, for 100\% UCN and $^3$He polarization.
Since the gyromagnetic ratios of the
$^{3}$He and the neutron ($\gamma_3, \gamma_n$ respectively) are quite close with 
$\gamma_{3}\sim 1.1\gamma
_{n}$, the scintillation rate will vary with an angular beat frequency $\omega^{\prime}=B_0\left(  \gamma_{3}-\gamma
_{n}\right)$. Because the EDM of the $^{3}$He atoms will be shielded by the atomic electrons, the $^3$He EDM is expected
to be $\sim 10^{-5}$ that of the neutron \cite{FLAM07}, and hence unobservable. Thus, one way to perform a neutron EDM measurement is to measure an electric field
dependent shift in the scintillation beat frequency. The neutron precession frequency shift, and hence the neutron EDM, can then be extracted if the precession frequency of the
$^{3}$He can be determined. Since, as discussed below, the density of $^3$He is much larger than the neutron density, SQUID sensors can be used to measure the magnetic field produced by the precessing $^{3}$He magnetic moments.

A second way to perform the measurement involves applying an AC magnetic field which forces the UCN and $^3$He to precess at the same frequency in the absence of an electric field, eliminating the need to directly probe the $^3$He precession frequency. This method, `critical spin dressing' can lead
to a factor of $\sim$ 2 increase in statistical sensitivity in comparison to the method in which SQUIDs are employed. Also by including both methods (spin-dressing and SQUIDS) possible unknown systematic effects can be uncovered, if the two methods do not agree. 

Thus, in this experiment, the polarized $^{3}$He can serve as co-magnetometer and polarization analyzer, resulting in the following design concept: A volume, bounded with good neutron reflecting walls and containing a dilute solution of polarized
$^{3}$He in superfluid $^4$He, is exposed to a neutron beam until the density of UCN
in the helium builds up to an equilibrium value. Parallel electric and magnetic fields are 
applied to the volume and the spins are rotated into a plane perpendicular to the fields. The Larmor precession of both the UCN and $^{3}$He spins along with the spin-dependence of the capture cross section can be used to search for a shift of the neutron's precession frequency correlated with the applied electric field. The capture is measured via the extreme UV scintillation light, monitored by means of a wavelength shifting dye that coats the surface of the walls. This dye must have minimal UCN absorption while maintaining polarization of the UCN and $^3$He. Thus by monitoring the scintillation intensity under different directions of the electric field a value, or upper limit, for the neutron EDM can be extracted. 

A major practical parameter affecting the design of the experiment is the storage time of UCN in the
cell. The longer this is, the longer the effective measuring time and the more
sensitive the experiment. The
$^3$He density can be chosen so that absorption by $^3$He  and $\beta$ decay dominate the neutron loss mechanisms. This is possible if the wall loss rate, due to neutron capture on wall materials, and the UCN upscattering rate, which allows the UCN to escape, are only a small fraction of the neutron $\beta$-decay rate. The polarization of the neutrons and $^3$He are also important
for optimizing the sensitivity of the method. Unpolarized $^3$He atoms reduce the amplitude of the spin-dependent scintillation signal, thus contributing to the background
and decreasing the signal.

The choice of operating frequency (magnitude of the $B_0$ field) is largely
determined by the linear-in-electric-field frequency shift or the so-called geometric phase effect, referred to here as the Bloch-Siegert induced 
false EDM effect (see Sec. \ref{sec:falseEDM}). It will be shown below that the effect for
$^3$He can be reduced by adjusting the collision rate of the $^3$He atoms with
phonons in the liquid $^{4}$He by adjusting the $^3$He mean free path, which is a strong function of temperature at low temperatures. However, the
only handle on the size of the effect for UCN is the operating frequency, since
any collisions would lead to the UCN having their energy increased so much
that they would be lost from the storage vessel. Since the false EDM produced by
this effect scales as $\sim 1/\omega_0^{2}$ ($\omega_0$ being the Larmor frequency in
the field $B_0$), higher operating frequencies are preferred. However both the false EDM effect (see Sec. \ref{sec:falseEDM}) and the spin coherence time, $T_2$ \cite{MCGREGOR90}, scale with the magnetic field gradient which suggests smaller $B_0$ field, assuming that the field gradient scales with $B_0$. Thus the operating frequency can be optimized as discussed below, which for this experiment leads to $B_0\sim 3~\mu$T. 

The choice of operating temperature also requires an optimization. If the temperature is too high ($\gtrsim$ 0.6 K), UCN scattering from phonons results in an energy gain and thus loss of UCN. However, the false EDM effect for $^3$He becomes large when $^3$He-phonon scattering is small (increasing its mean-free path, see Sec. \ref{sec:falseEDM}). This favors operating at higher temperatures ($\gtrsim$ 0.45 K). Fortunately, as discussed below, a satisfactory compromise can be found.

Optimization of electric field strength and uniformity as well as minimization of spin coherence loss leads to measurement cells that are approximately 40 cm long in the neutron beam direction, 7.5 cm wide in the magnetic and electric field direction, and 10 cm high vertically. The use of two cells with opposite electric field (with the HV electrode between the cells) allows for cross checks and reduction in systematic uncertainties. 

\subsection{Overview of Method}

\subsubsection{UCN Production}

UCN can be produced in superfluid $^{4}$He by downscattering of slow
neutrons \cite{GOL-PEN77}. This neutron energy loss occurs by generation of excitations, principally phonons, in the superfluid. Due to 
energy and momentum conservation, neutrons can only come to
rest as the result of a single scattering event if they have momentum $\hbar
q^{\ast}$ and energy $\hbar\omega$ which satisfies%
\begin{equation}
\label{omega_q_neutron}
\omega\left(  q^{\ast}\right)  =\frac{\hbar\left(  q^{\ast}\right)  ^{2}}{2m}%
\end{equation}
where $m$ is the neutron mass. 
Thus neutrons can only be scattered inelastically (i.e. with change of
energy) by single phonons in $^4$He if the dispersion relation for $^4$He - $\omega(q)$ crosses that for the neutron from Eq. \ref{omega_q_neutron}. 
This corresponds to a neutron with about
12~K of kinetic energy or a deBroglie wavelength of 0.89~nm.

The steady state UCN density in a $^4$He filled vessel exposed to a neutron
beam with flux $\Phi\left(  E_{1}\right)  dE_{1\text{ }}$ is given by%
\begin{equation}
\rho_{\scriptscriptstyle UCN}=R\tau_{\scriptscriptstyle UCN}%
\end{equation}
where $\tau_{\scriptscriptstyle UCN}$ is the lifetime of the UCN in the storage vessel including all
possible loss mechanisms and $R$, the production rate per unit volume, is given
by%
\begin{equation}
R=\int r\left(  E_{\scriptscriptstyle UCN}\right)  dE_{\scriptscriptstyle UCN}%
\end{equation}
where%
\begin{equation}
r\left(  E_{\scriptscriptstyle UCN}\right)  =n_{He}\int\Phi\left(  E_{1}\right)  \sigma\left(
E_{1}\rightarrow E_{\scriptscriptstyle UCN}\right)  dE_{1}%
\end{equation}
 with $\sigma\left(  E_{1}\rightarrow E_{\scriptscriptstyle UCN}\right)  dE_{\scriptscriptstyle UCN}$ is
the cross section for scattering from an energy $E_{1}$ to an energy between
$E_{\scriptscriptstyle UCN}$ and $E_{\scriptscriptstyle UCN}+dE_{\scriptscriptstyle UCN}.$ and $n_{He}$ is the number density of liquid helium in the target. Converting neutron energy to deBroglie wavelength, with an incident flux spectrum of $\left(
d\Phi/d\lambda\right)  $ in units of $\left( \mbox{cm}^{-2}\sec^{-1}\mbox{\AA}^{-1}\right)$
this leads to~\cite{GOL-PEN77}, \cite{BAKER03}%

\begin{equation}
R=2.2\times10^{-8}\left(  \frac{d\Phi}{dE}\right)  \mbox{cm}^{-3}\sec^{-1}%
\end{equation}
where the production of UCN up to the cut-off energy for UCN storage of $E^{max}_1 =$ 160~neV is assumed.  

\subsubsection{Detection of Scintillation Light}

Liquid $^4$He is a useful scintillator for detecting ionizing radiation.
It emits broad-band radiation in the extreme ultra-violet centered at 80~nm.
Ultra-violet (UV) radiation in this region is strongly absorbed by practically all materials so
there is no possibility of transmitting it through even the thinnest window. Since the first excited states $\left(  2S,2P\right)  $ in $^4$He are at a
higher energy than the UV photons, the liquid itself is transparent to this radiation. The method of choice for detecting the UV radiation is to coat the walls of
the chamber with a wavelength shifting dye that absorbs the UV and emits
visible light. One of the best materials for this purpose is tetra-phenyl
butadiene (TPB) which emits visible light near
400~nm with a quantum efficiency greater than unity. The TPB consists of a
linear chain of C-H bonds terminated by two benzene rings at each end. Materials
containing H atoms are very bad for UCN storage because the H atoms have a very
large cross section for up-scattering the neutrons. So, instead, one can 
work with deuterated TPB (dTPB) which will result in much smaller up-scattering of the neutrons. The dTPB can be applied by evaporation or by
mixing with a polymer. Using deuterated polystyrene as the polymer has been shown to
produce a smooth transparent coating, while deuterated
polypropylene might produce a surface with both better UCN reflection
probabilities and more efficient UV detection.
In the apparatus described here, 400~nm light emitted by the TPB will be captured by wavelength
shifting fibers attached to one side of the measurement cell and then
transported to a set of Silicon Photomultipliers (SiPM).

\subsubsection{UCN Storage Time}

Several processes contribute to the rate at which UCN are lost. These influence the effective lifetime of UCN which is given by
\begin{equation}
\label{tauucn}
\frac{1}{\tau_{\scriptscriptstyle UCN}}=\frac{1}{\tau_{cell}}+\frac{1}{\tau_{\beta}}+\frac
{1}{\tau_{up}}+\frac{1}{\tau_{3}}%
\end{equation}
where $1/\tau_{cell}$ is the loss rate due to material wall interactions and cracks in the measurement cell,
$1/\tau_{\beta}$ is the beta decay loss rate ($\tau_{\beta}
=880$~sec~\cite{PDG18}) , $1/\tau_{up}$ is the loss rate due to upscattering of the
UCN by excitations in the superfluid (phonons and rotons) and $1/\tau_3$ is the $^3$He-neutron 
capture rate.

The storage properties of a UCN container can be characterized by the phenomenological parameter - $\tau_{cell}$. The UCN wall interaction can be described by the
solution of the Schr\"odinger equation for a one dimensional potential barrier
(see Ref.~\cite{UCNBOOK}). 
During reflection, the wave
function penetrates into the material as an evanescent wave and thus suffers from absorption by the nuclei of the wall material as well as upscattering by the
thermal motion of the wall atoms. If the energy of the UCN is increased as a result of each  interaction with the wall, the probability that it will be 
reflected during each subsequent wall encounter decreases and eventually it will be lost from the system. This loss process can be characterized by a loss probability
per bounce; $f\left(  E_{\scriptscriptstyle UCN}\right) $. In general, the number of wall
collisions per second is given by $\left( vA/4V\right)  $ where $v$ is the
UCN\ velocity, $A$ the area of the storage chamber, and $V$ its volume. Thus
the contribution of wall losses to the storage time is a function of the UCN energy and given by
\begin{equation}
\frac{1}{\tau_{cell}}=f\left(  E_{\scriptscriptstyle UCN}\right)  \left(  \frac{vA}{4V}\right).
\end{equation}
The goal for the experiment is to have neutron cell wall loss times on the order of 2,000~s
with walls coated with materials capable of detecting the UV scintillation, as discussed above.

The lowest order neutron-phonon upscattering timscale, single phonon absorption, is strongly suppressed by a Boltzmann factor ($e^{-E^{\ast}/T}$) due to energy
and momentum conservation since it requires that it occurs only for phonons with energies $E^{\ast
}=12$~K, much higher than the operating temperature of less than $0.5$~K. Thus the
dominant process is then multi-phonon scattering which is shown \cite{GOLUB83} to result in
\begin{equation}
\frac{1}{\tau_{up}}=\frac{\left [T(\mbox{K})\right ]^7}{100~\mbox{s}}.
\end{equation}
For $T\lesssim 0.5$~K this is a small contribution relative to the other terms in Eq.~\ref{tauucn} 

The rate at which UCN are absorbed by $^{3}$He depends on the time-dependent angle $\theta_{n3}$ between
the magnetic moments. This leads to 
\begin{equation}
\frac{1}{\tau_{3}(t)}=\left (\frac{1}{\bar{\tau}_3}\right )\left(  1-P_{n}P_{3}\cos\theta
_{n3}\left(  t\right)  \right) \label{1aaa}%
\end{equation}
where $P_{n},P_3$ are the polarizations of the UCN and $^{3}$He respectively and $\bar{\tau}_3$ is the mean unpolarized $^3$He-n
absorption time
\begin{equation}
\frac{1}{\bar{\tau}_3}=N_{3}\sigma_{abs}v=2.4\times
10^{7}X_{3}\sec^{-1}%
\end{equation}
where $N_3$ is the $^3$He density, $\sigma_{abs}$ is the unpolarized thermal capture cross section,
$v$ is the UCN mean velocity and $X_{3}$ is the fractional concentration of $^{3}$He in the liquid
$^{4}$He. For $X_{3}=10^{-10}$ one finds $\bar{\tau}_3\sim 400$~s
which is a reasonable value for the proposed experiment. The optimized value, based on minimizing the statistical uncertainty, will be discussed below. 

The contribution of neutron capture and beta-decay to the time-dependent scintillation rate is 
\begin{equation}\label{bb00}
\dot N\left(  t\right)  =N_{UCN}(t)\left [\frac{1}{\tau_{3}\left(  t\right)} +\frac{1}{\tau_\beta}\right ]%
\end{equation}
where $N_{UCN}(t)$ can be obtained by integrating%
\begin{equation}\label{overview9}
\frac{dN_{UCN}(t)}{dt} =-\frac{N_{UCN}(t)}{\tau_{UCN}\left(  t\right)  }
\end{equation}
to give
\begin{equation}\label{overview10}
N_{UCN}\left(  t\right)  =N_0\exp\left[  -\left(  \frac{1}{\tau_{cell}
}+\frac{1}{\tau_{\beta}}+\frac{1}{\tau_{up}}+\frac{1}{\bar{\tau}_3}\right)
t+\frac{P_{n}P_{3}}{\bar{\tau}_3}\int_0^t\cos\theta_{n3}\left(  t'\right)  dt'\right]
\end{equation}
where $N_0$ is the initial number of UCN. Note that the scintillation signal from beta-decay reduces the 
sensitivity to the EDM. As the energy distribution of the scintillation signal is different for neutron-$^3$He capture compared to neutron beta-decay, this effect can be mitigated due with an energy analysis window having different detection efficiencies, $\epsilon_3, \ \epsilon_\beta$ respectively. This is discussed in Sec.~\ref{sec:Stat}.

\subsubsection{$^{3}$He Polarization}
\label{sec:polarization}
With a $^3$He concentration of $X_{3}=10^{-10}$ and two measurement
cells of 3 liters volume each, a total of $1.2\times10^{16}$ $^{3}$He atoms are required, which amounts to less than 1~torr-cm$^{3}$ at room temperature. As this experiment requires a high $^{3}$He
polarization, but very few $^{3}$He atoms,
a polarized atomic beam source (see e.g.~\cite{ABS}) has been chosen. This device, presently under construction, uses
inhomogeneous magnetic fields to focus one polarization state and defocus the other
one, giving rise to final state polarizations approaching 100\%. The device will be capable of producing about $10^{14}$ polarized $^{3}$He~atoms/s so that the number of atoms required to fill the measurement cells can be collected in a
filling in a time of about 160~s, which is much shorter than the anticipated measurement cycle time. The polarized $^{3}$He beam will impinge vertically
on a volume of liquid $^4$He called the injection volume. From there it will
be directed to the measurement cell by applying temperature gradients which
induce phonons that carry the $^{3}$He along in a manner described in Sec.~\ref{sec:He3S}.

\subsubsection{Detection of an Electric Dipole Moment}

As already mentioned, the experiment anticipates using two different methods to search for a
neutron EDM with a $^{3}$He co-magnetometer: one method involves direct detection of free $^3$He precession using SQUIDs while the other involves an experimental technique know as critical dressing.
At the start of each measurement cycle, both the UCN and $^{3}$He spins will initially be polarized along the magnetic
field ($B_0$) direction. By a suitable pulse ($\pi/2$ pulse) of alternating (AC) magnetic field perpendicular to $B_0$, both spins can be brought into the plane perpendicular to the $B_0$ field. Following this pulse the two detection methods have different approaches for measuring the neutron precession frequency.

\paragraph{Free precession method}

For the free precession method, the UCN and $^3$He spins will each precess independently at their own Larmor frequency, $\omega_{i}=\gamma_{i}B_0$ where $i=n$(UCN) or 3($^{3}$He). But due to the slightly different precession frequencies (recall that $\gamma_3/\gamma_n\sim 1.1$) the scintillation angular frequency will be given by
\begin{equation}\
\label{omega_def}
\omega=\left(  \gamma_{3}-\gamma_{n}\right)  \frac{\omega_{3}}{\gamma_{3}%
}\mp 2\frac{exE}{\hbar}%
\end{equation}
where the minus sign in the second term applies when the magnetic and electric fields are parallel. To extract the value of the neutron EDM $d(=ex)$ from $\omega$, $\omega_3$ will be measured directly by a system of SQUID sensors.

\paragraph{Critical Dressing method}

If a spin undergoing Larmor precession in a field $\vec{B}_0$ is
subjected to an alternating field perpendicular to $\vec{B}_0,$ the time-averaged
Larmor precession is modified so that the effective precession frequency is given by
$\omega_i=\gamma_{i,eff}B_0,$ where
\begin{align}
\gamma_{i,eff}  & =\gamma_{i}J_0\left(  x_{i}\right) \\
x_{i}  & =\gamma_{i}\frac{B_{1}}{\omega_{1}}%
\end{align}
where $B_{1}$ and $\omega_{1}$ are the magnitude and frequency of the applied
dressing field respectively (see e.g. Ref.~\cite{CT69}) and $J_0$ is the zeroth order Bessel function of the first kind. Using this technique, a value of $\frac{B_{1}%
}{\omega_{1}}$ can be chosen so that the UCN and $^{3}$He spins precess at the same time averaged rate in
the absence of a neutron EDM by requiring that%
\begin{equation}
\gamma_{n}J_0\left(  x_{n}\right)  -\gamma_{3}J_0\left(  \frac{\gamma_{3}%
}{\gamma_{n}}x_{n}\right)  =0.
\end{equation}
This equation has a first solution at $x_{n}\approx1.2.$ When this condition is satisfied, a state known as critical dressing is achieved and the two spin species will precess together at the same rate ($\dot{\theta
}_{n3}=0$). Moreover, if the spins are arranged to be perpendicular, the scintillation rate
will be constant and equal to one-half of the maximum rate, and will be maximally sensitive to changes in
$\theta_{n3}.$ If the two species are perfectly polarized in the same direction, the
scintillation (absorption) rate will be effectively zero.

In the presence of a non-zero EDM, the scintillation rate is not constant, with the neutron and $^3$He spins changing as   $\theta_{n3}=\phi_0
\pm\left(  2d'E/\hbar\right)  t$, where $\phi_0$ is the initial angle between the spins
and $d^{\prime}=dJ_0\left(  x_{n}\right)  $ is the effective electric dipole
moment ($d=|\vec d|$) under critical dressing conditions. Thus there will be a small change in the
scintillation rate due to the change in the $\cos\theta_{n3}\left(
t\right)$ term in equation (\ref{1aaa}). The angle $\phi_0$ can be chosen to optimize the sensitivity (see Sec.~\ref{Sec:CritStat}). In particular, if
the dressing parameter $x_{n}$ is modulated such that
\begin{equation}\label{bb01}
\theta_{n3}\left(  t\right)  =\phi_0+a_{m}\sin\omega_{m}t\pm\frac
{2d^{\prime}E}{\hbar}t,
\end{equation}
then the scintillation rate will be proportional to
\begin{equation}
\left[  \theta_{n3}\left(  t\right)  \right]  ^{2}=\frac {\left(  a_{m}%
\sin\omega_{m}t\right)  ^{2}}{2}\pm\left(  \phi_0+\frac{2d^{\prime}E}{\hbar
}t\right)  a_{m}\sin\omega_{m}t+\frac{1}{2}\left(  \phi_0\pm\frac{2d^{\prime}E}{\hbar
}t\right)  ^{2}.%
\end{equation}
Thus the
scintillation signal will involve a term proportional to the first harmonic of the modulation
frequency that grows linearly with time and is  proportional to the neutron EDM. The second harmonic term can be useful as a monitor of the system parameters. Other forms of critical dressing modulation are possible and will be discussed below.

\section{Sensitivity Reach of nEDM@SNS}

The ultimate sensitivity of a neutron EDM search is set by the statistical and systematic uncertainties of the experiment. For the proposed nEDM@SNS search the main improvements to the statistical sensitivity of the experiment, compared to the previous room temperature measurements, come from higher electric fields because the electrodes are immersed in liquid He, higher number of trapped neutrons because of the phonon down-scattering scattering of cold neutrons in liquid $^4$He and increased observation time because of longer neutron storage times. 

For the systematic uncertainties, providing experimental controls on key system parameters, e.g. leakage currents, magnetic and electric field non-uniformities, accounting for magnetic field fluctuations, etc is essential. In addition, the experiment should allow variation of key parameters and multiple measurement options to provide opportunities to search for systematically-induced false EDM signals. 
For the experiment discussed here, two techniques will be used to measure the neutron EDM: free precession of neutrons as measured by the time dependence of the n$-^3$He capture rate as well as the critical spin dressing conditions. As these two techniques have different sensitivities to systematic effects, comparison of the two results can help constrain possible unknown systematic effects. To limit the known systematic effects key experimental parameters must be controlled. The  experimental controls and measurements that influence the systematic uncertainty along with the main parameters that influence the statistical uncertainty for both techniques will be discussed below. 

\subsection{Statistical Uncertainty}
\label{sec:Stat}

The statistical uncertainty in the extracted neutron EDM arises from fluctuations in the number of observed scintillation events. The observed scintillation rate from neutron capture on $^3$He and neutron beta-decay can be determined from Eqs.~\ref{bb00} and ~\ref{overview10}. However,
The time dependence of the scintillation rate  $\dot{N}(t)$ is clearly different for the two anticipated experimental modes, and so free precession and critical dressing will be addressed separately. For free precession the
second term on the right-hand-side of 
Eq.~\ref{overview10} produces a small initial transient which will be ignored, as it otherwise averages to zero, giving
\begin{equation}\label{free-rate}
\text{Free Precession:} \\ \, \, \, \dot{N}(t) = \frac{\epsilon_\beta N_0}{\tau_\beta}e^{-t/\tau}+ \frac{\epsilon_3 N_0}{\bar{\tau}_3}\left [ 1-P_nP_3\cos\theta_{n3}(t)\right ]e^{-t/\tau}+\dot{N}_B
\end{equation}
where $N_0$ is the initial number of stored neutrons, $\epsilon_\beta$ and $\epsilon_3$ are the total detection efficiencies for $\beta$-decay and neutron capture by $^3$He, $P_3$ and $P_n$ are the $^3$He and neutron polarizations, which have their own time dependence, and the average neutron loss rate $1/\tau$ is given by 
\begin{equation}
\frac{1}{\tau}=\left (
\frac{1}{\tau_\beta}+\frac{1}{\bar{\tau}_3}+\frac{1}{\tau_{\rm cell}}+\frac{1}{\tau_{\rm up}}\right ).
\end{equation}
Equation~\ref{free-rate} includes a time-independent background $\dot N_B$ caused by ambient ionizing radiation incident on the cells. Of course there will also be some time-dependent background count rate caused by neutron activation which can be separately included in the fitting procedure. Monte Carlo calculations suggest that neutron shielding (i.e. Li-loaded plastic) can significantly reduce these backgrounds for the nEDM@SNS apparatus. There is minimal impact on the EDM sensitivity if these backgrounds can be reduced to $< 5 $ s$^{-1}$ compared with typical n$-^3$He capture rates of >500 s$^{-1}$ anticipated for nEDM@SNS.

The separate efficiencies $\epsilon_\beta, \epsilon_3$ are included because of the different scintillation energy distributions for n$-^3$He capture and neutron beta decay. With appropriate analysis cuts on the pulse height of scintillation light it is possible to reduce the contribution of the beta decay signal (which dilutes the EDM sensitivity). This is described in more detail in Sec.~\ref{light_coll_req}.

For the critical dressing measurement the second term on the right-hand-side of Eq.~\ref{overview10} is nominally constant which leads to
\begin{equation}\label{crit-rate}
\text{Critical Dressing:} \\ \, \, \, \dot{N}(t) = \frac{\epsilon_\beta N_0}{\tau_\beta}e^{-\Gamma t}+ \frac{\epsilon_3 N_0}{\bar{\tau}_3}\left [ 1-P_n P_3 cos\theta_{n3}\right ]e^{-\Gamma t}+\dot{N}_B
\end{equation}
In this case, the average neutron loss rate is given by 
\begin{equation}\label{crit-tau}
\Gamma =\frac{1-P_n P_3 \cos\phi_0}{\bar{\tau}_3}
+\frac{1}{\tau_\beta}+\frac{1}{\tau_{cell}}+\frac{1}{\tau_{up}}
\end{equation}
where $\phi_0$ is the now nominally constant relative angle between the neutron and $^3$He spins (see Sec.~\ref{Sec:CritStat}). Here again a constant background rate $\dot N_B$ has been included and the detection efficiencies $\epsilon_n, \epsilon_3$.
Eqs. \ref{free-rate}-\ref{crit-tau} are the basis for the sensitivity estimates presented below. 

\subsubsection{Statistical Uncertainty: Free Precession Mode}

For the free precession measurement technique, the statistical sensitivity for the extraction of the neutron EDM is based on a fit to the decaying oscillation signal parametrized by Eq. \ref{free-rate}. Rewriting this equation gives

\begin{align}\label{bb1}
\dot{N} (t) =& \dot{I}_0e^{-t/\tau}\left [1-F\cos{(\theta_{n3})}
\right ]+\dot{N}_B\\
=& \dot{I}_0e^{-t/\tau}\left [1-F\cos{(\omega t+\phi_0 )}
\right ]+\dot{N}_B\notag
\end{align}
with
\begin{equation}
\dot{I}_0=N_0\left (\frac{\epsilon_\beta}{\tau_\beta}+\frac{\epsilon_3}{\bar{\tau}_3}\right );\ \  F=\frac{\epsilon_3
P_3P_n}{\bar{\tau}_3\left (\frac{\epsilon_\beta}{\tau_\beta}+\frac{\epsilon_3}{\bar{\tau}_3}\right )}\label{bb2}
\end{equation}
where $\omega$ is defined in Eq.~\ref{omega_def} and $\phi_0$ is, again, the initial angle between the neutron and $^3$He spins following the $\pi/2$ spin-flip pulse. For purposes of estimating statistical sensitivities $F$ will be assumed to be independent of time which is a reasonable approximation since the polarization decay times are expected to be much longer than $T_m$, the time for a single measurement cycle. 

The uncertainty in the inferred EDM is then dominated by the uncertainty in the neutron angular precession frequency $\omega_n$ which is extracted from $\omega$, using $\omega_3$ determined from the SQUID monitors. As the SQUID noise (discussed below) leads to an uncertainty in angular frequency that is significantly less than the statistical uncertainty in $\omega$ from a given measurement cycle it should contribute negligibly to the uncertainty in the EDM. As discussed in 
Sec.~\ref{basics}, the neutron EDM is determined from the difference in neutron frequency for parallel $\omega_{\uparrow\uparrow}$ and anti-parallel $\omega_{\uparrow\downarrow}$ electric and magnetic fields via

\begin{equation}
d_n=\frac{\hbar\left ( \omega_{\downarrow\uparrow}-\omega_{\uparrow\uparrow}\right )}{4|\vec{E}|}\label{bb3}
\end{equation}
and the uncertainty in the EDM is then given by
\begin{equation}
\sigma_d=\frac{\sqrt{2}\hbar\sigma_{\omega}}{4|\vec{E}|}\label{bb4}
\end{equation}
where $\sigma_\omega$ is the uncertainty in the angular frequency measurement for a single configuration of electric and magnetic fields (assumed identical for both configurations). 

A simple estimate of the uncertainty in frequency can be made by assuming that the amplitude of the decaying exponential $\dot{N}_0$, the oscillatory amplitude $F$ and the constant background $\dot{N}_B$ are largely uncorrelated with both $\omega$ and $\phi$. This assumption has been confirmed in detailed Monte Carlo simulations. In this case the uncertainty in $\omega$ can be determined from the error/covariance matrix (see e.g. Ref.~\cite{EADIE71}) for $\omega$ and $\phi$ alone. 

The error matrix can be estimated using the least-squares function $\chi^2$ defined in terms of the observed number of counts $y_i(t_i)$ in time bin $t_i$ with width $\Delta t$, the  estimated variance of these counts $\sigma_i^2$ and the fitting function $\dot{N}(\beta_i,t_i)$ with fit parameters $\beta_i$ via
 
\begin{align}
\chi^2 =&\sum^n_{i=1}\left\{\frac{y_i(t_i)-\dot{N}(\beta_i,t_i)\Delta t}{\sigma_i}
\right\}^2.\label{bb5}
\end{align}
In this case the error matrix can be determined from the inverse of the curvature matrix (see e.g. Ref.~\cite{BEVINGTON69}) which is given by

\begin{align}
\alpha_{ij}=&\frac{1}{2}\frac{\partial^2\chi^2}{\partial\beta_i\partial\beta_j}\label{bb6}
\end{align}

For the case of only two free parameters and with $\beta_1 = \omega,\ \beta_2=\phi_0$, the components of the curvature matrix $\alpha_{ij}$ can be approximated as

\begin{align}\label{bb7}
\alpha_{11}&\simeq\frac{1}{2}\int^{T_m}_0\dot{N}_0e^{-t/\tau}F^2t^2dt=\frac{\dot{N}_0F^2}{2}[2\tau^3-(2\tau^3
+2\tau^2T_m+\tau T^2_m)e^{-T_m/\tau}]\notag\\
\alpha_{12}&=\alpha_{21}\simeq\frac{1}{2}\int^{T_m}_0\dot{N}_0e^{-t/\tau}F^2tdt=\frac{\dot{N}_0F^2}{2}[\tau^2-(\tau^2+\tau T_m)
e^{-T_m/\tau}]\\
\alpha_{22}&\simeq\frac{1}{2}\int^{T_m}_0\dot{N}_0e^{-t/\tau}F^2dt=\frac{\dot{N}_0F^2\tau}{2}[1-e^{-T_m/\tau}]
\notag
\end{align}
where $T_m$ is the total measurement time. The error matrix is the inverse of the curvature matrix and the variance of the two parameters are the diagonal elements of this matrix. Thus the variance of $\omega$ and $\phi_0$ are given by

\begin{align}
\label{bb8}
\sigma^2_\omega =&\frac{\alpha_{22}}{\alpha_{11}\alpha_{22}-\alpha^2_{12}}=\left (\frac{2}{\dot{N}_0F^2}\right )\frac{(1-e^{-T_m/\tau})}{\tau^3(1+e^{-2T_m/\tau})
-(2\tau^3+\tau T^2_m)e^{-T_m/\tau}}\\\label{bb9}
\sigma^2_{\phi_0} =&\frac{\alpha_{11}}{\alpha_{11}\alpha_{22}-\alpha^2_{12}}=\left (\frac{2}{\dot{N}_0F^2}\right )\frac{2\tau^2-(2\tau^2+2\tau T_m+T^2_m)e^{-T_m/\tau}}
{\tau^3(1+e^{-2T_m/\tau})-(2\tau^3+\tau T^2_m)e^{-T_m/\tau}}
\end{align}
Eqs.~\ref{bb8}, and~\ref{bb9} can be simplified in the limits $\tau<<T_m$ and $\tau>>T_m$ giving 
\begin{align}\label{2parmerr} 
\sigma^2_\omega = \frac{2}{\dot{N_0}F^2\tau^3} & ;\ \ \ \sigma^2_{\phi_0} = \frac{4}{\dot{N_0}F^2\tau}; \ \ \ \tau<<T_m\\\label{bb95}
\sigma^2_\omega = \frac{24}{\dot{N_0}F^2T^{3}_m}& ;\ \ \ \sigma^2_{\phi_0} = \frac{8}{\dot{N_0}F^2T_m}; \ \ \ \tau>>T_m\notag
\end{align}
Note that the result for $\sigma_\omega^2$ for $\tau>>T_m$ is consistent with Ref.~\cite{Chibane95}. 

It should be clear that the two parameters $\omega$ and $\phi_0$ are highly correlated, such that if one of them is known, a priori, then the variance in the other is reduced. For example, if there is sufficient knowledge of $\phi_0$ from knowledge of the $\pi/2$ pulse sequence then the error correlation matrix has only one element, resulting in 
\begin{equation}\label{bb10}
\sigma^2_\omega =\frac{1}{\alpha_{11}}=\frac{2}{\dot{N_0}F^2\left [2\tau^3-(2\tau^3+ 2\tau^2T_m+\tau T^2_m)
e^{-T_m/\tau}\right ]}
\end{equation}
Eq.~\ref{bb10} gives a smaller variance than Eq. \ref{bb8} because of the removal of the strongly correlated phase parameter as can easily be seen in the two limits:
\begin{equation}\label{bb105}
\sigma^2_\omega = \frac{1}{\dot{N_0}F^2\tau^3}  \ \ \ \text{for} \ \ \ \tau<<T_m \ \ \ \text{and} \ \ \ \sigma^2_\omega = \frac{6}{\dot{N_0}F^2T^{3}_m} \ \ \ \text{for}\ \ \ \tau>>T_m.
\end{equation}

In order to keep the uncertainty in the phase angle negligible, the initial phase angle must be reproducibly set to a level smaller than the statistical uncertainty in Eq.~\ref{2parmerr}. For the parameters discussed below, this requires $\sigma_\phi < \sqrt{2}\tau\sigma_\omega$ or $\sigma_\phi< $1 mrad, which represents a modest fractional reproducibility for the holding magnetic field $B_0$ and the $\pi/2$ pulse magnetic field and frequency of $\pm 10^{-3}$ (see below). In addition, there exist many pulse sequences whose spin-rotation results are robust against variations in the pulse parameters. Thus Eq.~\ref{bb10} can be used to estimate the statistical sensitivity. 

The nEDM@SNS experiment has adopted a set of design goals for the key parameters that influence the statistical sensitivity. These values are listed in Table \ref{tab:mag3}.  These parameters can used to evaluate Eq. \ref{bb10} and give e.g. $\sigma_\nu=\frac{\sigma_\omega}{2\pi} = 1.7 \mu$Hz for a single cell and a single measurement cycle. Thus in order to apply the single parameter estimate of sensitivity $\sigma_\phi$ needs to be significantly smaller than 5 mrad which justifies the above requirement of setting the initial phase to $\sim 1$ mrad. The estimate for experimental sensitivity for a total live time $T_L$ is then
\begin{equation}\label{bb11}
\sigma^{TOT}_d =\frac{\sigma_d}{\sqrt{m_{cycle}}}=\sigma_d\sqrt{\frac{T_f+T_m+T_d}{T_L}}
\end{equation}
where $T_{f}$ is the time to load UCN into the measurement cell and $T_{d}$ is the time to remove 
depolarized $^3$He and replace with highly polarized $^3$He and $m_{cycle}$ is the total number of measurement cycles. In this expression $\sigma_d$ is the uncertainty from a single measurement cell. 

Note that there are three parameters that have been chosen to optimize the total statistical sensitivity: $\tau, T_m$ and $T_f$. Using Eq. \ref{bb10}, the parameters of Table \ref{tab:mag3} and assuming a total live time of 300 days (which can be achieved after three calendar years of running), provides a 1~$\sigma$ statistical sensitivity from a free precession measurement of 

\begin{equation}\label{bb12}
\sigma^{TOT}_d= 3.3\times 10^{-28} \ \text{e-cm}.
\end{equation}

This sensitivity assumes that the magnetic field is stable at the level of 1 part in $10^7$ during the measurement time. While this is the goal for the experiment (with both room temperature and cryogenic magnetic shielding and a superconducting magnetic field coil operated in persistent mode), the $^3$He precession signal (discussed in Sec.~\ref{SQUIDS}) can likely reduce this requirement. Also note that the critical spin dressing mode of measurement does not have such a magnetic field stability requirement.

\begin{table}
\hspace*{-2cm}
\vspace*{-.5cm}
\center
\includegraphics[width=6.5in]{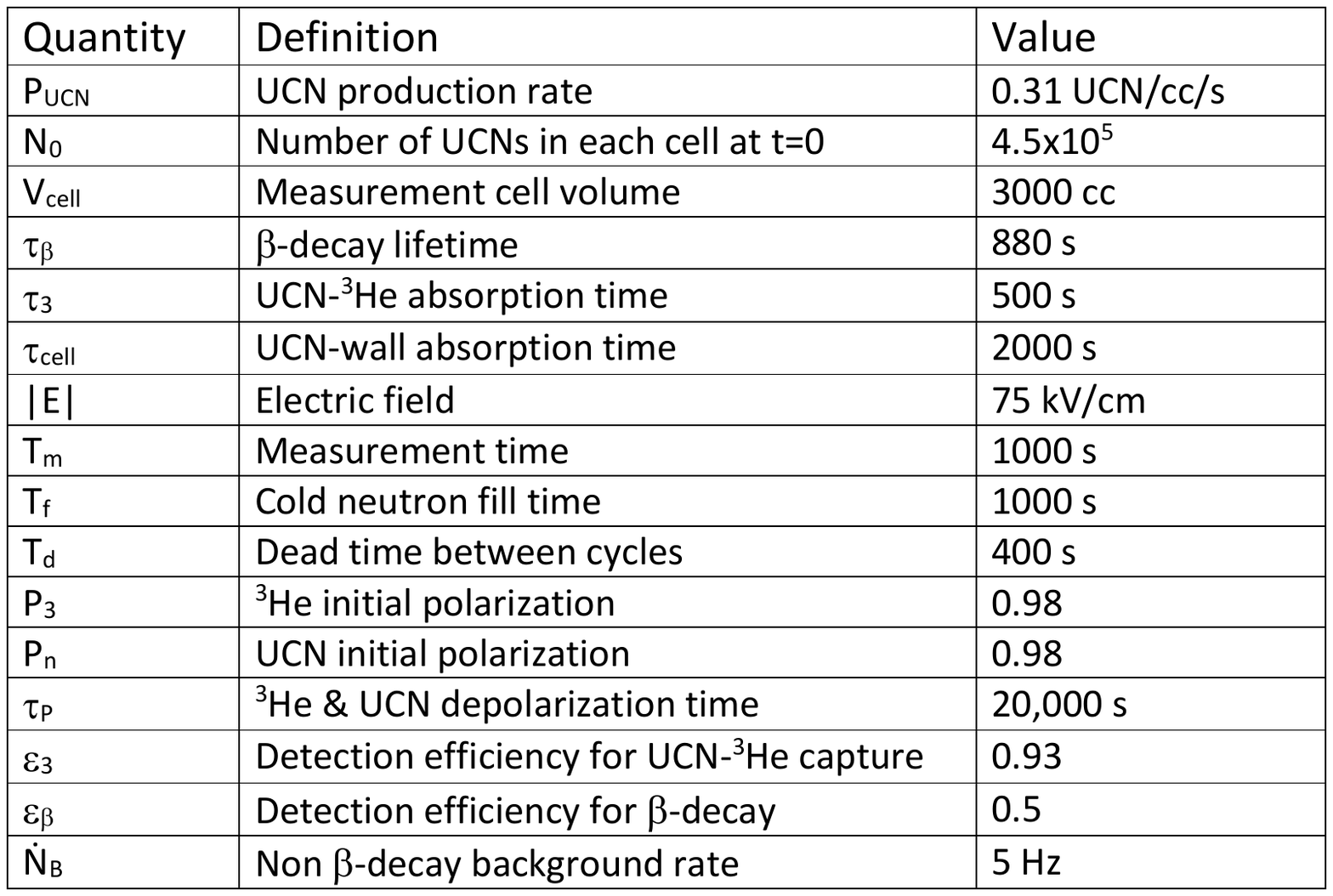}
\vskip -4.5in
\caption{Experimental design goals for the key parameters that influence the statistical uncertainty} 
\label{tab:mag3}
\end{table}
\subsubsection{Statistical Uncertainty: Critical Spin Dressing Mode}
\label{Sec:CritStat}

To estimate the statistical sensitivity for the critical spin dressing technique one can construct a rate asymmetry from which it is easy to 
estimate the statistical uncertainty. To do this, consider the case where the phase angle between the neutron and $^3$He spins in the plane perpendicular to the $B_0$ field is set to an 
initial value $\phi_0$ and the dressing parameter $x_n$ is set to the critical value such that, in the absence of an EDM or electric field, the angle between the two remains fixed and hence the scintillation rate is constant. After  
counting the scintillation events for some time $\Delta t$, the angle between the neutron and $^3$He is quickly changed to -$\phi_0$, again with the dressing parameter 
set to the critical value ($x_n=x_c$), and the scintillation events are counted for time $\Delta t$. Since the angle between the neutron and $^3$He spin direction has been changed from $\phi_0$ to $-\phi_0$ there is no change in the scintillation rate (again in the absence of an EDM or electric field) assuming that the times are kept short so that the losses of UCN and polarization between the two periods can be neglected. 

There are a variety of possible types of modulation, where the angle between the neutron and $^3$He spin directions are changed periodically from $+\phi_0$ to $-\phi_0$ (sine, sawtooth, square wave). It is easy to show for the case where counting statistics are dominant (the shot-noise limit) that modulation (as just outlined), with a minimum time to change the angle compared to the counting time, has higher statistical power than harmonic modulation (as discussed in relation to Eq.~\ref{bb01}).
This is primarily because the time when the count rate is low, with a small phase angle between neutron and polarized $^3$He nuclear spins, is minimized. 

Now, assuming that there is an EDM - $d_n$ - and an electric field $|\vec E|$ then, using Eq.~\ref{bb01}, the angle between the spins is
\begin{equation}
\theta_{n3}(t) =\pm\phi_0+\tfrac{2J_0(x_c)d_n|\vec E|t}{\hbar}
\end{equation}
which, for a small EDM, gives 
\begin{equation}
\cos\theta_{n3}\simeq\cos\phi_0\mp\frac{2J_0(x_c)d_n|\vec E|t\sin\phi_0}{\hbar}.
\end{equation}
Writing the total number of detected counts for the two phases of the cycle as $N_+$ and $N_-$ for the first and second periods respectively at time $t_i$ and using Eq. \ref{bb1} (ignoring the small change in rate due to the $\Gamma$ factor) gives
\begin{align}
N_\pm (t_i) =&\left (\frac{N_0\epsilon_\beta}
{\tau_\beta}\right )e^{-\Gamma t_i}{\Delta t}+\dot{N}_B\Delta t\\
+&\left (\frac{N_0
\epsilon_3}{\bar{\tau}_3}\right )\left (1-P\cos\phi_0 \mp\ \frac{2J_0(x_c)d_n|\vec E|t_iP\sin\phi_0}{\hbar}\right )e^{-\Gamma t_i}\Delta t
\end{align}
where
\begin{equation}
P=P_3P_n\ \ \text{and}\ \ \ \Gamma =\frac{1-P\cos\phi_0}{\bar{\tau}_3}
+\frac{1}{\tau_\beta}+\frac{1}{\tau_{cell}}+\frac{1}{\tau_{up}}
\end{equation}
and ignoring the decay in the polarizations. 
The time-dependent EDM asymmetry can now be formed as
\begin{align}
A_d(t_i)=&\frac{N_--N_+}{N_-+N_+}\\
 =&\frac{\left (\frac{N_0\epsilon_3}
{\bar{\tau}_3}\right )\left [\frac{2J_0(x_c)d_n|\vec E|t_iP\sin
\phi_0}{\hbar}\right ]e^{-\Gamma t_i}}{\left (\frac{N_0
\epsilon_\beta}{\tau_\beta}\right )e^{-\Gamma t_i}+\left (
\frac{N_0\epsilon_3}{\bar{\tau}_3}\right )(1-P\cos\phi_0)e^{-\Gamma
t_i}+\dot{N}_B}
\end{align}
Since the variance in the asymmetry is given by 
\begin{equation}
\sigma_A^2=\frac{1}{N_-+N_+}
\end{equation}
when $A_d << 1$
and since $d_n$ is proportional to $A_d$, the fractional variance in the EDM is
\begin{equation}
\left (\frac{\sigma_d}{d_n}\right )^2 =\left (\frac{\sigma_A}{A}\right )^2
=\frac{N_++N_-}{(N_--N_+)^2}.
\end{equation}
This provides the time-dependent variance in each measurement of the asymmetry
\begin{equation}
\sigma^2_d(t_i)=\frac{\left (
\frac{N_0\epsilon_\beta}{\tau_\beta}\right )e^{-\Gamma t_i}+\left (
\frac{N_0\epsilon_3}{\bar{\tau}_3}\right )(1-P\cos\phi_0)e^{-\Gamma t_i}+\dot{N}_B}
{2\left (\frac
{N_0\epsilon_3}{\bar{\tau}_3}\right )^2\left (\frac{2J_0(x_c)|\vec{E}|t_iP\sin\phi_0}{\hbar}\right )^2e^{-2\Gamma t_i}
\Delta t}
\end{equation}
The total uncertainty in the EDM from this method for a single measurement cycle is then given by the weighted average of 
all of the individual measurements at times $t_i$:
\begin{equation}
\sigma^{\rm tot}_d=\left [\frac{1}{\sum\frac{1}{\sigma^2_d
(t_i)}}\right ]^{\frac{1}{2}} =\left\{\sum\frac{2\left (
\frac{2J_0|E|}{\hbar}\right )^2\left (\frac{N_0\epsilon_3}{\bar{\tau}_3}
\right )^2\sin^2\phi_0[P^2t^2_ie^{-\Gamma t_i}\Delta t]}
{\dot{N}_B e^{\Gamma t}+N_0\left [\frac{\epsilon_\beta}{\tau_\beta}
+\frac{\epsilon_3}{\bar{\tau}_3}(1-P\cos\phi_0)\right ]}
\right\}^{-\frac{1}{2}}.\\
\end{equation}
For sufficiently short modulation times compared to the total measurement cycle time $T$, the sum can be turned into an integral (noting that the period of the modulation waveform is $dt=2\Delta t$) giving
\begin{equation}
\sigma^{\rm tot}_d=\frac{\hbar \bar{\tau}_3}
{\left (2J_0|E|N_0
\epsilon_3\sin\phi_0\right )
\left\{
\displaystyle{\int^T_0}
\frac{P^2t^2e^{-\Gamma t}dt}{\dot{N}_Be^{\Gamma t}+N_0\left [
\frac{\epsilon_\beta}{\tau_\beta}+\frac{\epsilon_3}{\bar{\tau}_3}
(1-P\cos\phi_0)\right ]}
\right \}^{\frac{1}{2}}}.
\end{equation}
Using the parameters listed in Table \ref{tab:mag3} (with the exceptions $\bar{\tau}_3$ = 100 s and $\phi_0$ = 0.48 which yield better sensitivity when spin dressing is employed) gives 

\begin{equation}
\sigma^{TOT}_d= 1.6\times 10^{-28} \text{e-cm}.
\end{equation}
With more sophisticated modulation schemes, one could anticipate additional small improvements in sensitivity.

\subsection{Systematic Uncertainties}

Unlike statistical
uncertainties, systematic uncertainties cannot be reduced by increasing the measurement
time. Systematic uncertainties are effects on the experimental results due to
physical processes that cannot be controlled to arbitrary precision. There are both known and unknown systematic effects. While the known ones can be dealt with to some extent by characterizing the precision of relevant physical quantities (e.g. magnetic field non-uniformity), the unknown
ones are often responsible for dramatic shifts in experimental results when
measurements are repeated using different techniques. In fact they can even be due to new
physics as was the case with early polarized electron scattering experiments, where
the measured asymmetry was later shown by Grodzins (see Ref.~\cite{GRODZINS59}) to be due to 
parity violation in the weak interaction. 

Below the systematic uncertainties are characterized in terms of standard uncertainties, that were 
appreciated for the first 50 years of study, and a more recently appreciated effect - the geometric phase effect. This uncertainty was only observed once experimental sensitivities reached a level where it could be resolved, but it is now a major factor in planning of particle EDM searches.

\subsubsection{Known Standard Systematic Uncertainties}

An EDM of $10^{-28}$ e-cm
in an electric field of 100 kV/cm will give rise to an interaction energy of
$10^{-23}$~eV while the neutron magnetic moment is $6\cdot10^{-8}$ eV/T. Thus thus to achieve this level of sensitivity, one needs to keep any magnetic fields associated with the electric field
reversal to less than about $\delta B_{sys}=(10^{-23}~\text{eV})/(6\cdot
10^{-8}$ eV/T)  = 0.2 fT. It is important to realize that this
constraint applies only to magnetic fields that vary coherently with a frequency component at the frequency of field reversal.

The situation can be improved by choosing the E field reversal scheme to be
other than a simple alternating sequence $\left(  +-+-+-...\right)$. For example a sequence%
\begin{equation}
A=\left(  +--+-++-\right)
\end{equation}
will eliminate drifts that are linear in time while a sequence%
\begin{equation}
B=\left(  +A,-A,-A,+A,-A,+A,+A,-A\right)
\end{equation}
will eliminate quadratic as well as linear drifts.
Randomly reversing the order of these sequences can help eliminate possible correlations with external periodic variations.  
\paragraph{Leakage currents}

A line source leakage current flowing in the insulating walls of the experimental chamber
or in the liquid Helium surrounding it will produce a magnetic field of
$B_{leak}(T)=2\times 10^{-7} I(A)/r(m)$ at a distance $r$ from the current.  
If the current
is exactly parallel to the electric field the magnetic field produced will be
perpendicular to the current, hence perpendicular to the electric and the
parallel magnetic field and will produce a change in the magnetic field
magnitude that is second order in the electric field and hence will not change
if the leakage current exactly reverses with the electric field. Of course
none of these assumptions is exact. The leakage current is not very likely to
flow in straight lines parallel to the electric field, there will be random
deviations in its direction and there will certainly be
some misalignment between the electric and magnetic fields. If the effective
displacement angle is $\theta$ then the effective magnetic field will be $\delta B\sim \theta B_{leak}$.
Then in order to keep $\delta B<\delta B_{sys}$, the leakage current must be
\begin{equation}
I  <8.0\cdot10^{-9}r(m)/\theta\sim 10 nA
\end{equation}
assuming that $r\sim 0.1$ m and $\theta\sim .01$ rad. 
\paragraph{Pseudomagnetic Field}
As discussed previously, the interaction between
the UCN and $^{3}$He is spin dependent, due to the tensor nature of the inter-nucleon force. UCN moving in a gas of polarized
$^{3}$He atoms will be subject to a potential%
\begin{equation}
V_{n3}=\alpha+\beta\vec{I}\cdot\vec{s}%
\end{equation}
where $\alpha,\beta$ are complex, $\vec{I}$ is the $^{3}$He nuclear spin
and $\vec{s}$ is the neutron spin. The imaginary part of the interaction corresponds
to absorption, while the real part acts on the
neutron spins as a pseudo-magnetic field $\vec{B}_{eff}=\operatorname{Re}%
\left(  \beta\vec{I}\right)$ \cite{ABRA82}. 
For 100\% $^3$He polarization and a fractional particle density of $10^{-10}$ compared to superfluid $^4$He (a typical density for the experiment presented here) the pseudo-magnetic field is comparable to a magnetic field of $\sim 23$ pT. 

For the free precession method of measuring the neutron frequency, if the two spins are rotated transverse to $B_0$, this effect will be cancelled out during the precession period. However, if the $^3$He spins are slightly longitudinal (i.e. along or against the $B_0$ direction), then the UCN precession frequency will change if this longitudinal component changes. However suitable $\pi/2$ pulses can be used that keep the longitudinal components of the two spins to < 1~mr, which should suppress this effect to below the statistical uncertainty.

If the UCN and $^{3}$He spin are transverse to $B_0$ and separated by a fixed angle
$\theta_{n3}$, as is the case for the critical spin dressing method, the UCN spin will precess around the $^{3}$He nuclear spin and develop a longitudinal component. This will reduce the sensitivity of the measurement and could introduce additional frequency noise.

In Ref. \cite{GL94} it was shown that this effect can be dealt with by
adjusting the value of the dressing parameter to eliminate the first harmonic term in the scintillation rate and
using the size of this adjustment as a measure of the EDM. An alternative approach, discussed below, is to choose the modulation frequency high enough so that the longitudinal component of spin remains negligible for the duration of the measurement period. This can be achieved because, as the neutron spin gains a longitudinal component during the first half of the modulation cycle, it is partially cancelled (only partially, since finite rotations don't commute) during the second half of the cycle. 

\paragraph{Motional magnetic fields}

According to special relativity a particle moving slowly through an electric field
will see a magnetic field in its rest frame of%
\begin{equation}
\vec{B_{v}}\simeq-\vec{\frac{v}{c^2}}\times\vec{E}
\end{equation}
which could interact with the particle's magnetic moment to produce a frequency
shift linear in $E$. Even with $v/c\sim 10^{-8\text{ }}$ applicable to
UCN, with the electric fields of interest $\gtrsim 50$ kV/cm this would mean a
field of $\sim$ 200 pT which is much
larger than the systematic limit $\delta B_{sys}=$ 0.2 fT.

Ideally $\vec{E}\Vert\vec{B_0%
}$, so that  $\vec{B_{v}}\bot\vec{B_0}$ and there
would be no first order effect, but in the presence of misalignment by an
angle $\theta$ there would be a linear effect $\sim \vec{B_{v}}\sin\theta\sim$ 2 pT if $\theta
\sim 10^{-2}$ rad.
One of the main reasons for abandoning beam experiments in favor of UCN 
to search for a neutron EDM
was the reduction of this effect due to the fact that the average velocity of
the UCN over the course of a measurement would approach zero. For a UCN
with an average distance from initial to final location in the storage cell $\delta r\sim 0.2$ m and a
measurement lasting $T\sim 500$ seconds, the time average velocity would be $\delta
r/T\sim 4\times 10^{-4}$ m/sec or a reduction by a factor of $10^{-4}$ with
respect to the average UCN speed. Now $\delta r$ will average to zero over
the ensemble of UCN and the final result will have an rms fluctuation
decreased by $1/\sqrt{N}$ with $N\sim 10^{9}$ the total number
of UCN measured in the experiment, which reduces the effect
$<<\delta B_{sys}$.

In addition, the component of $\vec{E}$ parallel to $\vec{B_0}$ will produce a $\vec{B_{v}}$ perpendicular to
$\vec{B_0}$ so that the magnitude of the total magnetic field
will change by%
\begin{equation}
\delta B=\frac{1}{2}\frac{B_{v}^{2}}{B_0}\sim 10 \, \text{fT}
\end{equation}
for a $B_0 = 3 \mu$ T.
While this does not change with reversal of the electric field, it means that
the variation of the magnitude of the electric field on sign reversal should be
smaller than $1\%$. This is often called the second-order vxE effect. 

\paragraph{Co-magnetometers}

With a need to control coherent magnetic field fluctuations to the level of
$\delta B_{sys}\sim 10^{-16}$ T it is clear that good control of the
magnetic fields in the measurement cell is crucial. The idea of a `co-magnetometer', that is a system capable of measuring the
magnetic field in the same volume occupied by the UCN, was proposed by Ramsey (Ref.~\cite{RAMSEY84}) in conjunction with the Sussex-ILL EDM
search. The initial idea was to inject polarized $^3$He gas into the UCN measurement cell.
However, due to technical details, it was decided to use Hg atoms as a co-magnetometer.

The use of a co-magnetometer, while reducing some sources of uncertainty, such as the
fact that leakage-current induced fields can influence different external
magnetometers in complicated ways, also introduces some new areas of concern: 

\begin{itemize}
\item Does the density distribution of the co-magnetometer atoms really mirror that
of the UCN?

\item What are the effects of the gravitational offset between the neutrons 
and atoms as well as possible atom density inhomogeneities?

\item Do the co-magnetometer 
atoms interact with the walls in such a way
that there could be an E field sensitive frequency shift? 

\item Do the
co-magnetometer atoms interact with the UCN in such a way as to shift their
resonance frequency (e.g. pseudomagnetic field)? 

\item Do the co-magnetometer atoms have an EDM, as then the experiment would be measuring the difference in EDM's of the atoms and UCN?
\end{itemize}

\paragraph{Other systematic uncertainties}

There are, of course, other potential systematic uncertainties. Table~\ref{Systematics}
lists some of the primary systematic uncertainties expected in the proposed experiment (Ref.~\cite{PREPROP}). 

\hspace*{-0.6cm}
\begin{table}
\begin{tabular}
[c]{|l|l|l|l|}\hline
Uncertainty Source & $%
\begin{array}
[c]{c}%
\text{Systematic}\\
\text{uncertainty (e-cm)}
\end{array}
$ & Comments & Key Parameters\\\hline
Linear (E x v) & $<1\times10^{-28}$ & $%
\begin{array}
[c]{c}%
\text{Uniformity of }\\
\text{B}_0\ \text{field}%
\end{array}
$ & $%
\begin{array}
[c]{c}%
\text{B field gradient}\\
\text{Temperature}%
\end{array}
$\\\hline
Quadratic (E x v) & $<0.5\times10^{-28}$ & $%
\begin{array}
[c]{c}%
\text{E field reversal}\\
\text{accuracy
$<$
1}\%
\end{array}
$  & \\\hline
$%
\begin{array}
[c]{c}%
\text{Pseudomagnetic}\\
\text{field effects}%
\end{array}
$  & $<1\times10^{-28}$ & $%
\begin{array}
[c]{c}%
\text{Modulation,}\\
\text{comparing two}\\
\text {cells}%
\end{array}
$ & $%
\begin{array}
[c]{c}%
^3\text{He density,}\\
\pi/2\text{ pulse, }\\
\text{modulation}%
\end{array}
$\\\hline
$%
\begin{array}
[c]{c}%
\text{Gravitational }\\
\text{Offset}%
\end{array}
$ & $<0.2\times10^{-28}$ & $%
\begin{array}
[c]{c}%
\text{with 1 nA}\\
\text{leakage current}%
\end{array}
$ & \\\hline
$%
\begin{array}
[c]{c}%
^3 \text{He inhomogeneity}\\
\text{due to leakage}\\
\text{current heating}%
\end{array}
$ & $<1.5\times10^{-28}$ & leakage $<1pA$ & $%
\begin{array}
[c]{c}%
\text{Temperature}\\
\text{B field gradient}%
\end{array}
$ \\\hline
\end{tabular}
\caption{Summary of key systematic effects, their corresponding uncertainties and key parameters that determine their contribution.}
\label{Systematics}
\end{table}

\bigskip

\subsubsection{Bloch-Siegert Induced False EDM}
\label{sec:falseEDM}

After more than 50 years of experimental work on searching for a neutron EDM,
it came as something of a shock when a new systematic uncertainty was observed~\cite{PEND04}.
The fact that it had remained unknown for so long was partly due to the fact
that it was generally smaller than the sensitivity of the then existing
experiments and partly due to a fortuitous result of using a co-magnetometer.
Without the latter its discovery might have been delayed for one or more
future generations of experiments, possibly leading to a false alarm from a non-zero
EDM. 

\begin{figure}
[ptb]
\begin{center}
\includegraphics[
width=7in
]
{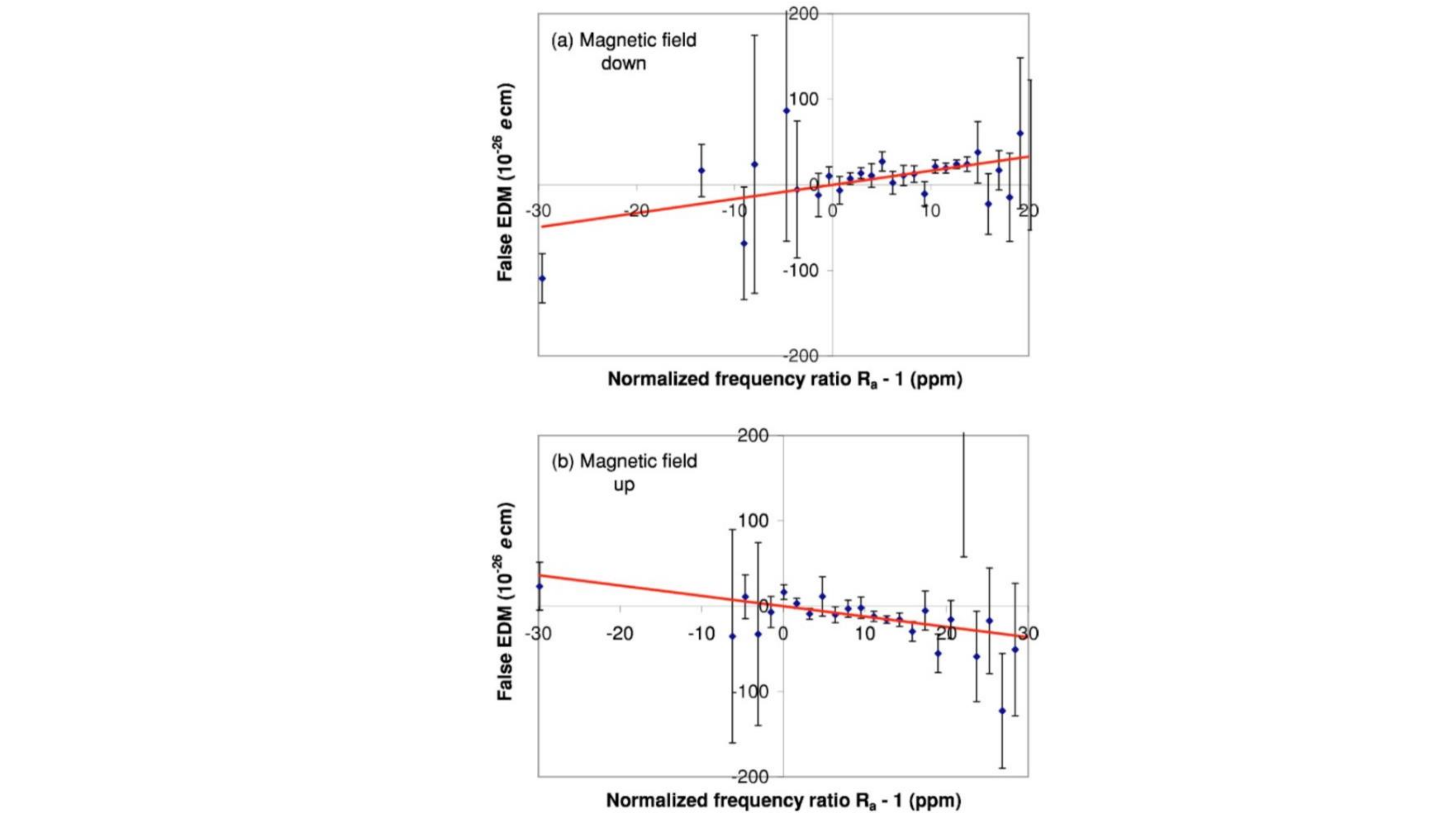}
\caption{Measured EDM as a function of frequency offset function $R-1$, which is
proportional to the gradient $\left(  \frac{\partial B_{z}}{\partial
z}\right)  $, from Ref.~\cite{PEND04}.}
\label{EDM-R}
\end{center}
\end{figure}

As mentioned above, the Sussex-ILL nEDM search was designed to use $^{199}$Hg atoms as a co-magnetometer at room temperature. As the thermal energies of the
atoms, $\sim 25$ meV, are much larger than the average energy of the UCN
$\sim 200$ neV, only the trajectories of the latter will be influenced by
gravity (with a potential energy $\sim 1$ neV/cm), so that the center of gravity of the UCN distribution will be
displaced with respect to that of the Hg atoms. In the presence of magnetic
field gradients $\left(  \partial B_{z}/\partial z\right)  $ with $\vec B_0 = B_0 \hat z$ and 
positive $\hat z$ directed upward, the two spin
species would see different average magnetic fields and hence exhibit
different Larmor frequencies. In the absence of gradients
$\omega_{UCN}/\omega_{Hg}=\gamma_{UCN}/\gamma_{Hg}$ where $\gamma_{i}$ are the
respective gyromagnetic ratios. In the presence of a gradient
$\omega_{UCN}=\left(  \gamma_{UCN}/\gamma_{Hg}\right)  \omega_{Hg}%
+\gamma_{UCN}\left(  \partial B_{z}/\partial z\right)  \Delta h,$ where
$\Delta h$ is the displacement of the center of mass of the UCN. Defining%
\begin{equation}
R_a=\frac{\omega_{UCN}}{\omega_{Hg}}\frac{\gamma_{Hg}}{\gamma_{UCN}}%
\end{equation}
then%
\begin{align}
R_a-1  & =\frac{\gamma_{Hg}}{\omega_{Hg}}\left(  \frac{\partial B_{z}}{\partial
z}\right)  \Delta h\\
& =\pm\frac{1}{B_0}\left(  \frac{\partial B_{z}}{\partial z}\right)  \Delta
h
\end{align}
where the plus sign corresponds to $B_0$ pointing down. During the data analysis of
the Sussex-ILL\ experiment, while attempting to extract a value for the $^{199}$Hg magnetic dipole moment, a strong correlation was observed between the extracted neutron EDM values and
the ratio $R$ as shown in the Fig.~\ref{EDM-R}%

This effect appeared because the field gradient had been
varied inadvertently every time the apparatus had been opened and the magnetic
shields had been reassembled and de-magnetized. After the discovery, a
small amount of data was taken with deliberately applied strong field
gradients [the large $\left(  R-1\right)  $ points with large uncertainties].
Fortunately, the co-magnetometer allowed the monitoring
of the volume average field gradient in the cell. 
Although it turned out that the physics had been discussed by Commins (Ref.~\cite{COMMINS91}) in
connection with a molecular beam EDM search, the ILL-Sussex group came to an initial understanding by independent means.

There are several ways of understanding the effect. The simplest is
to consider that the spins moving in an electric field will see an additional magnetic
field $\vec{B}_{v}=\left(  \vec{E}\times\vec{\frac{v}{c^2}}\right)  $ directed in the plane perpendicular to the $z$ axis,
the nominal $\vec{B}$ and $\vec{E}$ field directions.
Its direction will vary with time as the velocity does. It is well known that
a time varying field directed in that plane can cause a shift in the Larmor
frequency (so-called Bloch-Siegert shift~\cite{BS40}) of%
\begin{equation}
\delta\omega=\frac{\gamma^{2}B_{1}^{2}}{2\left(  \omega_0-\omega_{r}\right)
}%
\end{equation}
where this applies to the case of a perturbing field $B_{1}$ rotating with
angular velocity $\omega_{r}$ in the $\left(  x,y\right)  $ plane. If $B_{v}$
was the only field present this would be second order in the E field and would
not constitute an EDM signal. However if another field is present with a
component parallel to $B_{v}$ the cross term in the square will be linear in
$E.$ At first one might think that since $B_{v}$ depends on velocity
any shifts caused by it would \ average out as discussed above. However
this is not the case as $\omega_{r}$ will also change sign with the velocity
and the cancellation will not be complete. The additional field in the
$\left(  x,y\right)  $ plane can arise because of the fact that $\vec
{\triangledown}\cdot\vec{B}=0$ so that a non-zero $\left(
\frac{\partial B_{z}}{\partial z}\right)  $ implies a non-zero field in that
plane. Given cylindrical symmetry, then $\left(  \frac{\partial B_{x}%
}{\partial x}\right)  =\left(  \frac{\partial B_{y}}{\partial y}\right)
=-\frac{1}{2}\left(  \frac{\partial B_{z}}{\partial z}\right)  $ and there will
be a radial magnetic field $\vec{B}_{r}=-\frac{1}{2}\left(
\frac{\partial B_{z}}{\partial z}\right)  \vec{r}.$ Noting that
for a particle in a circular orbit, $\vec{B}_{v}$ will be radial
\begin{equation}
B_{1}^{2}=\left|  \vec{B}_{v}+\vec{B}_{r}\right|  ^{2}%
\end{equation}
and the cross term%
\begin{equation}
2\vec{B}_{r}\cdot\vec{B}_{v}=-\left(  \frac{\partial
B_{z}}{\partial z}\right)  \vec{r}\cdot\left(  \vec{E}\times\overrightarrow{\frac{v}{c^2}}\right)  \sim -\left(  \frac{\partial
B_{z}}{\partial z}\right)  \frac{E}{c^2}\omega_{r}R^{2}%
\end{equation}
assuming, for example, a circular orbit at radius $R$, corresponding to the radius of
the cylindrical container as in the Sussex-ILL experiment , with $
\omega_{r}=v/R$. Then for the term linear in $\vec{E}$
\begin{equation}
\delta\omega=-\frac{\gamma^{2}\left(  \frac{\partial B_{z}}{\partial z}\right)
\frac{E}{c^2}\omega_{r}R}{2\left(  \omega_0-\omega_{r}\right)  }.%
\end{equation}
For every velocity $\vec{v}$ there are an identical number of spins
with velocity $-\vec{v}$ so the two
directions of velocity, i.e. $\pm\omega_{r},$ must be averaged. Because of the term in the
denominator this does not vanish:%
\begin{align}
\overline{\delta\omega}=\delta\omega_+-\delta\omega_-  & =-\frac{\gamma^{2}\left(  \frac{\partial B_{z}%
}{\partial z}\right)  ER\omega_{r}}{2c^2}\left(  \frac{1}{\left(  \omega
_0-\omega_{r}\right)  }-\frac{1}{\left(  \omega_0+\omega_{r}\right)
}\right) \\
& =-\frac{\gamma^{2}\left(  \frac{\partial B_{z}}{\partial z}\right)
ER^{2}\omega_{r}^{2}}{c^2\left(  \omega_0^{2}-\omega_{r}^{2}\right)
}\allowbreak
\end{align}
as an estimate of the size of the effect. To go deeper into the effect one must consider the motion of the particles in more detail. Both the position
and velocity will be functions of time as the particles move around the cell
colliding with the walls and possibly other particles. Writing the
$\left(  x,y\right)  $ components of the perturbing fields as
\begin{align}
\label{eq:BX}
B_{x}  & =\frac{1}{2}\left(  \frac{\partial B_{z}}{\partial z}\right)
x+\frac{E}{c^2}v_{y}\\
\label{eq:BY}
B_{y}  & =\frac{1}{2}\left(  \frac{\partial B_{z}}{\partial z}\right)
y-\frac{E}{c^2}v_{x}%
\end{align}
they can be treated according to time-dependent perturbation theory. The
perturbation is then $\vec{B_1} = B_{x}\hat x +B_{y}\hat y$.

As usual the effect of a time-dependent perturbation depends on the
fluctuations of the perturbation at the frequency of the transition being
studied, in this case the Larmor frequency. Since the power spectrum of a
generalized function is the Fourier transform of the auto-correlation function
of the fluctuations, there will be terms in the auto-correlation
of $\left(  B_{x}\hat x+B_{y}\hat y\right)$ involving the correlation of $B_{x}$ and
$B_{y}$ and thus, from Eqs.~\ref{eq:BX} and \ref{eq:BY}, $x$ and $v_{x}$ as well between $y$ and $v_y$. Thus these terms, being linear in $E$, constitute a `false EDM'. Many detailed studies (Ref.~\cite{PEND04,GEOPHASE})
have been made concerning the effects of different motional regimes and
container shapes and analytic results have been obtained in many cases. One
key point is that the effect can be significantly reduced by collisions which
continuously re-randomize the direction of the velocity. The possibility of using this effect to study this false EDM is outlined in the next section. 

\subsubsection{Studying the False EDM in the nEDM@SNS Experiment}
\begin{figure}
[ptb]
\begin{center}
\includegraphics[
width=6in
]%
{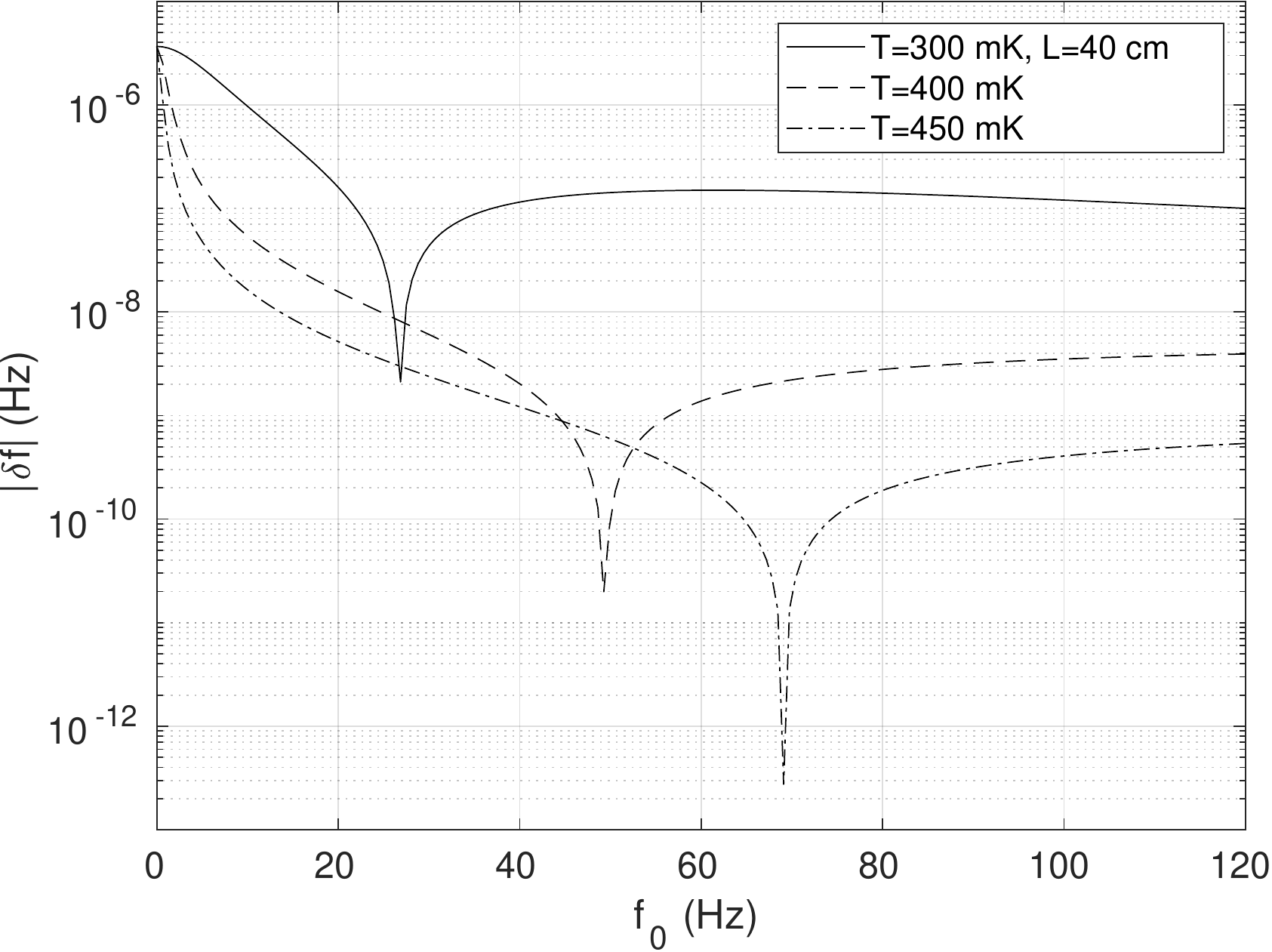}%
\caption{Variation of the magnitude of the Bloch-Siegert frequency shift for polarized $^3$He as a function of the $^3$He precession frequency in a gradient field of $5\times10^{-10}$ T/m plotted for several different temperatures (from Ref.~\cite{SWANK16}). The strong dependence on temperature allows this shift (false EDM) to be characterized experimentally.}
\label{temp-dep}
\end{center}
\end{figure}

In the experiment discussed here, the polarized $^3$He, which is used as a co-magnetometer, is in dilute solution $\left(  \sim 10^{-10}\text{ relative
concentration}\right)$ in $^{4}$He. Thus the
collision rate of the $^3$He with the phonons in the $^{4}$He can be controlled by adjusting
the temperature, and used to tune the behavior of the `false EDM' signal.
Fig.~\ref{temp-dep} shows the dependence of the absolute value of the $^3$He false
EDM signal on the $^3$He precession frequency for a cell 40 cm long at several temperatures. 
This is the contribution of this single dimension. The shorter dimension of
the cell will make a smaller contribution and the frequency shift will be the
sum of the two contributions. The results shown in Fig.~\ref{temp-dep} have been obtained from a recently
developed analytic calculation which allows for velocity changing
(thermalizing) collisions [Ref.~\cite{SWANK16}). As seen in the figure, it is possible to 
reduce the false EDM by raising the operating temperature. In addition, lowering the
temperature can increase the effect by
several orders-of-magnitude, allowing for detailed study of this contribution. 

As the UCN must operate in a collision-free regime, in order to avoid loss by
upscattering, the only way to control the analagous effect for UCN (at fixed gradient) is to
vary the Larmor frequency. As shown by the qualitative results above, the false
EDM goes as $1/\omega_0^{2}$ ($\omega_0$ being the Larmor frequency) which
is why the experiment is being planned to operate at the relatively large frequency of $100$ Hz. 
The technical details of the experimental design, with a focus on the individual sub-systems, is discussed in the following sections.  


\section{Apparatus}

\subsection{Overview of nEDM@SNS Apparatus and Infrastructure}%



\begin{figure}
[ptb]
\begin{center}
\includegraphics[width=\textwidth]{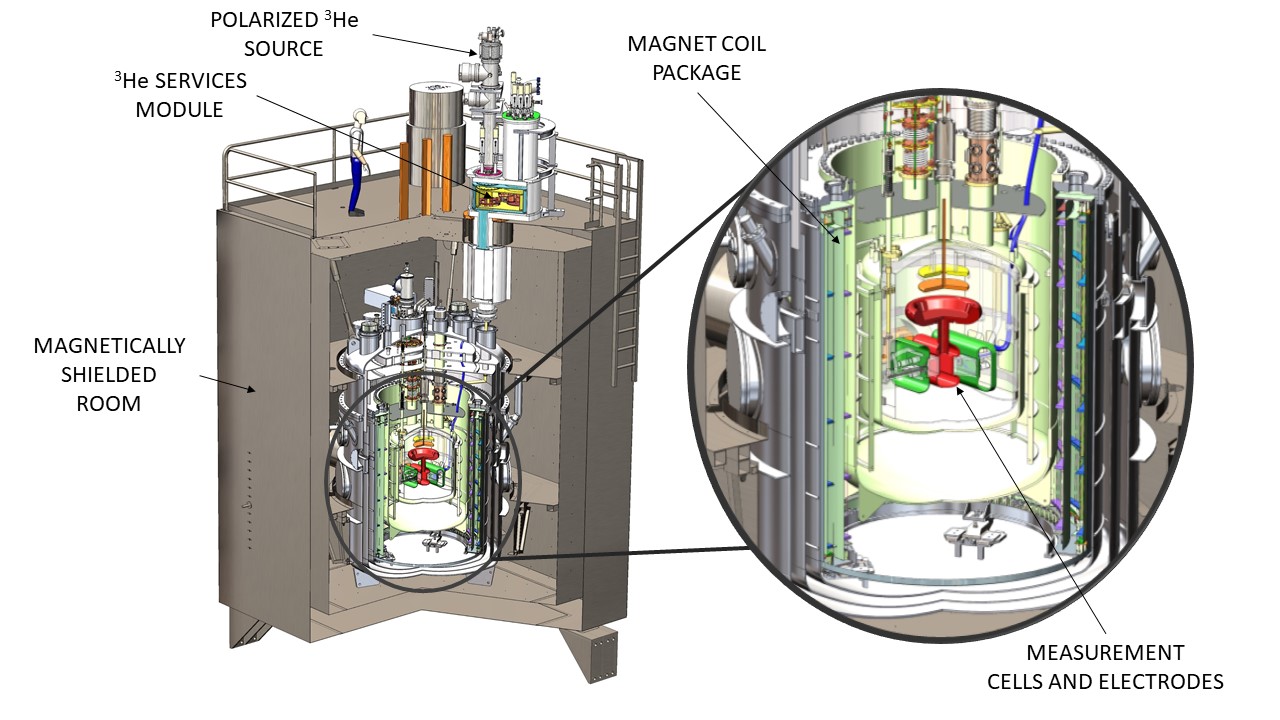}
\caption{Schematic of the nEDM apparatus. The inset shows the measurements cells, HV (red) and ground electrodes (green) and the Cavallo HV multiplier (orange/yellow).}
\label{fig:schematic}
\end{center}
\end{figure}

The main features of the apparatus are shown in Fig.~\ref{fig:schematic}. The construction
of the two measurement cells and the electrodes for producing the electric
field are shown in the lower right inset. The high voltage electrode, which can be operated at either positive or negative HV, is shown
in red and is connected to the lower plate of the voltage amplifying
system (Cavallo multiplier). The two ground electrodes are shown in green. For the Cavallo multiplier  (see Ref.~\cite{CAV95}), a relatively modest high voltage is fed in from the top of the main apparatus (HV feed) and
used to charge a capacitor. The charge is then transferred from this capacitor
to the electrodes multiple times in such a way that the voltage on the
electrode can be built up to be several times greater than the input voltage,
somewhat similar to a van der Graaff  generator.
The polarized neutron
beam (0.89~nm) passes between the electrodes and enters through a series of
windows in the various magnetic and thermal shields. Optical fibers, which carry the scintillating light signal to the silicon
photo-multipliers are located only on the electrode ground side of the measurement cells. 

The magnetic fields are
produced by a set of cylindrical coils coaxial to the vertical direction and
implemented as a module (magnet package) which can be removed as a whole from
the apparatus. There are coils for producing the main horizontal $B_0$ DC field and the AC
dressing field as well as gradient fields and shimming coils. There is also a magnetic flux return (thin layer of highly permeable material) and superconducting Pb shield to both improve the field uniformity as well as shield against external magnetic field variations. 

The system for handling the polarized $^3$He is shown above the main cryostat. 
$^3$He atoms, polarized by passing as an
atomic beam through a strong magnetic field gradient ($^3$He atomic beam
source discussed above in Sec~\ref{sec:polarization}), produced by permanent magnets, are incident on a surface of isotopically pure liquid
$^{4}$He (injection module) to which they are attracted by a relatively strong binding energy
of 2.8~K. They are then transported by heat flush (Ref.~\cite{FLUSH}) to the two
measurement cells. After
each measurement cycle, partially depolarized $^3$He are removed from the
measurement cells and concentrated and recycled by means of heat
flush and evaporation (discussed in detail below).

Most of the apparatus is contained within a room temperature magnetic field enclosure (based on two or three layers of mu-metal) to reduce the ambient field and minimize magnetic field gradients. Inside this enclosure, but outside the vacuum vessel, is an additional magnetic field coil with field = $\vec{B_0}$ to maintain the $^3$He polarization during transport to the measurement cells. 

\begin{figure}[btp]
\begin{center}
\includegraphics[width=\textwidth]{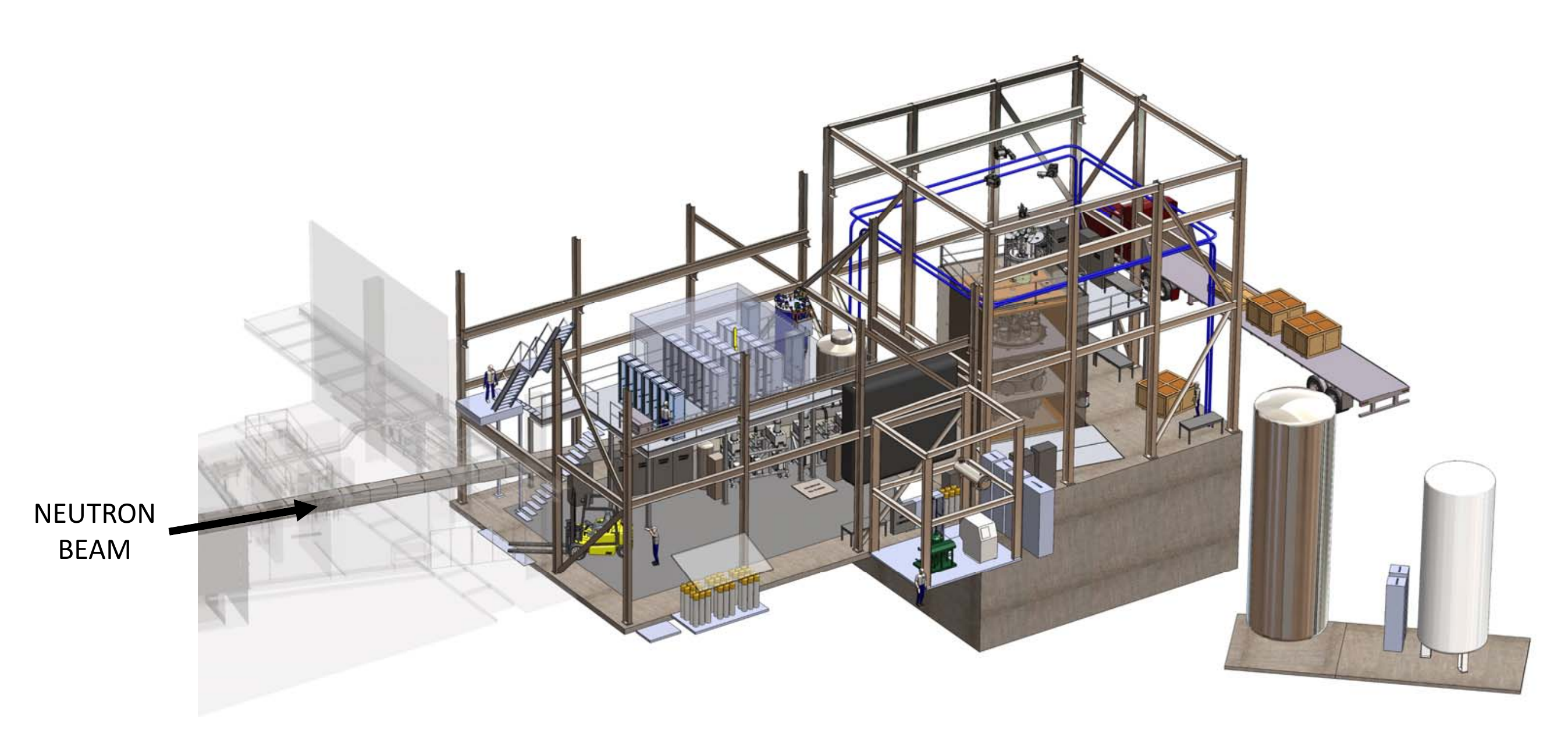}
\caption{General layout of the nEDM apparatus and associated infrastructure at the SNS. External magnetic field coils are also shown outside of the Magnetic Shield Enclosure (MSE) to reduce the ambient Earth's magnetic field.}
\label{fig:overview}
\end{center}
\end{figure}
\begin{figure}[htbp]
\begin{center}
\includegraphics[width=0.6\textwidth]{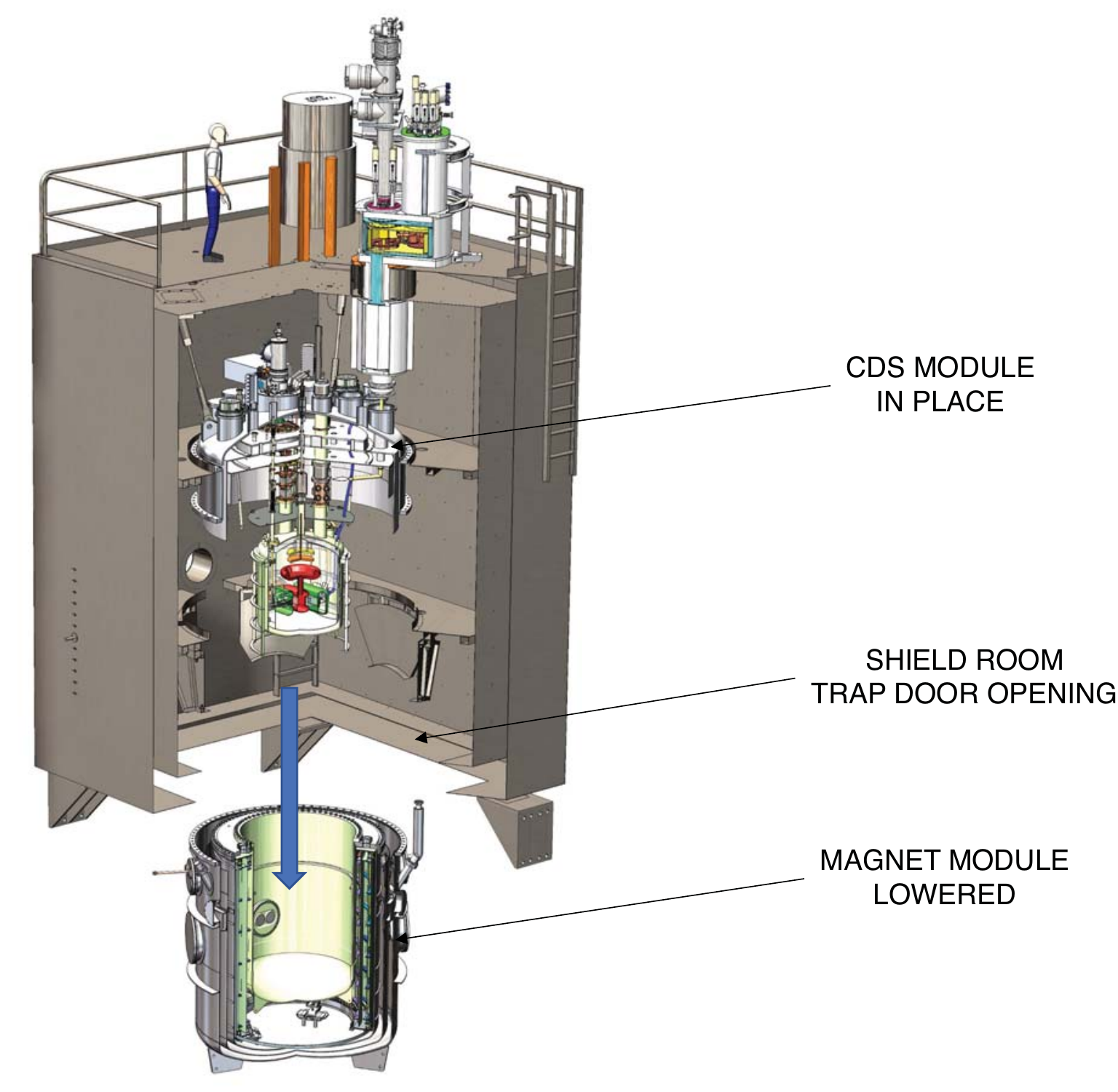}
\caption{Example of the modular design of the apparatus showing the magnet module lowered for servicing.}
\label{fig:modules}
\end{center}
\end{figure}


The large size, cryogenic nature, and materials constraints of the apparatus requires a modular design of the major components.  To large extent, these functional components are separated into cryogenic modules with well defined interfaces, both physically and scientifically.  These cryogenic modules, the Central Detector System (CDS), Magnetic Field Module and $^3$He services (3HeS), are being constructed and tested independently before being assembled into the final apparatus. An example of how the magnetic field module can be separated from CDS is shown in Fig.~\ref{fig:modules}.

The experiment will be located in two satellite buildings adjacent to the Spallation Neutron Source (SNS) target building as shown in Fig.~\ref{fig:overview}.  These buildings house the entire experiment including the apparatus itself, the neutron beam, magnetic shield enclosure, and the cryogenic system. 

The following sections provide a brief introduction to the main experimental components, including the cryogenic modules, neutronics, magnetic shield enclosure, and cryogenic system. The detailed requirements and mechanical/cryogenic design of the experimental components will be presented in the later sections.

\subsection{Central Detector System Module}  
\begin{figure}[htbp]
\begin{center}
\includegraphics[width=\textwidth]{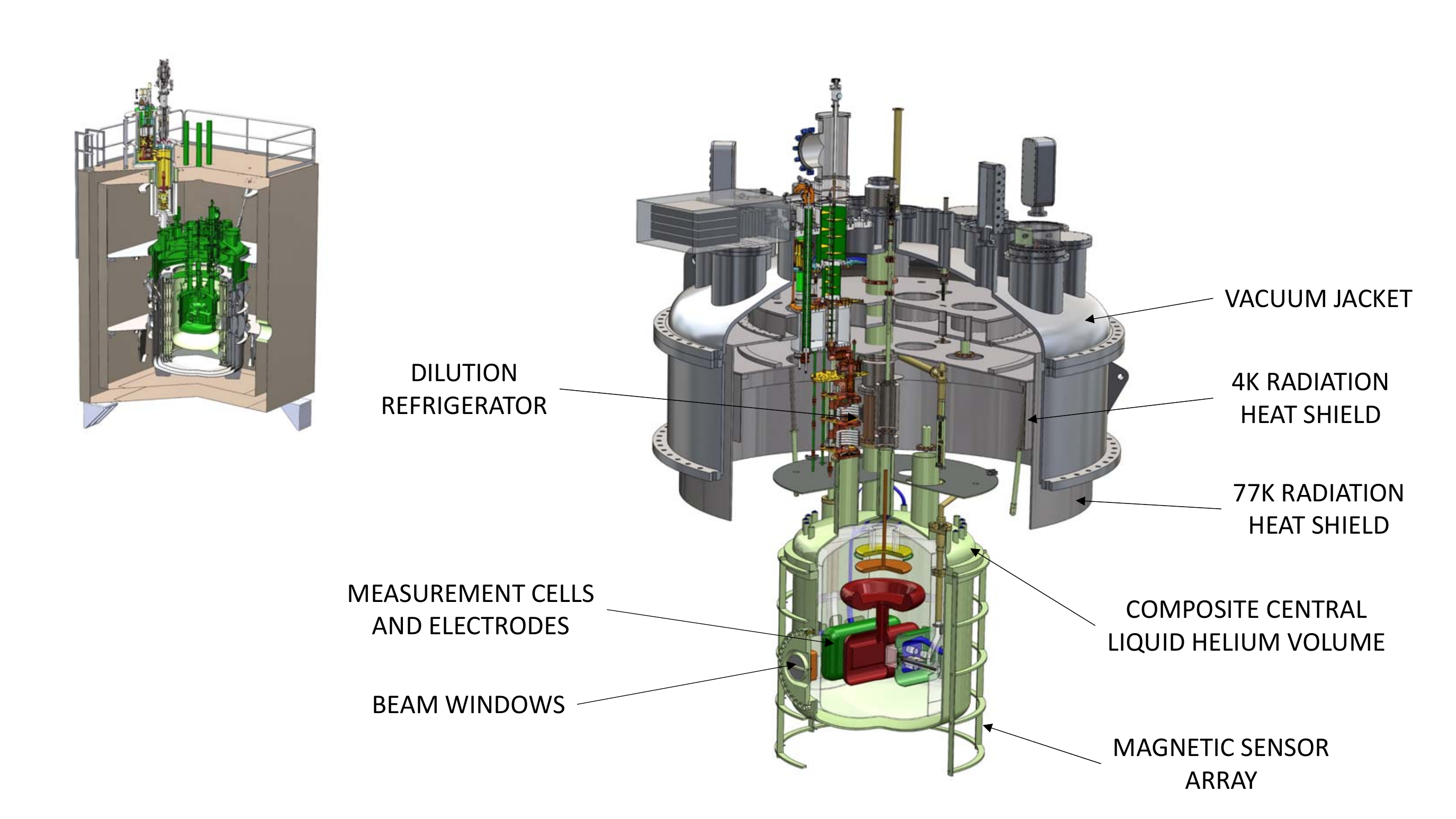}
\caption{Overview of the central detector system. Individual components are described in the text.  The upper left insert shows the location of the CDS (highlighted in green) relative to the larger-scale apparatus.}
\label{fig:CDS}
\end{center}
\end{figure}

The Central Detector System (CDS) houses the measurement cells, high voltage system with Cavallo multiplier, light collection system, and \textsc{squid} sensor arrays in an approximately 1600~liter bath of liquid helium cooled with a dilution refrigerator below 0.5~K.  A cutaway view of the CDS design is shown in Fig.~\ref{fig:CDS}. 

The location of the CDS components imposes significant constraints on their construction.  For example, materials in and around the neutron beam must be non-activating to minimize backgrounds in the nEDM signal.  Also, materials within the superconducting shield must largely be both non-magnetic and non-metallic (due to eddy current heating).  These constraints necessitate non-standard construction of various components such as the liquid helium volume, which will be fabricated from a G-10 composite, and the electrodes, which will be fabricated from acrylic with implanted coatings. 

The overall layout is vertical in design with all components suspended from the approximately 3.5~m diameter top flange of the vacuum jacket to allow for maintenance.  As can be seen in Fig.~\ref{fig:modules}, the outer vacuum can and magnet module can be lowered, providing access to the entire CDS.  The top flange of the vacuum jacket is suspended from the magnetic shield enclosure (MSE).

An external liquid nitrogen supply and a helium liquefier provide cooling for the 77~K and 4~K radiation shields respectively.  Additional cooling is provided by a $^3$He-$^4$He dilution refrigerator that allows the central helium volume to be cooled to below 0.5~K.  This refrigerator is being constructed in-house in order to minimize magnetic contamination. Safety measures, such as large external vents for the helium in case of a vacuum failure, are also included. Additional details on the CDS are given below in section~\ref{sec:CDS}.

\subsection{Magnet Field Module and Cryogenic Field Monitors} 
\begin{figure}[htbp]
\begin{center}
\includegraphics[width=\textwidth]{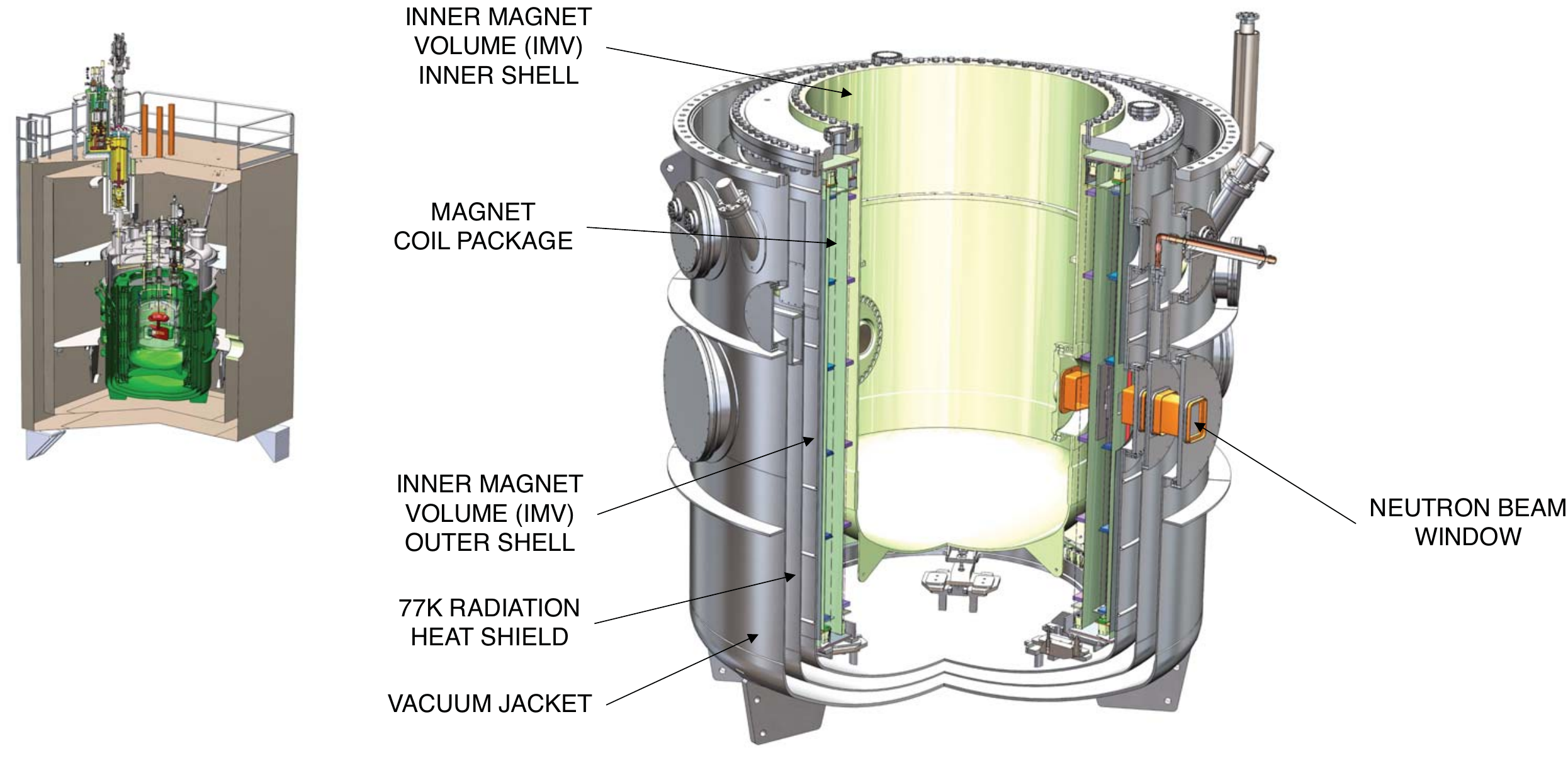}
\caption{Overview of the magnetic field module. Individual components are described in the text.  The upper left insert shows the location of the module (highlighted in green) relative to the larger-scale apparatus.}
\label{fig:magnet}
\end{center}
\end{figure}

The magnetic field module houses the current-carrying coils and shields that provide the required magnetic environment. This
includes a ferromagnetic shield, a superconducting shield, and the coils for the uniform DC holding field as well as spin rotation and other AC fields for spin manipulation. This module, which includes the lower cryovessel, liquid nitrogen shield (LN-shield) and inner magnet volume (IMV) is positioned around the CDS and schematically is shown in Fig.~\ref{fig:magnet}. 

The housing for the magnet system, the Inner Magnet Volume (IMV), is hybrid in structure, with the outside shell being aluminum and the inner shell fabricated from a G-10 composite. The outer shell is actively cooled (to < 6~K) while the interior volume is filled with low pressure He as heat exchanger. Thus the inner shell also serves a low-temperature shield for the CDS. The IMV is suspended within the lower cryovessel by three G10 composite struts that pass through the LN-shield. The top flange has two cryogenic seals for the outer and inner shells.  All of the shields and coils are mounted from a kinematic mount on the floor of the IMV. 

The magnetic field module has similar design requirements as the CDS.  Components surrounding the neutron beam must not activate and interior components must be non-metallic and non-magnetic.  The coils and shields reside in a cryogenic (<6~K) environment, requiring a liquid nitrogen radiation shield.  All components are cooled using an external nitrogen supply and the helium liquefier system using cryogenic feeds that are independent of the CDS cooling system.  The magnet housing itself is filled with helium exchange gas and cooled primarily though conduction.
Additional details on the individual components within the coil package are given below in section~\ref{sec:magnet}.

\subsection{$^3$He Services (He3S) Module}
\begin{figure}[htbp]
\begin{center}
\includegraphics[width=\textwidth]{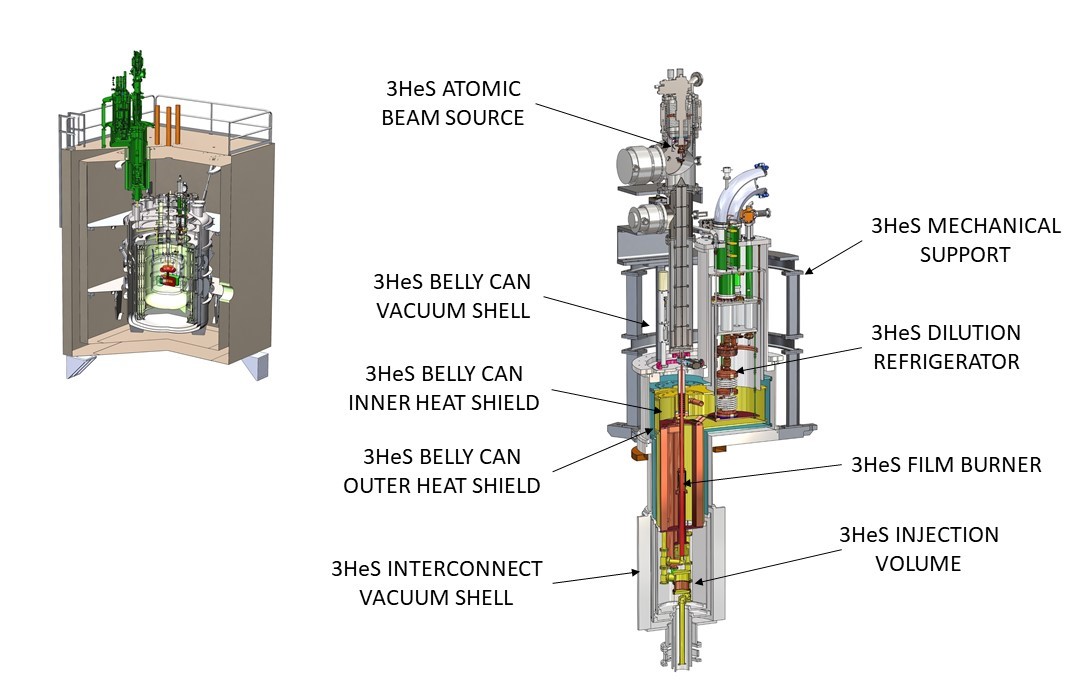}
\caption{Overview of the $^3$He Services system. Individual components are described in the text.  The upper left insert shows the location of the He3S (highlighted in green) relative to the larger-scale apparatus.}
\label{fig:He3S}
\end{center}
\end{figure}

The He3S system provides the polarized $^3$He used as the co-magnetometer. This includes the atomic beam source (ABS) that polarizes the $^3$He, the cryogenic components to collect and move this polarized $^3$He in the liquid $^4$He from the collection region to the measurement cells, and finally remove the depolarized $^3$He from the system at the end of data collection, and a second $^3$He-$^4$He dilution refrigerator that provides cooling for the system.  As can be seen in Fig.~\ref{fig:He3S}, the system attaches to the top vacuum flange of the CDS system, extending into the vacuum space shared with the CDS.  The ABS and dilution refrigerator reside above the magnetic shield enclosure.

The three primary components of the He3S system, ABS, injection system, and purifier, are described in detail in section~\ref{sec:He3S}.  The dilution refrigerator will be of a similar design to that of the CDS and also constructed in-house.  Cooling will be provided by the external nitrogen supply and helium liquefier system.  A small helium reservoir will provide helium for the 1~K pot of the dilution refrigerator.   

\subsection{Cold Neutron Transport}
\begin{figure}[htbp]
\begin{center}
\includegraphics[width=\textwidth]{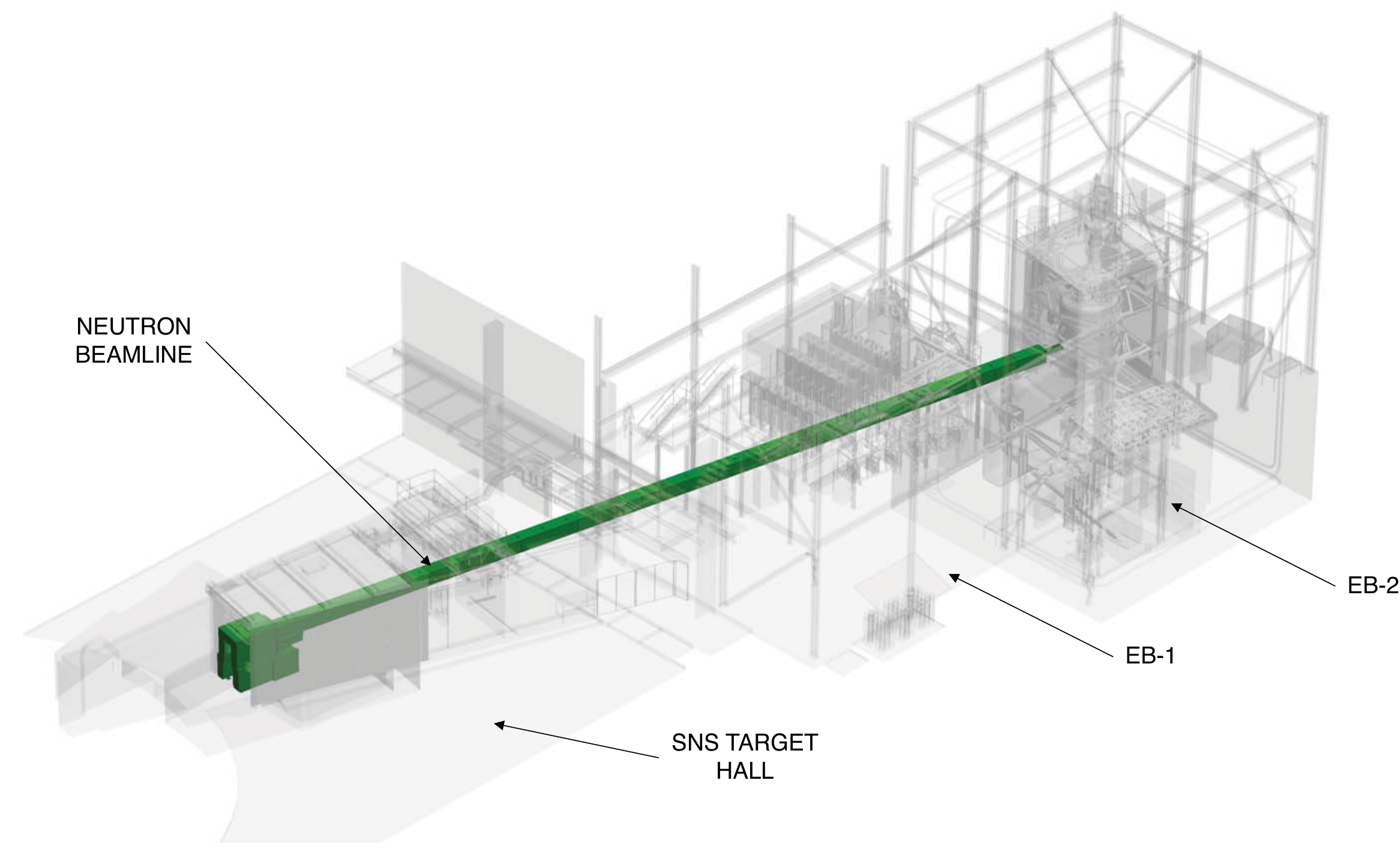}
\caption{Overview of the neutron beamline (highlighted in green) extending from the polarizer in the SNS target hall to the apparatus. EB-1 and EB-2 refer to external buildings to the SNS target hall.}
\label{fig:neutronics}
\end{center}
\end{figure}

The neutron beam extends along SNS beamline BL-13 from the cold source to the experiment.  It is comprised of supermirror guides, choppers to select the neutron energy, a supermirror polarizer for the beam, magnets for spin transport, a splitter to guide the beam into the two measurement cells, biological shielding, and supports and is shown in Fig.~\ref{fig:neutronics}. Sec.~\ref{sec:neutronics} provides a detailed design. 

\subsection{Magnetically Shielded Enclosure (MSE)}
\begin{figure}[htbp]
\begin{center}
\includegraphics[width=\textwidth]{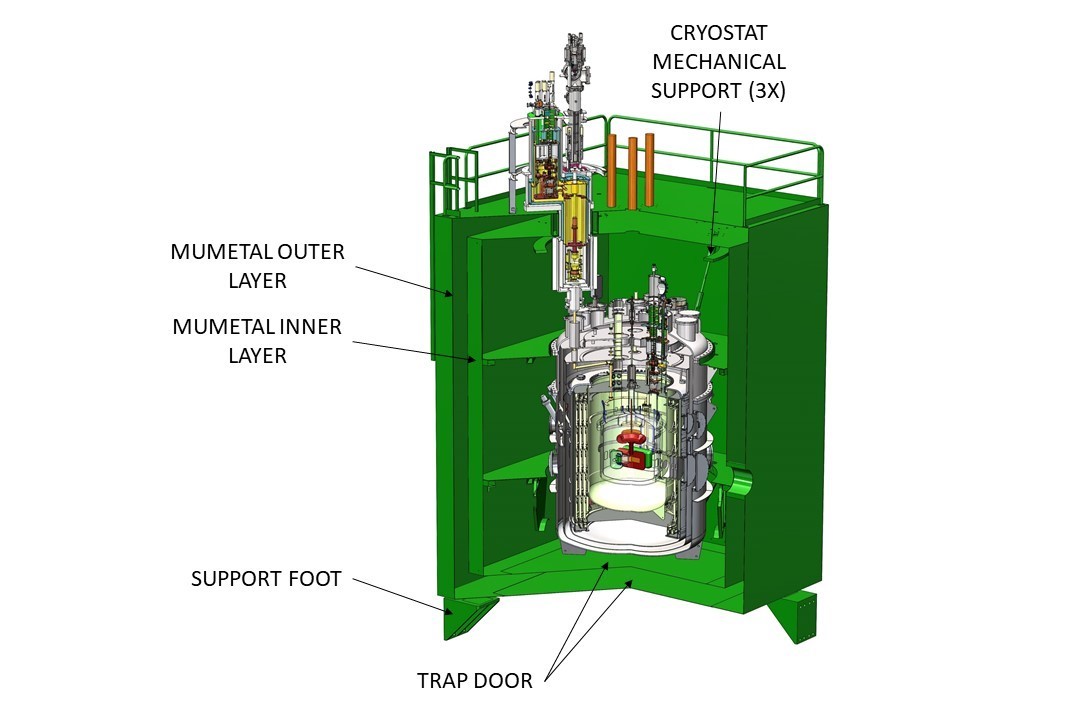}
\caption{Magnetically shielded enclosure (MSE) (highlighted in green) relative to the measurement apparatus.}
\label{fig:ShieldRoom}
\end{center}
\end{figure}

The MSE, as shown in Fig.~\ref{fig:ShieldRoom}, is a large enclosure surrounding the apparatus constructed from two or three layers of $\mu$-metal.  The room is large enough so that the apparatus can be serviced in-place. Inside the MSE, two platforms will allow individuals access to the apparatus.  As shown in Fig.~\ref{fig:ShieldRoom}, the bottom panels of the MSE can be removed to permit lowering the magnet package providing access to the CDS.  Personnel can enter the shield house though a door on the side wall. See Sec.~\ref{sec:shieldroom} for more detail. 

\subsection{Cryogenics}

Each of the cryogenic modules requires LN and LHe for cooling. The magnetic field module must be cooled to $\sim 6$K to maintain the magnetic coils and superconducting shield below the critical temperature for the materials. This is achieved by flowing cold He through tubes attached to the IMV. In addition to LN and LHe cooling, the CDS and He3S require dilution refrigerators to reach their operational temperatures. Additional details are provided in Sec.~\ref{sec:cryogenics}

\section{Central Detection System}
\label{sec:CDS}
\subsection{Overview}
\begin{figure}[htbp]
\begin{center}
\includegraphics[width=1.\textwidth\vskip -.0in]{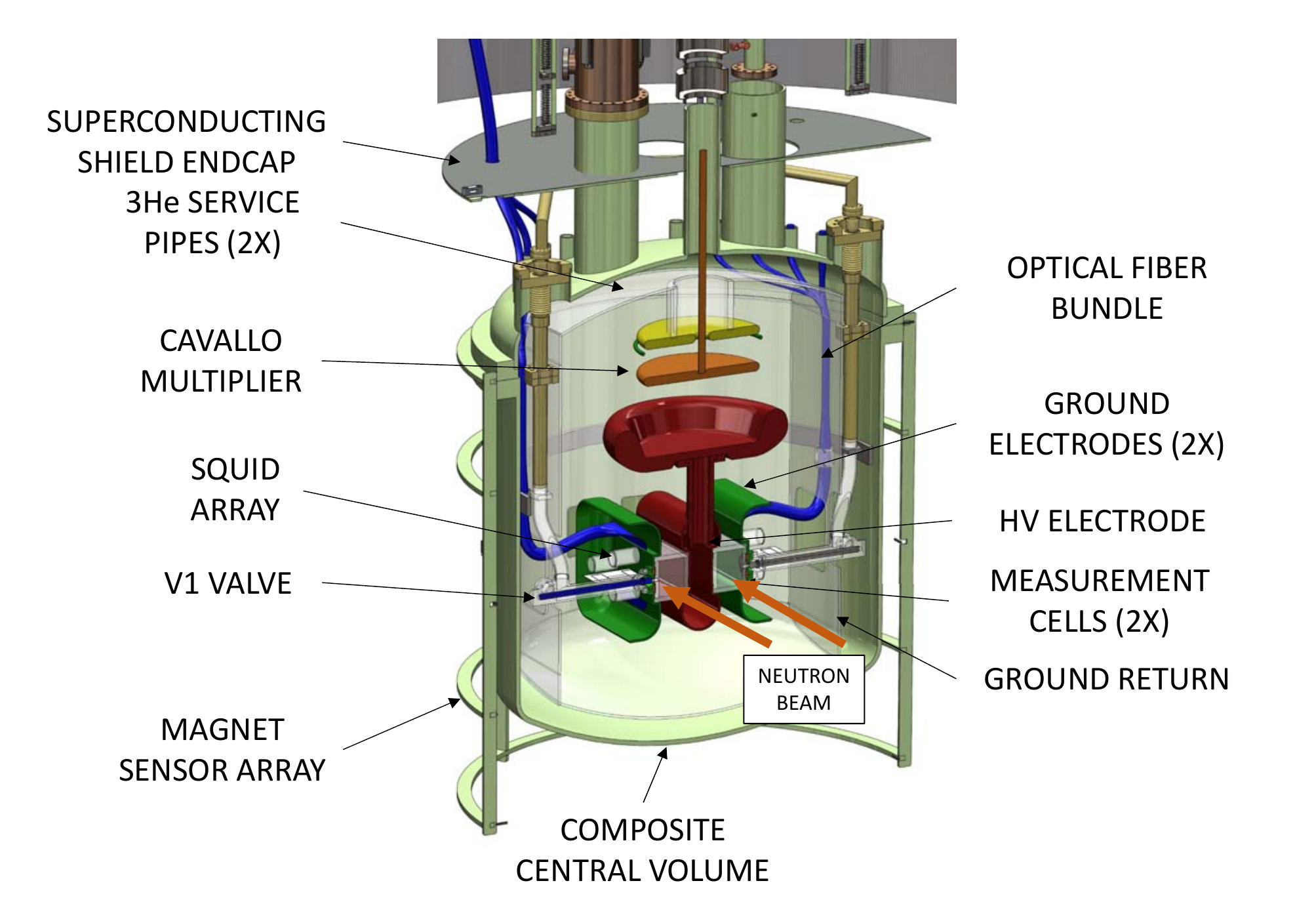}
\vskip -.1in\caption{Detailed view of the components of the central detector system within the composite central volume.}
\label{fig:CDSDetail}
\end{center}
\end{figure}

Immersed in the 1600~L 0.4~K LHe in the Central Volume (CV) is the
Central Detection System (CDS). Its functions are:
\begin{enumerate}
\item Produce UCN from the 0.89~nm cold neutron beam, store the UCN and maintain their spin polarization.
\item Allow polarized $^3$He atoms to be introduced in the region in
  which UCNs are stored and maintain their spin polarization.
\item Apply an electric field in the region in which UCN are stored.
\item Detect LHe scintillation light produced as a result of
  $^3$He$(n,p)^3$H events.
\item Detect the change in the magnetic field caused by the rotating
  magnetization of the $^3$He atoms, using SQUID (superconducting
  quantum inference device)-based magnetometers, to determine the spin
  precession frequency of the $^3$He atoms.
\end{enumerate}
In order to realize these functions, the CDS consists of several
components that are closely integrated with each other. They are: 1)
measurement cells, in which UCN are produced from cold neutrons, and
UCN and $^3$He atoms are stored, 2) high voltage system, which
provides the necessary electric field to the volume inside the
measurement cells, 3) light collection system, and 4) SQUID
system. Figure~\ref{fig:CDSDetail} shows the CDS as it is
currently designed.

The CV vessel, the cylindrical vessel that contains the 1600~L
superfluid LHe in which the CDS is immersed, is oriented
vertically. This is an important design feature to allow for access to
the CDS components for maintenance and repair without breaking multiple 
cryogenic connections. There are two measurement cells in order to
control and assess certain classes of systematic effects (see Sec.~\ref{Systematics}). The two measurement cells are each sandwiched
between HV and ground electrodes, with the HV electrode common to both
measurement cells. A HV amplifier based on Cavallo's
multiplier~\cite{CAV95} is mounted on top of the HV electrode in
order to fit in the vertical LHe containing vessel. The light
collection system transports the scintillation light generated inside
the measurement cells to photo-detectors located outside the
cryostat. The measurement cells and the ground electrodes accommodate
valves that allow introduction of spin polarized $^3$He atoms into the
measurement cells and removal of depolarized $^3$He atoms from the
measurement cells. Mounted behind each ground electrode are
gradiometer loops to pick up the magnetic field oscillation due to the
spin precession of $^3$He atoms. The signal from the gradiometer loops
is sent to SQUID detectors mounted in the upper part of the CV.

The entire CDS needs to made of nonmagnetic material to meet the
requirements on the magnetic field uniformity (see Sec.~\ref{sec:magnet}
). Because of
this, the vessel that contains the 1600~L superfluid LHe will be made
of a composite material. For the same reason, the use of superconductors in large volumes needs to be avoided. For example, vacuum seals
made of indium, a standard practice in cryogenic systems, cannot be used because the superconducting transition
temperature is 3.7 K.  In addition, to keep Joule heating from
the spin dressing field low, the use of conducting materials is severely
limited. The CDS system is cooled by a dedicated dilution
refrigerator with a design heat budget is 80~mW. In the remainder of this section, each of the components will be described.

\subsection{Measurement Cells and $^3$He Entrance Valves} 
\subsubsection{System Requirements}
The measurement cells must meet the following requirements:
\begin{enumerate}
\item UCN will be produced from a 0.89~nm cold neutron beam inside the
  measurement cells. The cell walls that the beam traverses must
  be made of materials with high cold neutron transmission.
\item UCN will be stored in the measurement cells. The inner surface
  of the cell walls need to be made of, or coated with, materials with a high
  Fermi potential. The design goal of the SNS nEDM is to let UCN
  precess for $\sim$ 1000~s in each measurement cycle
  with a concentration of polarized $^3$He atoms
  optimized for EDM sensitivity. This translates to an average UCN loss lifetime
  in the measurement cells\footnote{The mean time a UCN would live in the
    measurement cell in the absence of free neutron $\beta$ decay and
    $^3$He atoms.}  of 2000~s to ensure cell-related losses are negligible, corresponding to a loss per bounce
  of $10^{-5}$.
\item As a result of the spin-dependent $^3$He$(n,p)^3$H reaction, LHe
  scintillation light will be produced at 80~nm. Since LHe is the only
  material that such short wavelength EUV light can be transmitted
  through, the inner wall of the measurement cells must convert
  the 80~nm light to longer wavelength light that can be transported
  to photo-detectors placed elsewhere in the
  experiment.\footnote{Detecting the 80~nm light directly in the
    measurement cell would eliminate this requirement. However, a
    detector that can detect 80~nm extreme ultra-violet (EUV) light
    directly in the measurement cell in a manner compatible with all
    other experimental requirements has not been identified.}
    \item The inner walls of the measurement cells must be made of, or
  coated with, a material that retains the polarization of $^3$He
  atoms.
\item Polarized $^3$He atoms must be introduced into the
  measurement cells and will need to be removed once they become
  depolarized. In order to achieve this, each measurement cell needs to be equipped with a valve
  that allows for introduction and removal of $^3$He atoms. These
  valves (called the V1 valves), when closed, must seal sufficiently to provide a $^3$He loss rate $< 5\times 10^{-4}$~Hz. In addition, these
  valves need to be sufficiently robust so that they can be cycled
  over 10,000 cycles without failure.
\item The material choice and construction of the measurement cells
  and the V1 valves must be compatible with the requirements from
  electrostatics mentioned below.
\item Materials for CDS should be chosen to minimize activation from direct and scattered cold neutrons.
\end{enumerate}

\subsubsection{Design Concept}
The measurement cells will be made of poly(methyl methacrylate)
(PMMA). They will be 10.16~cm$\times$ 12.70~cm $\times$ 42~cm in outer
dimension with a wall thickness of 1.2~cm. The front and back walls
will be made of deuterated PMMA (dPMMA) to allow for transmission of
the 0.89~nm neutron beam. The side wall will be made of less expensive
regular PMMA. The inner walls will be coated with deuterated
tetraphenyl butadiene (dTPB) in a deuterated polystyrene (PS) matrix,
which will provide a sufficiently high Fermi potential to UCN as well
as allow for conversion of the 80~nm EUV light to blue light. The TPB must be deuterated (dTPB) in order to maximize UCN storage time. The
choice to use PMMA as the cell wall material was driven by the cell
wall's function as part of the light collection system as well as its high purity to minimize neutron activation. TPB film in a
PS matrix has been shown to be an efficient converter of EUV light from
LHe scintillation~\cite{MCK97}. The conversion efficiency and the
emission spectrum of dTPB have been shown to be the same as those for
protonated TPB~\cite{GEH13}. dTPB has also been shown to have a
sufficiently long $^3$He depolarization time~\cite{YE08,YE09,YOD10}.
Prototypes of the measurement cell have been tested for UCN storage
time~\cite{TAN18a,LEU18} using UCN from the UCN source at Los Alamos National
Laboratory~\cite{SAU13,ITO17}.

The current design for the V1 valve is to make both the valve seat and
stem from dPMMA. Prototype valves made of PMMA have been shown to
be sufficiently tight at 4~K and have survived over
$\sim$10,000 cycles at 4~K with no degradation in
performance~\cite{TAN18b}. A prototype valve made of dPMMA has been
shown to work well to store UCN~\cite{TAN18a,LEU18}. 

\subsection{High Voltage System} 
\subsubsection{System Requirements}
The design goal of the SNS nEDM experiment is to create a stable
electric field of 75 kV/cm in the region inside the measurement cells,
thereby giving an almost order of magnitude gain in sensitivity
from a larger electric field alone. This goal is based on the
expectation that {\it LHe is a better electrical insulator than
  vacuum}. The bulk of HV-related R\&D has been focused on
experimentally demonstrating that it is possible to apply a stable
electric field greater than 75~kV/cm in conditions approximating those
to be encountered in the SNS nEDM experiment. So far, the
collaboration has demonstrated that the required electric field can be
achieved in a system that is about a factor of 5 smaller in each
dimension compared to the SNS nEDM experiment's measurement cell
electrode system~\cite{ITO16}.  

There are various requirements on the materials used for the
electrodes. They are:
\begin{enumerate}
\item The measurement
  cells will be made of PMMA, which shrinks $\sim$1\% when cooled from room
  temperature to 0.4~K. As a result the electrodes need to be made of a material
  that has similar thermal contraction characteristics to PMMA. 
\item The material cannot have too high an electrical
  conductivity. This requirement comes from the requirement on Johnson
  noise on the superconducting quantum interference device
  (SQUID)-based magnetometer to measure the precession frequency of
  spin polarized $^3$He atoms and also from the requirement on joule
  heating from eddy currents due to the radio frequency (RF) field for
  dressed spin measurement. The allowed surface resistivity is
  100~$\Omega/\text{square} < \sigma < 10^8$~$\Omega/\text{square}$ at the
  operating temperature of $\sim$0.4~K.
\item The material must be non-magnetic. The static magnetic field
  in the region inside the measurement cells, which is approximately
  3~$\mu$T, must be uniform to $\sim 1\times 10^{-4}$ and needs to have
  field gradients smaller than $\sim$10~pT/cm in the direction of the
  static field and $\sim$5~pT/cm in the direction perpendicular to the
  static field. Because of this stringent requirement, many of the
  so-called ``non-magnetic'' technical materials, such as stainless
  steel and inconel, are disallowed. Also, materials that become
  superconducting cannot be used because the field expelled due to the
  Meissner effect would disturb the field uniformity inside the
  measurement cells.
\item The material must not have large neutron absorption
  properties, as such materials would become radioactively activated
  due to the exposure to a high flux neutron beam and become a source of
  background radiation. 
\end{enumerate}

In addition, the leakage currents along the cell walls must be
minimized. This requirement comes from the following considerations.
\begin{itemize}
\item These currents produce magnetic fields correlated with
  the direction of the electric field and therefore can produce
  effects that mimic the signal of nEDM (this applies to all nEDM
  experiments that use stored UCN).
\item The necessary HV will be generated inside
  the LHe volume using a Cavallo amplifier with the HV electrode disconnected from the
  HV power supply. As a result, leakage currents lead to a reduction of the
  electric field over time.
\item These currents produce heat, generating phonons in
  superfluid LHe and modifying the spatial distribution of
  the $^3$He atoms via the $^3$He-phonon interactions.
\end{itemize}

\subsubsection{Design Concept}
\noindent
{\it Measurement cell electrodes}\\
\noindent
The current design is to use PMMA as the electrode substrate material
and make the electrode surface conducting using methods including:
(1)~coating with appropriate materials and (2)~implanting conducting
materials into the surface layer. One promising candidate is copper
implantation. In tests using a system that can accommodate 
electrodes that are about 20\% scale in each linear dimension~\cite{ITO16}, it was 
demonstrated that it is possible to stably apply an electric field
exceeding 75~kV/cm even in the presence of an
object made of PMMA sandwiched between electrodes (simulating the
presence of the measurement cells) in a wide range of pressures
between a few torr to 600~torr at 0.4~K~\cite{ITO18}.

This apparatus was also used to study leakage currents. At a maximum field of 40 kV/cm, it was
demonstrated that leakage currents flowing between the two
electrodes through the surface of the PMMA dummy cell sandwiched between
the electrodes can be kept to less than 1~pA by proper design of the
electrode shape and proper cleaning of the PMMA dummy cell~\cite{ITO18}.

\begin{figure}
\centering
\includegraphics[width=\textwidth]{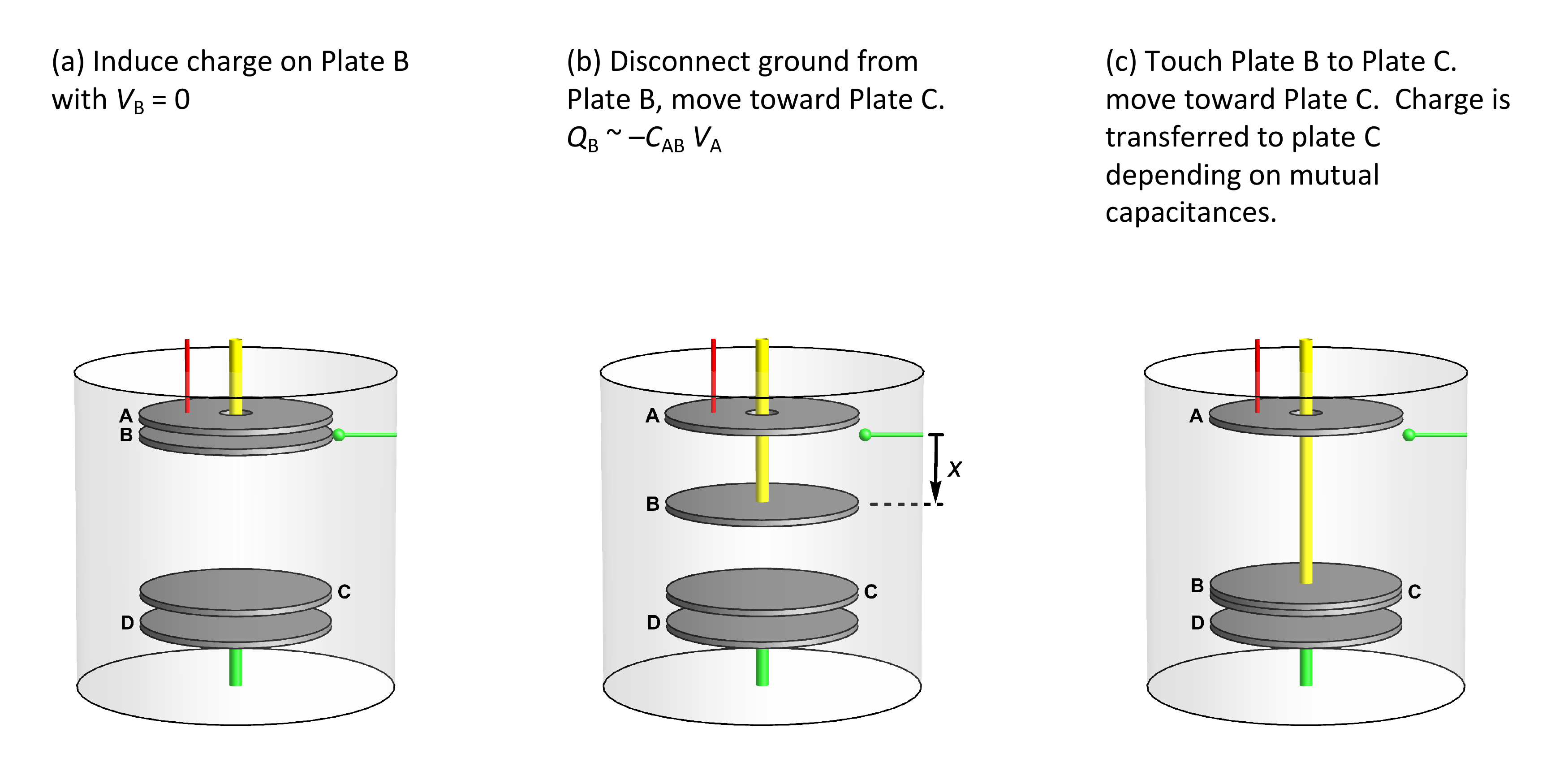}
\caption{A schematic showing the principle of Cavallo's multiplier. }
\label{fig:Cavallo}
\end{figure}

{\ }\\
\noindent
{\it HV generation}\\
\noindent
The application of 75~kV/cm across the 10.16~cm measurement cell (of
which, 7.62~cm is LHe and 2.54~cm is PMMA) requires an electric
potential of 635~kV. 
The necessary 635~kV will be generated inside the LHe volume within
the CV using a method based on Cavallo's
multiplier~\cite{CAV95}. A schematic of the principle of this
method is shown in Fig.~\ref{fig:Cavallo}. In the figure, electrode
$C$ represents the HV electrode, which must be charged up to
635~kV, and electrode $D$ represents the ground electrode. Electrode
$A$ is connected to a high voltage power supply with modest high
voltage ($\sim$50~kV). Movable electrode $B$ is initially grounded
(Fig.~\ref{fig:Cavallo}~(a)). A charge $Q_B \simeq -C_{AB} V_A$, where
$C_{AB}$ is the capacitance of the capacitor formed by electrodes $A$
and $B$, is induced on electrode $B$ by electrostatic
induction. Electrode $B$ is now moved toward electrode $C$, being
disconnected from the ground (Fig.~\ref{fig:Cavallo}~(b)). When
electrode $B$ comes in contact with electrode $C$, a fraction of the
charge on electrode $B$ is transferred to electrode
$C$. The fraction of the charge transferred from $B$ to $C$ is
determined by $C_{CD}$, $C_{BG}$, and $V_C$, where $C_{CD}$ is the
capacitance of the capacitor formed by electrodes $C$ and $D$,
$C_{BG}$ is the capacitance of the capacitor formed by (the back of)
electrode $B$ and the surrounding ground, and $V_C$ is the potential
of electrode $C$. The process can be repeated to increase the charge
on $C$, thereby $V_C$, until no more charge can be transferred   from
$B$ to $C$, which occurs when $Q_B = C_{BG} V_C$. The system can be
discharged or the HV of the opposite polarity can be applied by
repeating the same process with the voltage of the opposite polarity
applied on $A$.

As seen above, the maximum attainable voltage depends heavily on the
geometry of the system. Calculations~\cite{CLA18} have shown that potentials in excess of 635~kV can be generated using a realistic
geometry for the SNS nEDM experiment. A room temperature prototype has
been constructed, which showed the expected voltage
amplification~\cite{CLA18}.

Note that by generating the necessary voltage in LHe within the CDS
volume, it is possible to avoid dealing with feeding 635~kV directly
from outside the cryostat into the electrodes inside the 0.4~K LHe
volume. Such a task would be a formidable technical challenge, as it
requires a HV cable at 635~kV with a minimum heat
load and a superfluid-tight, non-magnetic, high-voltage feedthrough
that provides a minimum heat load (80~mW corresponds to a surface
leakage current of 0.1~$\mu$A at 635~kV).

\subsection{Light Collection System} 
\subsubsection{System Requirements}
\label{light_coll_req}
As described in Section~\ref{sec:NEDM_Concept} the neutron precession
frequency $\omega_n$ is obtained by monitoring the $n+\rm{^3He}$
capture rate, with its strong dependence on the angle between the
neutron and $\rm{^3He}$ spin directions. Each $n+\rm{^3He}$ capture
event deposits a fixed amount of kinetic energy (764\,keV) into the
superfluid helium used to generate UCN, resulting in a burst of
several thousand extreme ultraviolet (EUV) photons $(\lambda \approx
80\,\rm{nm})$~\cite{FLE59, THO59, SIM61, ADA95, BAN96, MCK03}. The
light collection system must contend with the following
considerations:

\begin{itemize}
\item Suppression of background events (primary sources are neutron
  $\beta$ decay, $\gamma$ rays resulting from long-lived neutron
  activation products, and cosmic rays) is achieved by applying a cut
  on the detected $n+\rm{^3He}$ capture signal size, which is peaked
  by virtue of being mono-energetic. The efficiency and effectiveness
  of this cut improves with improved energy resolution, which is
  primarily driven by photon collection efficiency.

\item The EUV photons must be converted to optical wavelengths at the
  surface of the measurement cell since they are absorbed within a few
  tens of nanometers in any material other than helium. In addition to
  good light yield, the wavelength shifter coated on the surface of
  the measurement cell must also have long time constants for loss of
  $\rm{^3He}$ polarization and UCN polarization and density.

\item The optical photons resulting from EUV conversion on the
  measurement cell walls must be extracted from a region at low
  temperature ($T < 0.5\,\rm{K}$) and within a high electric field
  ($E=75\,\rm{kV/cm}$), without introducing an unacceptable heat load
  or leading to electric sparks, and then transported a significant
  distance ($> 1$\,m) through at least one cryogenic vacuum
  interface to photosensors held at a warmer temperature.

\item Pulse shape analysis may help with some sorts of
  background
  and event position information may
  prove useful to identify backgrounds and/or diagnose systematic
  effects. Thus preserving both types of information is a design goal.

\item Finally, the scintillation yield from $\alpha$ interactions in
  liquid helium has been shown to depend on the magnitude of the
  electric field, with a reduction of 15\% at the operating field
  (75\,kV/cm)~\cite{ITO12}. A comparable reduction is expected for
  scintillation from neutron capture products~\cite{ITO13},
  providing a way to monitor the electric field in the measurement
  cell if the optical gain (the product of photon detection efficiency
  (PDE) and electronic gain) can be monitored to 0.5\% between $E=0$
  calibration runs.

\end{itemize}

\subsubsection{Design Concept}
These considerations have led to the following design choices:

\begin{itemize}

\item Coatings incorporating deuterated tetraphenyl butadiene (dTPB)
  are applied to the measurement cell walls to convert the EUV
  scintillation photons into blue light. Deuteration is required to
  minimize UCN wall losses. Evaporated dTPB coatings produce the most
  light~\cite{MCK04} but the resulting surface roughness is
  expected to lead to relatively small time constants for UCN storage
  and $\rm{^3He}$ depolarization. Coatings made by co-dissolving dTPB
  and polystyrene (dTPB+dPS) have been shown to have all the desired
  properties, as mentioned above.

\item For each measurement cell the blue light produced by the dTPB
  coating is captured by an array of 176 wavelength shifting (WLS) plastic
  optical fibers mounted on the cell wall adjacent to the ground
  electrode. This allows the fibers to be pulled through slots in the
  ground electrodes into a low-field region without producing a localized regions of increased electric field. 
  The other cell walls are
  covered by a dielectric mirror film\footnote{Vikuiti
    VM2000. Measured reflectivity $>90$\% at 6K~\cite{PAT14}} to
  allow efficient collection of all the light. By mounting the fibers
  perpendicular to the long axis of the cell modest position
  resolution (few cm rms) is obtained.

\item From behind the ground electrode the fibers are then routed to
  the top head of the central volume (see Fig.~\ref{fig:CDSDetail}). The
  flexibility of the fibers is a great advantage given the complicated
  optical path. Modern WLS fibers have fairly long attenuation
  lengths~\cite{BUG14}, so optimum light collection efficiency may
  be obtained by routing the WLS fibers all the way to the
  photosensors. However, efforts to fabricate a reliable cryogenic
  vacuum-tight multi-fiber feedthrough have been unsuccessful. Instead
  the WLS fibers are potted\footnote{Stycast 1266} into an array of
  precisely located blind holes in an acrylic block, which is
  subsequently diamond-machined to expose and polish the fiber ends. A
  mating array of clear fibers is mounted on the opposite side of the
  vacuum break 
  with somewhat wider fiber diameter to
  minimize sensitivity to fiber positioning error.

\item The clear fibers are read out by arrays of silicon
  photomultipliers (SiPMs). Modern devices have excellent Photon Detection Efficiency - PDE -  (
  $>50$\% at green wavelengths~\cite{OTT17} ) and each fiber is
  aligned with an individual SiPM, eliminating inter-sensor geometric
  inefficiency.

\item As the SIPM over-voltage ($V_{\rm over}$, the difference between
  the applied voltage $V_{\rm app}$ and the breakdown voltage $V_{\rm
    br}$) is increased PDE increases, but so does optical cross-talk
  (when a single photon generates multiple photoelectrons). The former
  effect improves the energy resolution, the latter worsens it. Dark
  rate also increases with increased $V_{\rm over}$. A photon counting
  technique, discussed below, is used to eliminate cross-talk and the SiPM dark rate is
  suppressed by cooling. Together these strategies relax upper limits
  on $V_{\rm over}$ and therefore improve PDE and energy resolution.

\item The photon counting technique takes advantage of the fact that
  the expected number of photons hitting an individual SiPM in a
  $n+\rm{^3He}$ capture event is well below one. As a result it is
  safe to assume that every signal corresponds to a single
  photon. This assumption is implemented by discriminating the analog
  output of each individual SiPM, which provides a digital signal
  time-correlated with the photon arrival. By measuring the arrival
  time of every individual photon the event waveform is preserved
  without needing to record waveforms.

\item An additional benefit of operating at higher $V_{\rm over}$ is
  that the SiPM quantum efficiency plateaus. As a result, PDE
  sensitivity to $V_{\rm over}$ variation (due to $V_{\rm app}$ drift, or
  temperature drift which changes $V_{\rm br}$, or sensor-to-sensor
  $V_{\rm br}$ variation) is greatly reduced. Furthermore the photon
  counting technique essentially eliminates electronic gain
  variation. Any remaining time-dependence in the optical gain is
  monitored by illuminating the WLS fibers, near the measurement
  cells, with two ``standard candles'': 1) an $\alpha$ source plated on
  a wire and embedded in a layer of scintillator paint, and 2) a
  pulsed LED~\cite{HAN04} which has its light split between a
  room-temperature photodiode (used to normalize the LED output) and
  an optical fiber used to deliver the light to the WLS fibers near
  the measurement cell.

\end{itemize}

The performance of the light collection system has been optimized and
evaluated using a fairly realistic prototype~\cite{CIA18} that uses
$^{210}$\,Po $\alpha$ particles to produce EUV scintillation in cells
coated with TPB-based materials and filled with liquid helium
(obtained by condensation of helium gas passed through a liquid helium
cold trap). Based on these studies, the expected light yield for $n+\rm{^3He}$ capture events
is $\approx$\,20 photoelectrons.

\subsection{SQUID System}
\label{SQUIDS}

\subsubsection{System Requirements}
In the ``free-precession'' measurement mode of the nEDM apparatus, the $^3$He co-magnetometer precession frequency will be measured by direct detection of the magnetic field created by the polarized $^3$He filling the measurement cells. At a $^3$He concentration of $\approx 10^{-10}$ relative to $^4$He, the concentration optimized for statistical precision of the nEDM measurement in terms of $n + ^3He$ capture event yield and free-precession time, the signal outside the cell is of order a few fT. With a nominal $B_0$ field of 3~$\mu$T, the $^3$He precession frequency is $\approx$100~Hz.

Superconducting quantum interference devices (SQUIDs) are extremely sensitive
detectors of magnetic field and are a natural choice to read out the
$^3$He magnetization in the SNS nEDM experiment, due to the
electromagnetically-shielded and cryogenic environment in the central
detector of the apparatus.
Magnetic field sensitivities of $\approx$1~fT/$\sqrt{\rm Hz}$ are routinely achieved
in ultra-low field nuclear magnetic resonance (NMR)
experiments~\cite{Matlashov2004,Burghoff2005}.
If a similar ultra-low noise performance is achieved by the co-magnetometer readout
for SNS nEDM, with reasonable assumptions for the signal amplitude,
this system will have negligible contribution to the overall
uncertainty in the nEDM measurement~\cite{Kim2013}.

Constraints on local RF and magnetic shielding within the nEDM central volume
prevent some of the usual best practices for ultra-low field, SQUID-based NMR,
such as wrapping the pickup loops with gold-coated mylar and keeping the leads
short.  The size and complexity of the nEDM apparatus also present challenges
in preventing noise from interfering with the SQUIDs.
Although it has been shown that a candidate SQUID and pickup loop combination
has sufficiently low intrinsic noise~\cite{Kim2015}, this is the lower
limit on the noise floor and will likely be higher in the nEDM apparatus.
High frequency electromagnetic interference (EMI) presents as a broadband increase
of the SQUID noise~\cite{Cantor2004}.  Unwanted low frequency magnetic signals,
for example from pickup loop vibrational motion in a magnetic field gradient,
can couple to the SQUID via the pickup loop the same as the desired
$^3$He magnetization signal.  High frequency EMI is addressed by careful
systems engineering of the entire apparatus, defining the cryostat vacuum
vessel and external shielded racks for electronics as the Faraday cage.
In the most conservative approach, all electronics
inside the Faraday cage are battery powered, avoiding any connection to
building mains power (this requirement can be relaxed later during commissioning,
if heavily-filtered mains power turns out to be acceptable).

Unwanted magnetic signals coupled into the pickup loops, with frequency near
the $\approx$100~Hz $^3$He precession frequency, are addressed by the
design of the pickup loops.  The original pickup loop arrangement, shown
in Ref.~\cite{Kim2013}, is an array of large-area planar gradiometers
behind the ground electrode.  Gradiometers, composed of two counter-wound
loops separated by some distance (the gradiometer ``baseline''),
are preferred, since they are much less
sensitive to a far-away magnetic field source than a single loop (magnetometer)
while retaining good sensitivity to nearby sources.
Figure~3 of Ref.~\cite{Kim2013} shows results of a calculation of sensitivity
to a magnetic field source versus position of the source, which falls
with distance from the pickup loops with characteristic length related to
the gradiometer baseline.  Since real gradiometers have some amount of imbalance
(sensitivity to a uniform field), and even with ideal gradiometers there
could be sources of gradient noise (e.g., vibrational motion in a static
gradient), additional magnetometer channels can be added to help ``tune''
away these unwanted signals in offline data analysis.

The Cavallo multiplier presents an additional concern with respect to
the SQUIDs: the possibility of sparks during Cavallo multiplier operation
causing trapped magnetic flux in the SQUIDs.
Exposure to strong magnetic fields is known to cause SQUIDs to become
inoperable due to excessive trapped magnetic flux at the
Josephson junctions~\cite{Uchida1983}, persisting even after the
strong field is turned off.
This is typically solved by sending a heat pulse to the SQUID chip to briefly heat
the junctions above the superconducting transition temperature (about 9.3~K for Nb).
This injects 0.5~J of heat per SQUID into the central volume, which will
raise its temperature and require a period of time before the dilution
refrigerator restores normal operating temperature to the nEDM central volume.
Due to this recovery time, it may not be practical to perform the SQUID
thermal cycling before each measurement cycle.
If a Cavallo charging cycle is executed between each nEDM measurement to
top off the high voltage, the SQUIDs need to be robust to trapping flux
in this case.  Fortunately, the spark energy from topping off the voltage
is not expected to be as high as in the first cycles.
The approach to this problem will be two-fold:
1) The vulnerability of the SQUIDs to trapping flux due to high voltage
discharges will be tested with candidate pickup loops and electrodes, with
particular attention to the spark energy that tends to cause trapped flux;
2) A new idea to remove trapped flux with negligible heat input will be
developed based on Ref.~\cite{Matlashov2016,Matlashov2017}, such that the SQUIDs could be defluxed
before every nEDM measurement period if necessary.

\begin{figure}
\centering
   \includegraphics[width=0.40\textwidth]{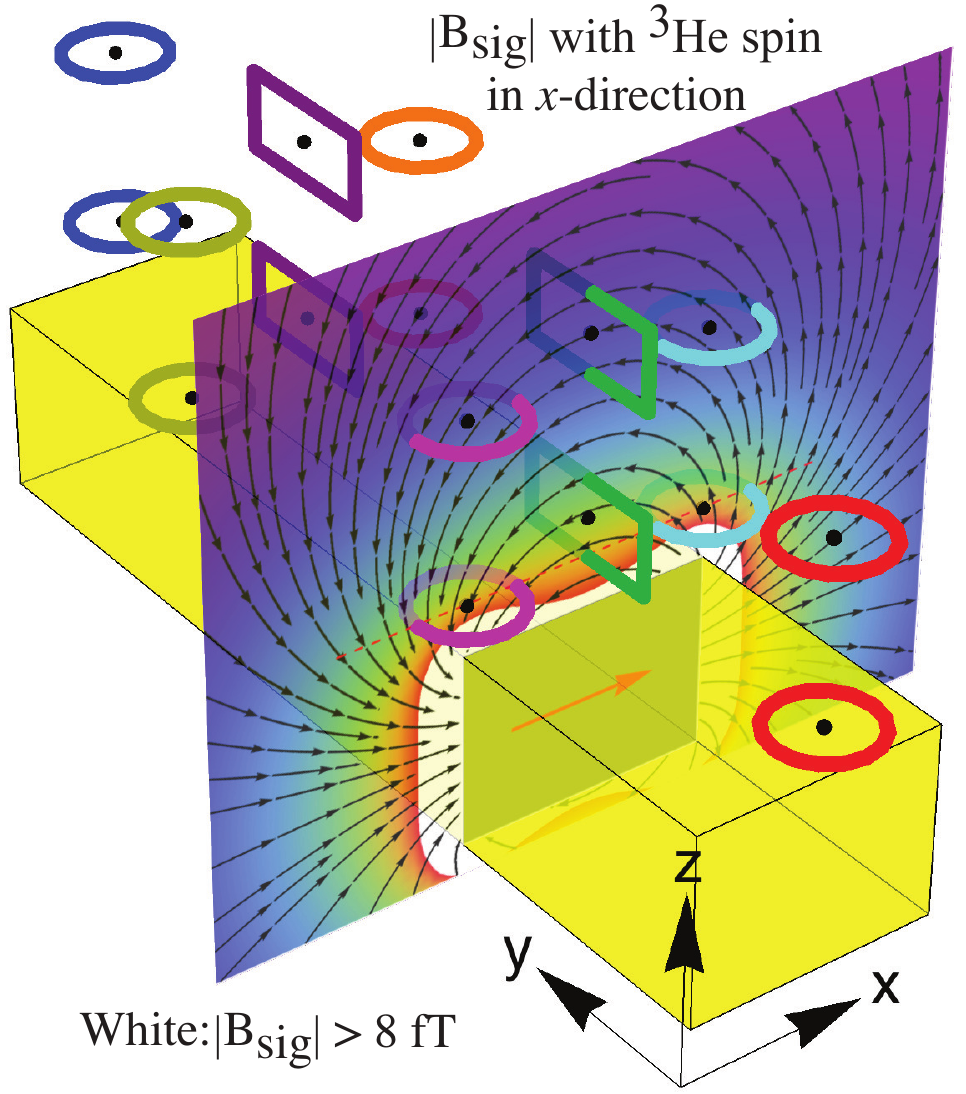}
   \includegraphics[width=0.40\textwidth]{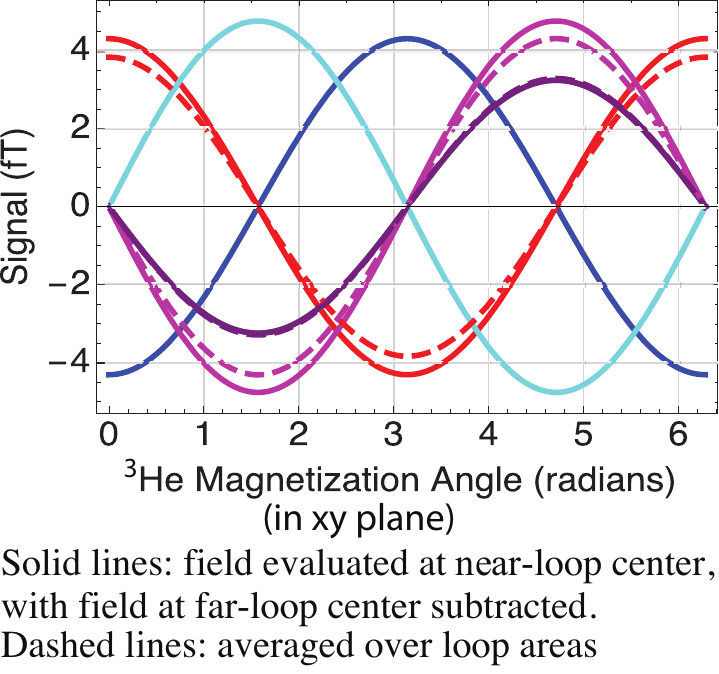}
\caption{Pickup loop arrangement using both axial (circles) and planar (rectangles)
gradiometers, with the nearest part of the pickups spaced 2.0~cm from the
inside cell wall and behind the ground electrode. Also shown is the expected field from 100\% polarized,
$2.2\times 10^{12}$~$^3$He atoms/cm$^3$ in the cell. Here, $\vec{B}_0$ and $\vec{E}$
are along the $z$-axis. The colors of the pickup loops refer to the various signals plotted in the right figure. 
%
}\label{fig:pickups_drawing}
\end{figure}

\subsubsection{Design Concept}
The pickup loop arrangement is
shown in Fig.~\ref{fig:pickups_drawing}, in which loops of the same color
comprise gradiometers (i.e., the loops further from the cell are wound in opposite direction).
This placement of the axial pickups (circular coils wound on a cylindrical form)
allows for both
differential and quadrature (90$^\circ$ phase-shifted) readout
of the $^3$He precession signal, by combining
symmetrically-placed pickup loops in offline data analysis.
The placement of the axial gradiometers takes advantage of the field
enhancement at the edges of a uniformly magnetized block.  Here, magnetometers
(loops closest to the cell only) may actually be preferred rather than gradiometers,
since a gradiometer can be formed offline from pairs of channels.
Additional channels to act as reference magnetometers may not be necessary,
reducing the overall number of SQUIDs.  This pickup loop arrangement
could increase the signal-to-noise ratio of the $^3$He precession signal
readout by a factor of $\sim$2 or better, compared to simple gradiometers, reducing the risk of this measurement
significantly increasing the overall uncertainty of the nEDM result.
The large-area planar gradiometers are retained in this design, as they may allow
additional opportunities to reject unwanted signals when combined with other
channels.


\section{Magnetic Field Module}
\label{sec:magnet}
\subsection{System Requirements}

The magnetic field system is designed to provide a precisely controlled magnetic environment. This includes the uniform DC holding field (called $B_0$) for spin precession and other AC fields (up to several kHz) for spin manipulation. Specifically the AC field from the spin dressing coil will be used for the $\pi/2$ pulse as well as spin dressing. $B_0$ is a 3 $\mu$T field with uniformity at the level of several ppm/cm. The nominal $B_0$ field must produce neutron and $^3$He coherence times due to transverse relaxation ($T_2$) greater than 10$^4$~s. In addition, the false EDM due to frequency shifts linear in E that arise due to field gradients must be below $d_f<2\times10^{-28}$~e-cm for both $^3$He and neutrons. Magnetic field uniformity constraints for neutron EDM experiments are discussed in detail in Ref.~\cite{Abel19}. These goals require uniform gradients in $B_0$ to be $<3$~ppm/cm, and uniform gradients in transverse fields to be $<1.5$~ppm/cm. The limits for the uniformity of the spin dressing field are increased by roughly a factor 15, or $<45~$ppm/cm. However, there is no constraint on transverse spin dressing gradients~\cite{Swank2018}. Furthermore, ambient environmental magnetic fields need to be be shielded to better than 1 part in 10$^4$ and magnetic drifts over the measurement time should be controlled to 1 part in 10$^7$. 

\subsection{Design Concept Overview}

\begin{figure}
\centering
\includegraphics[width=\textwidth]{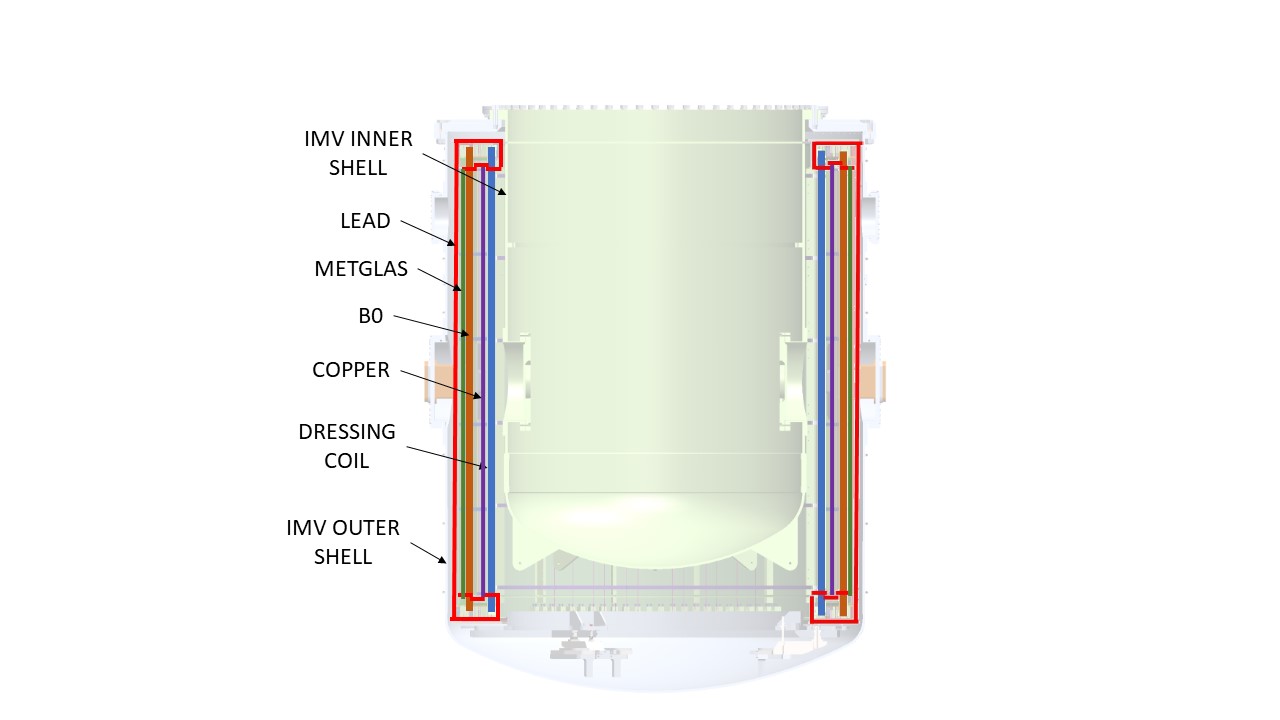}
\caption{Section view of the main magnetic field generating components of the Magnetic Field Module.} 
\label{fig:mag1}
\end{figure}

The main components of the magnetic field system include the lower cryovessel, lower LN2 shield and the Inner Magnet Volume (IMV) housing the main magnetic field generating components. These include, begining at the smallest radius, a spin dressing coil, gradient/shim coils, an eddy current shield, the main field coil $B_0$, a ferromagnetic flux return, and a cylindrical superconducting magnetic shield and superconducting endcaps for magnetic field shaping and reduction of magnetic field noise. These are shown in Figure~\ref{fig:mag1}. The magnetic system must simultaneously solve a series of complex design challenges, including magnetic optimization, neutronics, and cryogenics. The details of the mechanical design of the magnet system are presented next along with the key magnetic components.

\subsection{IMV Mechanical Design}

\begin{figure}[htbp]
\begin{center}
\includegraphics[width=\textwidth]{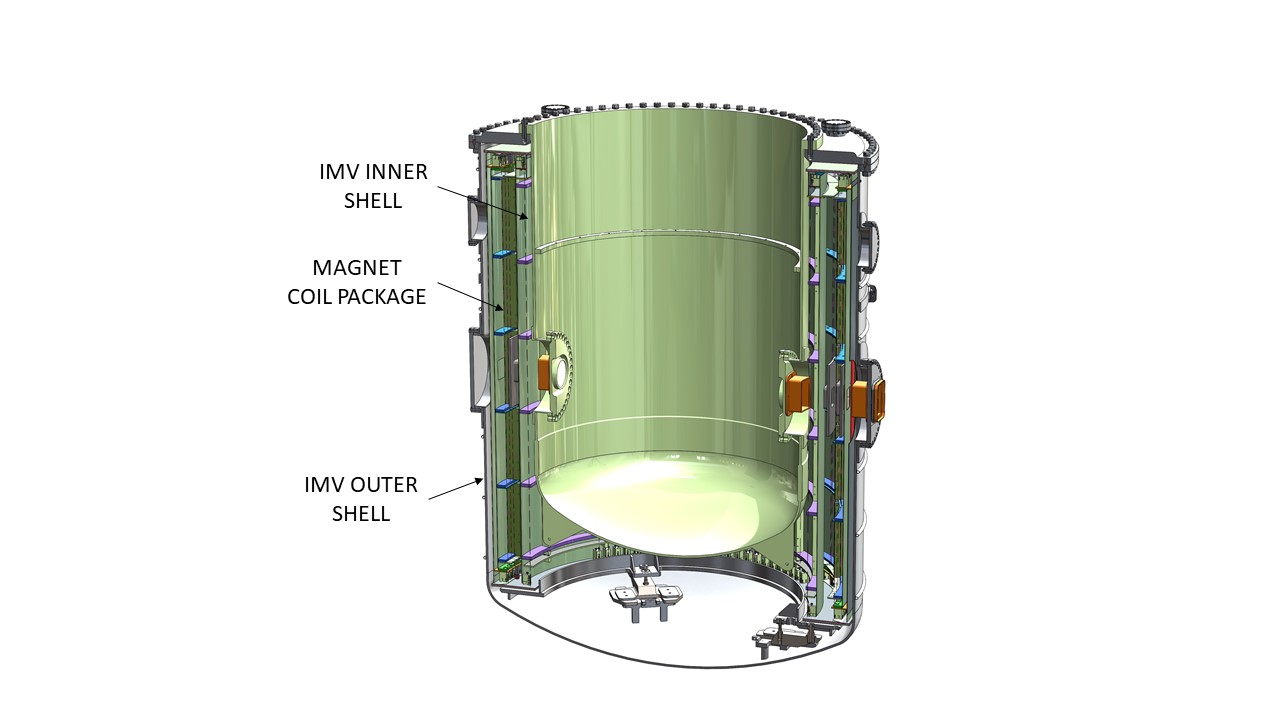}
\caption{The Inner Magnet Volume (IMV) which houses the components of the cryogenic magnetic field package. The outer wall and top flange are Al, while the inner wall is made from Glass-Fiber Reinforced Polymer (GFRP).}
\label{fig:magnetdetail}
\end{center}
\end{figure}

The mechanical design of the Inner Magnet Volume (IMV) is shown in Figure~\ref{fig:magnetdetail}. It is basically an annular vacuum volume that is separate from the vacuum of the main cryostat. This volume has an aluminum outer wall directly cooled with cold He. Internal components are cooled via a small amount of He exchange gas added to the magnet volume. The pressure in the inner magnet volume (IMV) must be large enough that the mean free path of the gas is shorter than the dimensions, but not so large as to add significant heat capacity. This is achieved with a pressure $\sim 10^{-2}$~torr. The inner wall of the IMV must be composed of non-metallic material to prevent eddy currents induced by the spin dressing coils from both distorting the magnetic field and providing an additional heat source to this cold volume. As the $B_0$ coil and dressing coils are meant to be superconducting ($B_0$ for current stability, dressing coils for reduced heat load) and the superconducting Pb shield must be below $T_C = 7.2$~K, the inner components must be easily cooled below 7~K. Specifically, the endcaps near the dressing coil must remain below 6.5~K, in order to keep the local magnetic field below $H_C$ for Pb. This constraint is due to the relatively large field generated by the saddle return of the spin dressing coil, which sit right above the endcap. 

The IMV is supported by the lower cryovessel, so that it can be removed as a package when the lower cryovessel is separated from the upper cryovessel and lowered to provide access to the components of the CDS. 

\subsection{Magnetic Design}

\subsubsection{Dressing Coil}
Interior to the $B_{0}$ coil, an optimized modified $cos(\theta)$ coil provides the oscillating dressing field to achieve the dressed spin configuration of the experiment~\cite{CT69} and the $\pi$/2 spin-flip to rotate the neutron and $^3$He spins by 90$^\circ$. The coil, which provides a uniform field to a large volume, must be operated at $\sim$ 1000 Hz with a current of $\sim$ 5-15 A, so care is needed to reduce losses that result in heating. In particular, commercial superconductors, which are clad in resistive copper, can generate large amounts of heat when the inductive impedance of the dressing coil becomes comparable to the resistance of the cladding, so that sizable currents can flow in the cladding. 

With this consideration only bare type I superconductors are adequate to build the spin dressing coil. 
The two best available bare type I superconductors, namely 50\%/50\% SnPb Solder and Ostalloy 203 (52.5\% Bi/32\% Pb/15.5\% Sn) are both adequate superconductors for the expected operating conditions (specifically critical field and temperature). 
If the superconductor remains below the critical field at the operating temperature, heating in the wire should be minimal due to the extremely short London penetration depth of the material, which is of order $10^{-7}~$m. When the solder is coated in Teflon for electrical isolation, it is found that the coating provides significant structural reinforcement as well. 

\subsubsection{Shim/Gradient coils} 
A set of shim/gradient coils are mounted onto the dressing coil structure to improve the uniformity of the main $B_0$ coil during normal operation and to provide known gradients during systematic studies. A minimal set of such coils includes two diagonal gradient coils (e.g. $dB_x/dx$) and three off-diagonal coils (e.g. $dB_x/dy$). The other possible linear gradients are then constrained by Maxwell's equations. Additional higher-order coils can be added based on initial measurements. 

\subsubsection{Eddy Current Shield} 
Eddy currents induced by the dressing coil would produce unacceptable heating in the Metglas flux return. To mitigate this, a Cu-clad G10 sheet is placed between the dressing and $B_{0}$ coils. A Cu thickness of 34 microns is chosen as a compromise between the effectiveness of the shielding to the Metglas and the additional Johnson noise generated by the shield as seen by the SQUID pickup loops (when the dressing coil is not energized). 
\subsubsection{$B_0$ Magnet}
It is critical for the nEDM measurement to have a uniform static magnetic field due to the linear-in-E frequency shift systematic and the coherence time, which are proportional to the inhomogeneities and to the square of the inhomogeneities, respectively.
The $B_0$ field is generated by a modified saddle-shaped cos$\theta$ coil, which is optimized to provide a sufficiently uniform $B_{0}$ field. The $B_0$ coil is a modified $\cos\theta$ saddle coil wound on a vertical cylindrical frame, producing a horizontal magnetic field. The coil spacings are optimized compared to a pure $\cos\theta$ coil in order to improve the magnetic field uniformity in the presence of the flux return (see~\cite{Galvan2011,Slutsky2017} for a discussion of this concept). Because of the small aspect ratio (length:diameter) of this coil, which is dictated by vertical space limitations, significant fringe fields at both the top and bottom of the coil could be expected. These fringe fields would produce a significant non-uniformity of the magnetic field if not for the superconducting endcaps which act to shield the saddle currents and create a magnetic mirror extending the effective length of the coil. This will be discussed below. 
 The magnitude of $B_{0}$ is chosen to be 3 $\mu$T so that the Larmor precession of the neutrons and $^3$He is near 100~Hz.  To maintain the polarization of the neutrons and $^3$He atoms, the magnetic field uniformity of $B_{0}$ should be $\sim 1.3\times10^{-4}$ averaged over each cell volume. The $B_0$ coil will be wound from superconducting wire (NbTi) and operated in persistent current mode. A separate requirement on the volume-averaged field gradients in the direction perpendicular to $B_{0}$ ($<$1~nT/m) is necessary to minimize the false EDM signal induced by motional magnetic field effects (see Sec.\ref{sec:falseEDM}).

\subsubsection{Ferromagnetic Shield} 

A cylindrical ferromagnetic layer placed just outside of the $B_{0}$ coil is included as a flux return and to improve field uniformity, mitigating the effect of errors in wire placement and reducing field distortions due to the cylindrical superconducting shield just outside of this shield.  Several layers of Metglas 2826M were chosen because, due to its amorphous nature, its permeability is preserved at low temperatures. In addition, this material has minimal cobalt content thus reducing neutron activation, as discussed in Section~\ref{section:neutronactivation}.

\subsubsection{Superconducting Shield}
A cylindrical superconducting shield made of 0.8 mm-thick high-purity Pb surrounds the magnets and Metglas flux return. This shield provides excellent magnetic shielding against external time-varying fields. The superconducting shield is a closed cylinder, with endcaps covering the two ends of the cylinder. The endcaps are dual purpose as they provide significant improvement to the overall shielding and act as magnetic mirrors. With perfect hermeticity and flatness the endcaps would provide magnetic uniformity equivalent to a coil with infinite length. Small penetrations in the lead will distort the field, but must be present for the operation of the experiment. However these distortions are small and decrease as $1/r^3,$ where $r$ is the distance from the penetration. 

Large annular gaps in the endcaps, due to the IMV inner wall, however, are not negligible. This is because the magnitude of the distortion is characterized by the size of the outer diameter of the annular gap, and not the size of the gap itself. However, with the addition of a field outside the IMV with the same magnitude as $B_0$, minimal distortions are generated to the internal field. With the addition of a magnetic shield enclosure coil (MSE coil), the infinite length approximation for $B_0$ is recovered. 

\subsubsection{Magnetic Cloak}
A material with sufficiently tuned magnetic permeability can be coated or wrapped on the outside of the lead. When the amount of flux attracted by the permeability of the cloak is equivalent in magnitude to the amount of flux repelled by the perfectly diamagnetic superconductor, then no distortions to the field will arise, and the field outside the cylinder of the cloak is nearly identical to a MSE coil field in the absence of the superconducting cylinder.  

One solution for a material with a tunable overall magnetic permeability is to space strips of Metglas with gaps calculated between them, wrapped circumferentially around the lead cylinder, forming rings. Allowing freedom in the spacing between the rings results in an effective permeability that can be tuned by changing the spacing. The effective permeability can be found by averaging over the total surface of the cylinder. This was determined to be sufficient from simulations in COMSOL over a range of spacings. The simulation achieved better than required results for a broad number of Metglas 2826M rings, and thus spacings. Adequate cloaking for the MSE field is found with 44 to 50 rings.


\subsubsection{MSE Coil}
A coil wound inside the magnetic shield enclosure coil (MSE coil), such that the MSE is the flux return for the MSE coil, naturally solves two difficult problems. It provides a field external to the IMV, mitigating any distortions that arise from penetrations and annuli in the endcap, and with sufficient uniformity it can provide the holding field for the polarized $^3$He transport and heat-flush.
The MSE coil is designed to provide a uniform field in the volume contained within the MSE, however interactions with the superconducting cylindrical shield inside the IMV will spoil the uniformity. Because the superconductor is perfectly diamagnetic, a large dipole will be generated when the magnetic shield enclosure coils are energized, unless the cylinder is magnetically cloaked as described in the next section.

\subsection{Cryogenic Design}
The magnetic requirements mentioned above place strict structural requirements on the coils. It was determined in COMSOL that random deviations in the wire positions of $\pm$ 0.5 mm would not lead to unacceptable magnetic gradients, so this was chosen as the design engineering tolerance. The frames for the coils and other components must be designed of non-metallic material, and must be cooled below 7 K, so that they do not present too large a thermal load. Additionally, with the IMV diameter of 2.3 m and other components of comparable size, thermal contraction is not negligible. 

\begin{figure}
\centering
\includegraphics[trim={4cm 0 0 0},clip, width=6cm]{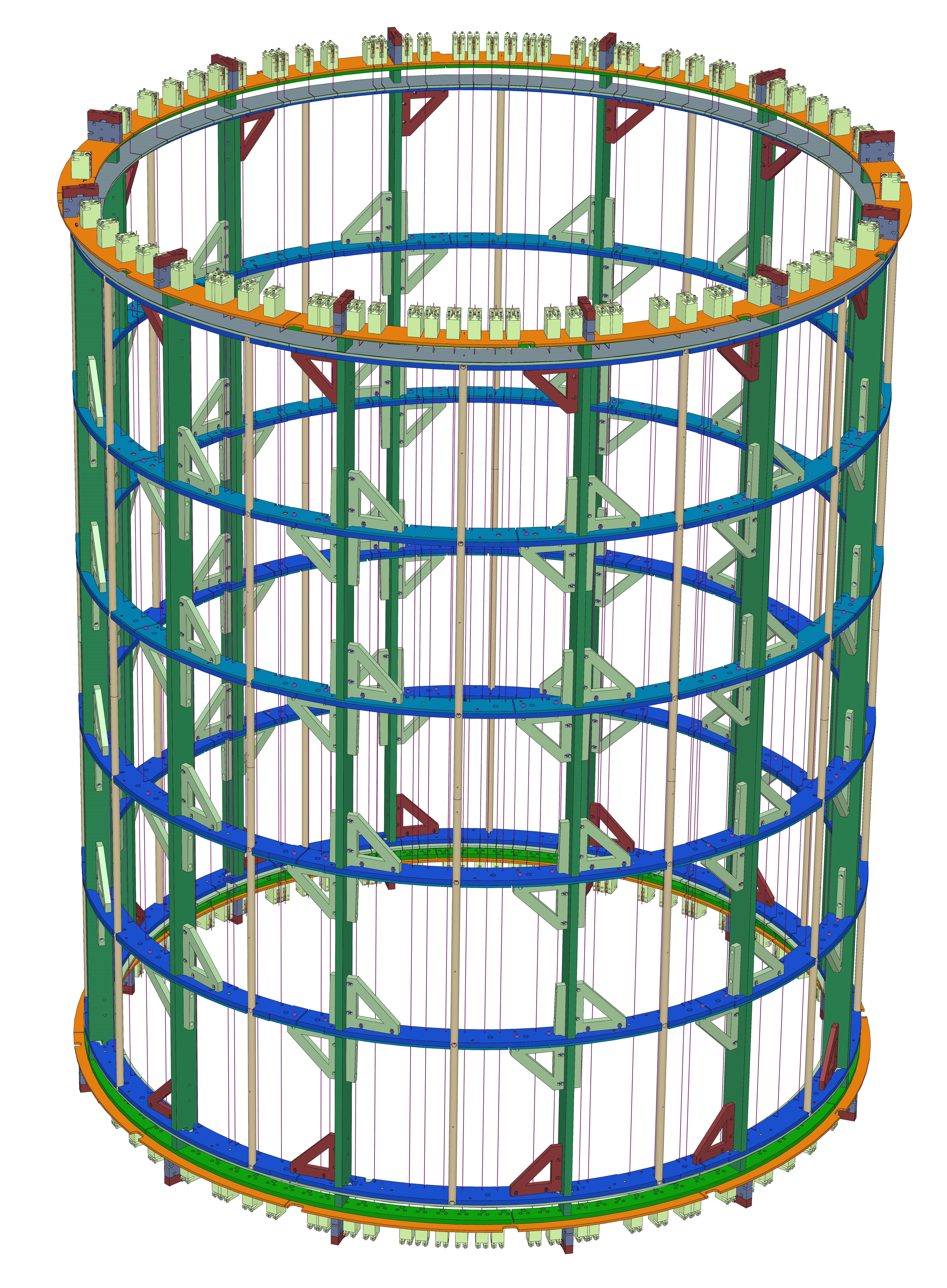}
\caption{Mechanical design of the $B_{0}$ coil.} 
\label{fig:B0CoilCAD}
\end{figure}

G10-CR was chosen as the material for the coil frames, as it is well characterized at cryogenic temperatures and has low thermal contraction. It will be used to construct rigid space frames from hoops connected by rectangular rods, with supporting gussets, as shown in Figure \ref{fig:B0CoilCAD}. Such frames minimize the thermal mass while maintaining the structural requirements. The frames also support sheets of G10-CR acting as skins on which to mount the lead, Metglas and copper shields. 

The coils will mount to a docking plate at the bottom, which itself mounts to pads in the IMV floor, ensuring sufficient clearance to avoid interference. The mounts of the dock plate are slotted to allow for radial contraction of the entire assembly. The IMV itself is expected to contract about 3 mm more than the coils.  


Tension will be maintained in the coil wires by pulley-like assemblies at the top and bottom containing a PEEK spring. The PEEK springs have been tested to 77 K and found to maintain spring behavior. By a fortunate coincidence, the total contraction of the $B_{0}$ NbTi(Cu) superconducting wire is very similar to that of the G10-CR, with a differential contraction of about 0.5 mm. 

The largest concern regarding thermal contraction is the differential contraction of lead and G10-CR, which differ by almost a factor of three. In many places, lead will rest on or be wrapped around G10-CR. In the former case, dimensions are chosen so that the lead does not overhang the G10-CR when cold. In the latter case, the lead may pull against the G10-CR. Epoxying the lead to the G10-CR seems to mitigate both issues, as shown by liquid nitrogen immersion tests. DP-190 gray epoxy successfully bonds G10-CR to lead through multiple thermal cycles, and the assembly can be bent when cold without tearing or breaking either material. 

\begin{figure}
\centering
\includegraphics[width=6.5in]{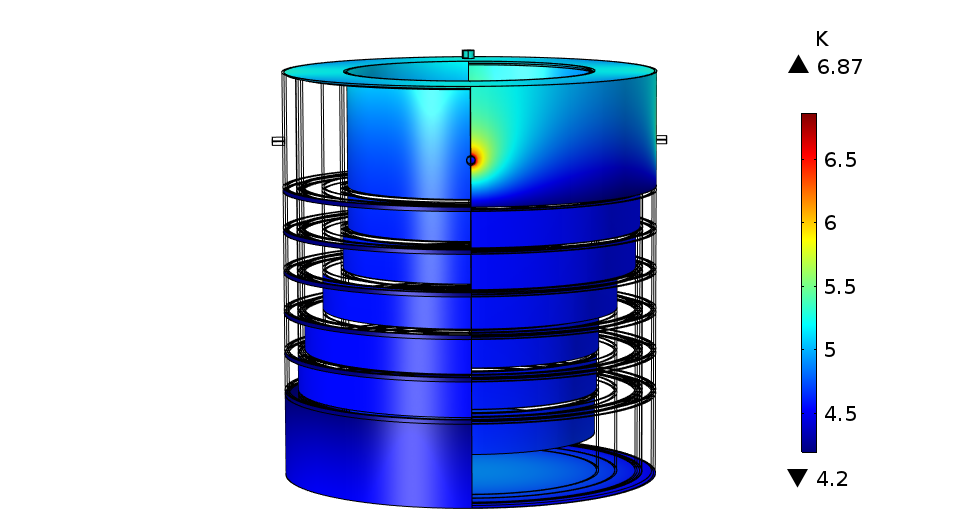}
\caption{Visualization of the steady state solution of the COMSOL thermal model for the inner magnet volume. The model contains the approximate geometry and the expected mass and materials determined from the design. Various shells (representing different parts of the coil package) have been selected in an attempt to show the temperature variation throughout the volume. The regions of the highest temperature are due to the G10 supports, which are in direct contact with 77 K.} \label{fig:IMVcold}
\end{figure}

Cryogenic modeling is completed in COMSOL~\cite{COMSOL}, as shown in Figure \ref{fig:IMVcold}. 
When a steady state is achieved ($\sim $6~K), approximately 3 W of cooling are required.  

\subsection{Magnetic Optimization}
The process to reach a final magnetic design was iterated several times. The initial design was based on a magnetic optimization of approximate positions of coils and shields. The results were used to generate a more detailed mechanical model, which lead to a more precise optimization, etc. This process was continued until a fully detailed mechanical and magnetically optimized system was completed. 

Magnetic optimization is addressed in COMSOL Multiphysics~\cite{COMSOL}. The magnetic diffusion equation is used to model the system, which includes the external MSE coil field, magnetic cloak, superconducting shield, superconducting endcaps, ferromagnetic flux return, Cu eddy current shield, spin dressing coil, and $B_0$ coil. The optimization uses the cryogenic positions of the design model, which are imported into COMSOL. The optimization of the cloak is completed first, to match the strength of the $B_0$ field and the cloaked field at the position of the annular gaps in the endcap. Then, within the IMV, each wire position along the spin dressing and $B_0$ circumferences is optimized so that $T_2$ is maximized for both coils. The model has octant symmetry so there is no reason (or ability) to optimize for the false linear-in-$E$ frequency shift, as this shift requires heterogeneities of odd spatial symmetry for it to manifest.

Despite the extensive effort to design a perfect coil, there will inevitably be defects in fabrication. Therefore shim coils that can reduce linear and quadratic inhomogeneities of the coil will be wound on the existing magnetic structure. To assist with initial setting for these shim coils, there will be a set of cryogenic fluxgate magnetometers that can provide an estimate of the field shape, this is discussed in Sect.  \ref{section:cryogenicprobes}. Furthermore, $T_2$ relaxation and $B^2$ frequency shifts are very sensitive measures of residual gradients, which will be zeroed by the shim coils when they have been characterized. In addition, the temperature dependence of the $^3$He false EDM can be used to confirm the suitability of magnetic field gradients. It should be noted that, while gradients in the $B_0$ field lead to the false EDM effect, gradients in the spin dressing coil do not~\cite{Swank2018}.

\subsection{Cryogenic Magnetic Field Monitor System}
\label{section:cryogenicprobes}

The purpose of the field monitor system will be to provide approximate information on the field gradients, $\partial B_i/\partial x_j$, present within the measurement cell region as reconstructed (via inversion of Maxwell's equations) from  measurements of the vector components of the field at a number of locations exterior to the measurement cells \cite{Nouri2014}.  Results from prototyping exercises \cite{Nouri2015} have informed the design, consisting of $\sim 30$--40 single-axis cryogenic-compatible fluxgate magnetometer probes mounted to the exterior of the central detector cryostat, with all of the probes located at least 25-30 cm from the measurement cells on account of each fluxgate magnetometer probe's intrinsic magnetization.

\subsection{Neutron Polarization and Transmission through Magnet System}
\label{section:neutronactivation}

In order to maintain an optimal magnetic field profile, the neutrons pass through both the Pb shield and at least one layer of the ferromagnetic flux-return made from Metglas. The magnetic properties and associated fields from these materials may lead to some loss of neutron polarization which should be kept below a few percent. In particular there may be losses before the Pb shield and between the Pb and Metglas layers, as well as losses in passing through the superconducting Pb and the Metglas.

For both the Pb superconducting shield and the Metglas, a simple approximation would be to assume that each surface behaves as a current sheet with different (and in some cases very different) magnetic fields on each side of the sheet. Generally this can be viewed as a non-adiabatic transition as the spin crosses the interface with too little time to allow any spin re-orientation. Thus one might expect a small depolarization in such a foil. A variety of Metglas samples were tested in neutron beams at both the Fundamental Neutron Physics Beamline at the Spallation Neutron Source and at the LENS facility at Indiana University. These tests included studies of the low-cobalt Metglas which is needed due to the large neutron capture cross section for cobalt and the long lifetime of the activation.  While these studies indicated some modest depolarization, this was attributed to the magnetic fields leaking out the top and bottom of these small foils due to a relatively strong applied vertical holding field. 

Measurements of both neutron depolarization and absorption in the magnet system will be performed 
before the full experiment is assembled and additional magnetic guide fields can be added if needed. In addition, the polarized $^3$He in the measurement cells also partially polarizes the incident neutrons during production of the UCN, e.g. for 90\% initial polarization, the polarization rises to 96\% by the end of the filling time.



\section{Polarized $^3$He System}
\label{sec:He3S}
\subsection{Overview}

\begin{figure}
\centering
\includegraphics[width=5in]{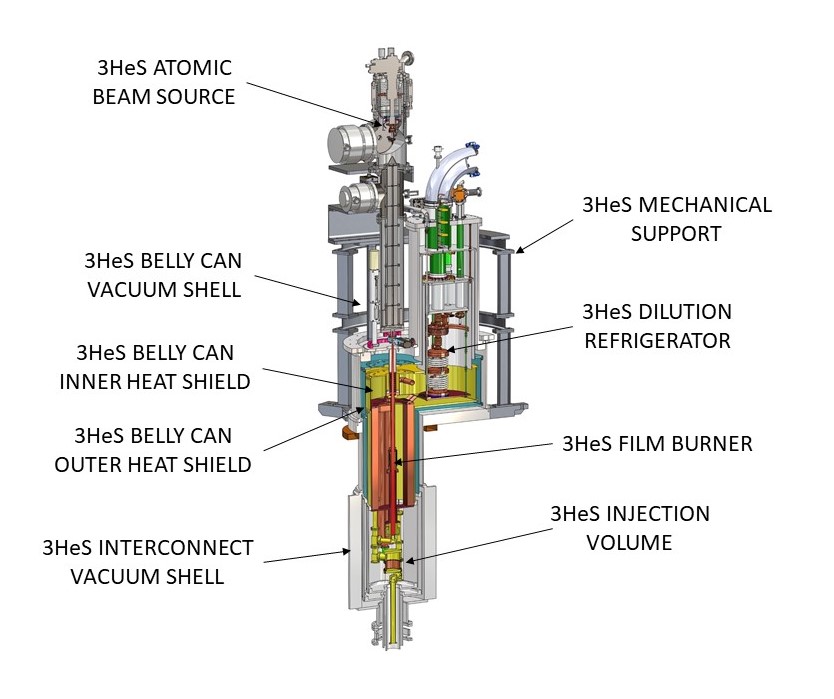}
\caption{Overall view of the $^3$He service subsystem with some of the major components identified.  For scale, the ``belly" flange has a diameter of 1.3~m and the overall height is about 4.4~m. Shields, thermal links and mechanical supports have been removed for clarity.}
\label{fig:HE3S_schematic}
\end{figure}

This system comprises hardware for three main tasks in the experiment cycle.  First, it injects polarized $^3$He from an atomic beam source (ABS) into a free surface of superfluid $^4$He.  Second, using a flux of thermal phonons (``heat flush''), it moves the polarized $^3$He from the injection system to the measurement cells.  Last, also using heat flush, it moves the partly depolarized $^3$He from the measurement system to the purifier system where it is removed and replaced with pure $^4$He. Fig.~\ref{fig:HE3S_schematic} shows the main components of this system.

The system sits atop the main cryostat, partially inside and partially outside the magnetically shielded enclosure (MSE).  Specially designed coils guide the polarized $^3$He beam from the ABS through the shield and into the MSE where a large coil provides a suitable holding field for the injection system. Below, an introduction to the heat flush concept is presented followed by the conceptual design of the three key components of the $^3$He services module. 

\subsection{$^3$He Transport via Heat Flush}

In order to move the polarized $^3$He from the injection volume to the measurement cells through connecting pipes with a total length of about 5~m, a ``heat flush'' technique~\cite{Golub07} is employed.  Here phonons, created in the superfluid $^4$He using a heater, flow from hotter to colder regions, scattering from, and therefore transferring momentum to the $^3$He with no effect on its polarization.  For the low $^3$He concentrations in the experiment, $X_3\equiv n_3/n \le 10^{-10}$, where $n_3$ and $n$ are the $^3$He and total fluid densities, the process~\cite{Baym15,Baym15a} can be characterized with a convection-diffusion equation
\begin{equation}
\frac{\partial x_3}{\partial t} + \vec v_{ph}\cdot \nabla X_3 = \nabla \cdot \left ( D_{3,ph} \nabla X_3 \right ),
\label{eqn:heatFlush}
\end{equation}
where $\vec v_{ph}$ is the phonon drift velocity, and $D_{3,ph}$ is the diffusion constant associated with $^3$He-phonon scattering
\begin{equation}
D_{3,ph} = \frac{3 T^2}{\Gamma} \sim \frac{0.88}{T_K^7}\hbox{\ cm$^2$/s}
\label{eqn:diffusionConstant}
\end{equation}
where $T$ is the temperature, $\Gamma$ is related to the phonon-$^3$He scattering rate, and $T_K$ is the temperature in K (see also~\cite{LANLFLUSH}).  The phonon drift velocity is defined by the heat flux, $\vec Q$,
\begin{equation}
\vec Q = T S_{ph} \vec v_{ph}
\label{eqn:vPh}
\end{equation}
where the phonon entropy density is
\begin{equation}
S_{ph} = (2 \pi^2/45)(T/s)^3
\label{eqn:phononEntropy}
\end{equation}
and $s = 2.38\times 10^4$~cm/s is the phonon sound speed.
In steady state and for one dimension,
\begin{equation}
v_{ph} X_3 = D_{3,ph} \frac{\partial X_3}{\partial z}.
\label{eqn:ssHeatFlush}
\end{equation}
In this equation, the diffusion constant is averaged over the temperature range associated with heat flush in the experiment (typically $\leq 100$~mK at the operating temperature of 450~mK).  In this case, for a pipe of length $L$, the solution is
\begin{equation}
X_3(z) = \frac{X_{3,0}}{e^{L/h}-1}\frac{L}{h}e^{z/h}
\label{eqn:x3Ofz}
\end{equation}
where $h\equiv D_{3,ph}/v_{ph}$ is the scale height.  

For a simple pipe, with an initially uniform density $n_{3,0}$, the time dependent solution is given by~\cite{Baym15a}
\begin{equation}
   n_3(z,t) = n_{3,0}\,e^{z/2h}\sum_{\nu=0}^\infty c_\nu \phi_\nu \left( z \right ) e^{-t/\tau_\nu},
   \label{expansion}
\end{equation}
where 
\begin{eqnarray}
c_\nu &=& \int_0^L \left(n( z,0 \right)/n_3^0)\phi_\nu\left( z \right)dz\nonumber \\
&=& \left\{
\begin{array}{lrr}
\frac{{\displaystyle L}}{{\displaystyle \left[ h\left( e^{L/h} - 1 \right )\right]^{1/2}}},& \hspace {.5in} &  \nu = 0\\ \\
\frac{{\displaystyle 8h \alpha_\nu}}{{\displaystyle 1+(2hk_\nu)^2}} \left( 1+ \left( -1 \right )^{\nu + 1} e^{-L/2h} \right).& \hspace {.5in} & \nu \ge 1
\end{array}
\right. 
\label{eq:c}
\end{eqnarray}
The spatial parts of the mode functions are the complete orthonormal set 
\begin{equation}
\phi_\nu(z) =  
\left\{ 
\begin{array}{lrr}
\frac{{\displaystyle e^{z/2h}}}{{\displaystyle \left[ h\left( e^{L/h} - 1 \right )\right]^{1/2}}},&\hspace {.5in} & \nu = 0\\ \\
\alpha_\nu \left ( \cos k_\nu z + \frac{1}{2hk_\nu}\sin k_\nu z \right ),& \hspace {.5in} & \nu \ge 1
\end{array}
\right. 
\end{equation}
with  $k_\nu =\pi \nu/L$ and $\alpha_\nu = \left[ \left(L/2\right) \left( 1+ 1/(2hk_\nu)^2 \right) \right]^{-1/2}$.  The time dependence of the modes is $e^{-t/\tau_\nu}$ where
\begin{equation}
   \frac{1}{\tau_\nu} = \left\{ 
\begin{array}{lrr} 0,& \hspace {.5in} & \nu = 0\\ \\
  k_\nu^2 D_{3,ph}+v_{ph}^2/4D_{3,ph}  = \left(k_\nu^2 + 1/4h^2 \right)D_{3,ph}, & \hspace {.5in} & \nu \ge 1.
\end{array}
\right.  
\label{eq:tauNu}
\end{equation}
The two terms in $\tau_\nu$ correspond to diffusion and phonon driven transport, respectively.  Typically, in the geometry of the 
experiment, the first term in the sum of Eq.~\ref{expansion}, $\nu=1$, dominates after the first few seconds.

The heat flush mechanism is used primarily to transfer polarized $^3$He from the injection system to the measurement cells prior to the accumulation of UCN and subsequent precession measurement, and to transfer the slightly depolarized $^3$He from the measurement cells to the purification system.  During the UCN accumulation and the precession measurement, the measurement cells are isolated from the injection and purification systems with a superfluid-tight valve, allowing for the removal of the remnant $^3$He in the rest of the system as described below.

\subsection{Key Components}

\subsubsection{Atomic Beam Source}

This experiment requires high $^3$He polarization (> 95\%), particularly for the spin-dressing measurement mode as discussed previously. However, because of the low polarized $^3$He density ($X_3 \sim 10^{-10}$) the required flux ($\sim 10^{14}$ atoms/s) is modest. Spin exchange~\cite{Coulter88} and metastability exchange~\cite{Milner89} optical pumping techniques are well-developed techniques to produce polarized $^3$He, but both tend to have a maximum value of around 90\%.  Brute force polarization of $^3$He at mK temperatures can also produce high polarization, but the necessary apparatus is complicated~\cite{Vermeulen87}. Thus an atomic beam source (ABS) has been built that utilizes a simple permanent magnet quadrupole implementation of the standard magnetic field gradient method of filtering spin states to produce a polarized atomic beam~\cite{Torgerson11}. 
The absence of ground state electronic polarization and the small nuclear magnetic moment of $^3$He requires that the atoms spend a relatively long time in the gradient field, hence the source is operated at a temperature of about $1$~K and use a $1.3$~m long quadrupole.  

The energy of a magnetic dipole $\vec{\mu}$ in a magnetic field $\vec{B} (
\vec{r})$ is given by
\begin{equation}
  U ( \vec{r}) = - \vec{\mu} \cdot \vec{B} ( \vec{r})
\end{equation}
and the force imposed on the dipole if the field is static is given by
\begin{equation}
  \vec{F} ( \vec{r}) = \mu ( \hat{s} \cdot \nabla) \vec{B} ( \vec{r})
  \label{F}
\end{equation}
where $\hat{s}$ is the direction of the spin and $| \hat{s} | = 1$. For
$^3$He (spin = $1 / 2$), $\mu = - \hbar \gamma_3 / 2$ where $\gamma_3 = 2.04 \cdot
10^8 /$T-s is the $^3$He gyromagnetic ratio.

The magnetic field in the rest frame of an atom will change in magnitude and
direction as the atom follows its trajectory through the polarizer. If these
changes are too fast, the atom's spin will not maintain its relationship to
the magnetic field and the atomic beam will lose polarization. To maintain an
atom's polarization throughout its trajectory, its spin must be able to
adiabatically follow the direction of the field. Explicitly, the required
relationship is
\begin{equation}
  \frac{| \dot{B} |}{|B|} \ll | \gamma_3 B|, \label{adiabatic}
\end{equation}
where $\dot{B} \equiv dB / dt$ and $\gamma_3 B$ is the Larmor frequency.
An additional concern is that the magnitude of the field is theoretically zero
at the center of the polarizer. Polarized atoms traveling through this region
of zero field may be unpolarized and reduce both the net polarization and
polarizer throughput.  In practice, the small phase space associated with such particles is such that there is a very small affect on the overall beam polarization


The quadrupole filter portion of the source is constructed from rare earth permanent
magnets. The magnetic field of each magnet
was measured to be 0.75 T at the surface of the magnets closest to the central
bore. No effort was made to control the return flux on the edges farthest from
the bore. The magnets are held in place with grooved aluminum end pieces and a
stainless steel center tube with an inner diameter of 1.0 cm. The complete polarizer uses
eight sections, each of which is about 16 cm
in length.

The angular dependence of the intensity of an effusing source is given by~{\cite{Ramsey56}}
$dI_0 / d \Omega = n \bar{v} A \cos (\theta) / 4 \pi$ where
$n$ is the source density, $\bar{v}$ is the mean particle velocity, $A$ is the aperture area, $\theta$ is
the azimuthal angle from the source aperture normal and $d\Omega$ is the solid
angle. Integrating between $0 \le \theta \le \theta_0$ yields
\begin{eqnarray}
  I_0 & = & \frac{1}{4} n \bar{v} A \sin^2 (\theta_0) \nonumber\\
  & \approx & \frac{1}{2} \frac{p}{\sqrt{mk_B T}} A \sin^2 (\theta_0)
  \nonumber\\
  & \approx & \frac{1}{8} \frac{\mu B_0}{\sqrt{m (k_B T)^3}} pA, 
  \label{intensity}
\end{eqnarray}
where $p$ is the source pressure. In the last equation above, the magnetic potential energy, $\mu B$, has been equated with the transverse kinetic energy $\frac{1}{2}m(v_x^2+v_y^2)$. For $^3$He and the parameters discussed
above, Eq.~\ref{intensity} yields $I_0 / pA \approx 1 \cdot 10^{16}
/$s$\cdot$mtorr$\cdot$cm$^2$. The source aperture has an area of about
$1${\hspace{0.25em}}cm$^2$.

To produce a cold effusive source of $^3$He, a
multistage refrigerator based on
a commercial helium
cryocooler provides the bulk of the cooling power. The entire apparatus is
wrapped in a copper shield anchored to the cryocooler's 50K reservoir. The
cryocooler's 4~K coldhead cools and liquifies a supply of $^4$He which is
subsequently used to create a nominally 1~K evaporative refrigerator. This 1~K
refrigerator is used to liquify a supply of $^3$He for the nozzle reservoir.

The pressure at which the source can be operated depends upon the specific
geometry of the source nozzle and there are several concerns that led to the present nozzle design. First, the gas pressure in the volume
outside the nozzle must be kept much lower than the source pressure. This
depends on several items such as the geometry of that volume, the capacity of
the pumps acting on that volume and, of course, the flow of $^3$He from the
nozzle. Second, the amount of $^3$He in the system will be
relatively small and will need to be used efficiently. Fortunately, the
forward flow and hence the quality of the vacuum outside the source can be
enhanced by building the aperture from a collection of small tubes of radius
$\rho_s$ and length $L_s$. The forward flow remains the same while the
integrated intensity is reduced by $8 \rho_s / 3 L_s$~{\cite{Ramsey56}}.

The mean-free-path in the nozzle aperture given by $\lambda_s \approx 1 /
\sqrt{2} n \sigma$, where $\sigma = 1.0 \cdot 10^{-
14}${\hspace{0.25em}}cm$^2$ is the scattering cross-section of He. This should
not be much smaller than $\rho_s$ to ensure that the lowest velocity atoms are
not scattered out of the beam. The mean-free-path for helium can be expressed
as $\lambda_3 / p \approx 4.4 \cdot 10^{- 3}${\hspace{0.25em}}cm/mTorr. With
$\rho_s = 3.5 \cdot 10^{- 1}${\hspace{0.25em}}mm, $p$ should be less than a
few $10^{- 2}${\hspace{0.25em}}mTorr and
\begin{equation}
  I_0 \approx 1 \cdot 10^{14}/{\rm s,} \label{inotbase}
\end{equation}
given that only half of the $^3$He enter the polarizer in the spin state where
$\hat{s} = \hat{B}$.

The nozzle is composed of 1~mm outside diameter thin wall stainless hypodermic tubing
packed into a tube of about 1~cm diameter. A skimmer is used about 10~cm from the
output of the nozzle to reduce the flow of $^3$He into the polarizer section
of the source. The nozzle side of the source is pumped with a 1900 l/s turbo
pump specially designed to maximize its efficiency for helium. The polarizer
section is pumped with a 950 l/s turbo pump similarly optimized for helium.
The output of these pumps and their dry roughing pumps is collected in the
$^3$He storage volume to form an almost closed $^3$He recovery system. The only
$^3$He that escapes recovery is that which exits the
quadrupole magnet polarizer through the aperture at the end of the polarizer vacuum chamber. 

Preliminary tests have been performed with the prototype ABS.
The flux was measured using a long tube mounted downstream of the polarizer.
Using an RGA on the output side of a tube 15.24~cm long and 1.19~cm
inside diameter, the $^3$He pressure into a volume that was pumped with a known
pumping speed was measured and indicated a flow rate of $3.8 \cdot 10^3$cc/s at
$1.6 \cdot 10^{- 8}$ Torr. This implies a current of $8 \cdot 10^{11}$/s.
Correcting for the 200~cm distance between the exit of the polarizer and the
tube, this implies a polarized flux of $1.5 \cdot 10^{14}$/s, which exceeds
the required value by a modest factor and provides the required $\sim
10^{16}$ atoms in about 100 s, a time short compared to the neutron accumulation and measurement times.

The prototype source has been used to show that the effective magnetic
moment of the $^3$He can be manipulated by applying a transverse oscillating
magnetic field \cite{Esler07}. The results showed that the effective magnetic moment can
be well described by classical calculations using the Bloch equation as well
as by a quantum approach. 

For this implementation, relatively simple shaping coils to taper the $\sim 1$~T fields of the ABS magnets to the 3~$\mu$T holding field inside the MSE will be used. The magnetic requirement is again that the adiabaticity condition is met, Eq.~(\ref{adiabatic}), as the spins evolve from a quadrupolar pattern at the exit of the MSE, to longitudinal polarization, to polarization in the horizontal plane and perpendicular to the neutron beam direction.  Along the $^3$He beamline, relatively short solenoidal and cos$\theta$ coils are used, making use of their fringe fields to meet the adiabaticity criterion.  In the space between the layers of the MSE a longer cos$\theta$ coil will be used to make the final transition to the 3~$\mu$T internal horizontal magnetic field of the enclosure.

\begin{figure}
\centering
\includegraphics[width=\textwidth]{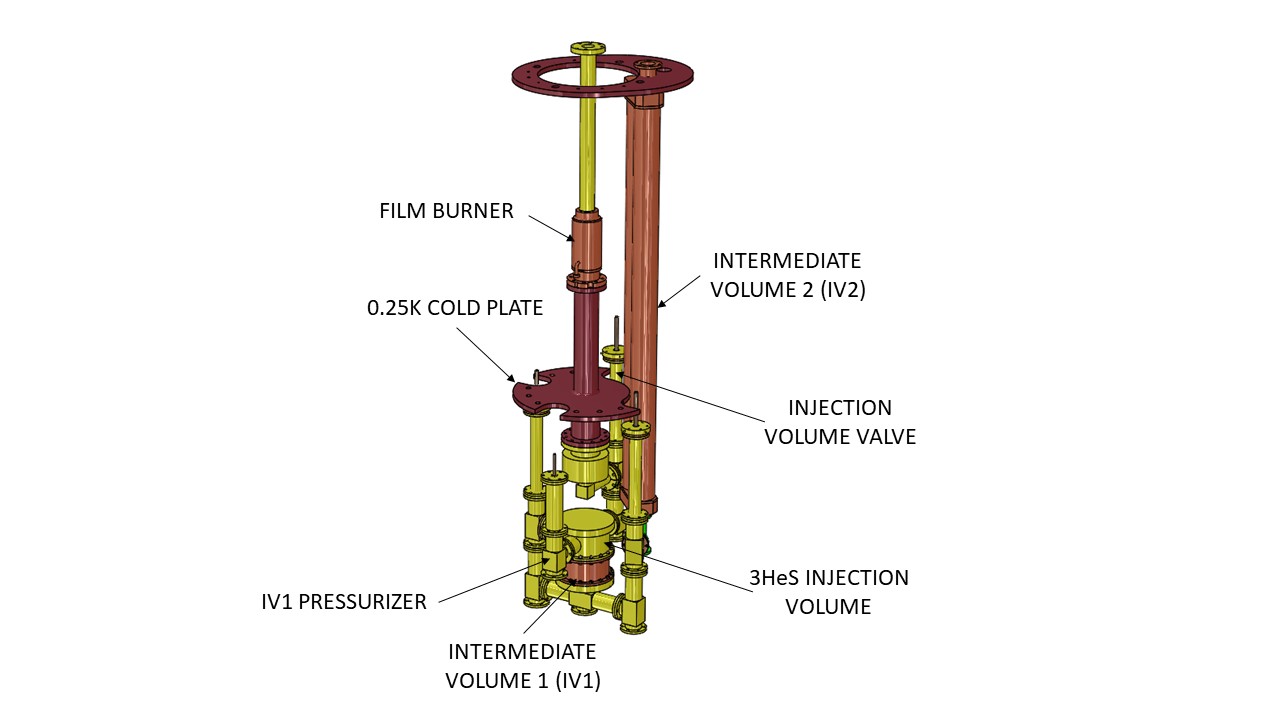}
\caption{Model of the injection system.  The Atomic Beam Source (ABS) and its interface plumbing are connected at the top flange shown here.  Only the key components are shown; thermal links and shields, actuators, and mechanical supports are also part of the full system.}
\label{fig:HE3S_injection}
\end{figure}
\subsubsection{Injection System}

Polarized $^3$He from the ABS enters the injection system through a free surface of superfluid $^4$He and is accumulated initially in a small ($\sim 100$ cm$^3$) ``injection volume'' with a diameter of about 3~cm.  The injection system is shown in Fig.~\ref{fig:HE3S_injection}. This step is performed at the same time as the precession measurement to minimize experimental deadtime.  

The superfluid film above the injection volume is controlled by a film burner~\cite{Beck16} to prevent vapor from the eventual film evaporation from interfering with the $^3$He beam from the ABS.  It utilizes a heater and a baffle to ensure that most of the evaporated film is recondensed away from the path of the $^3$He beam.  Based on tests with a prototype film burner, an operating temperature $T\sim0.3$~K for the injection volume is important to maximize the transmission of the beam.  Because of the possibility that the liquid helium in the measurement cell might need to be pressurized to reduce electrical breakdown (associated with bubble formation), the $^3$He-loaded helium in the injection volume is transferred first to an intermediate volume, IV1, where it can be isolated from the both the free surface in the injection volume and the rest of the measurement system.  The isolation is accomplished with a 2.5~cm diameter, superfluid tight valve~\cite{Williamson16}.  The initial part of the measurement cycle comprises loading the injection volume from the ABS, depressurizing IV1, and then simply allowing the polarized $^3$He to diffuse throughout the combined injection-IV1 volume.  The $^3$He diffusion constant increases rapidly as the temperature decreases ($\propto T^{-7}$) and is about 250 cm$^2$/s at $T=0.45$~K, so the timescale for diffusion is a few seconds.  The valves are also thermally isolating, allowing the temperature of the injection volume to return to 0.3 K prior to the next injection cycle.

\begin{figure}
\centering
\includegraphics[width=6.5in]{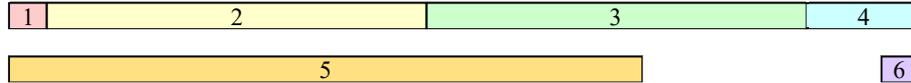}
\vskip -5.5in
\caption{Representative standard measurement cycle for the experiment, where the measurement cell is denoted by MC and the sequestration volume by SV.  The total cycle is about 2400 s in length, with precession measurement comprising 1000 s. The dump/refill of the SV and the filling of the injection volume with polarized $^3$He can take place in parallel with the neutron loading and/or during the precession measurement, as the relevant parts of the system are independent of the MC (and, in fact, of each other).}
\label{fig:Cycle}
\end{figure}

After IV1 is re-pressurized (using a small bellows), the polarized $^3$He atoms are moved by heat flush to the two measurement cells in the central detector system.  The measurement cells, immersed in the ``central volume'' (a bath of approximately 1600~l of superfluid helium at $T = 0.45$ K),  remain at that constant temperature, therefore, the temperature gradient necessary for this heat flush step is achieved by turning on a heater in IV1.  With a heater power of 10 mW, and a transfer line 3.8~cm in diameter and 5~m in length, the phonon drift velocity is 15~cm/s and the scale height is about 15~cm; the temperature difference between the ends of the pipe is about 9~mK.  Because the measurement cell and part of the transfer line are immersed in the constant temperature central volume, the temperature gradient and hence phonon drift velocity that drives the heat flush is substantially reduced (some of the heat escapes through the walls of the transfer line and target cell into the bath). Therefore, the overall transfer time has a significant contribution from diffusive transport of the $^3$He in the region of the measurement cells, Eq.~(\ref{eq:tauNu}).  Based on finite element simulations of the process~\cite{COMSOL}, including the full temperature dependences of the transport coefficients ($D_{34} \propto 1/T^7$; $v_{ph} \propto 1/T^4$, etc.), the transfer from the injection volume to the measurement cells is estimated to take a total of  about 100~s.  Because the measurement cell valve must be open during this transfer, it must be performed prior to the accumulation of neutrons for a given measurement cycle. A schematic of the components and times for a measurement cycle is shown in Fig.~\ref{fig:Cycle}.

As described above, the magnetic field environment of the polarized $^3$He is established by a pair of coils at the exit of the ABS and a low-field matching coil between the layers of the MSE.  Subsequently, at the injection volume and for the upper section of the injection system, the holding field for the injection system is provided by the MSE coil; this field joins smoothly with that of the main holding coil, $\vec B_0$, surrounding the central volume.  Relative to the requirements for $\vec B_0$, those for the field of the MSE coil are reduced because of the shorter period of time the polarized $^3$He spends in them, which is roughly an order of magnitude less (100~s vs. 1000~s).  However, even with the relaxed requirements, the injection system is still constructed excluding magnetic and superconducting components.

\subsubsection{Purification System}

After completion of a precession measurement, because the $^3$He loses some polarization due to wall collisions in the measurement cells, it is removed using heat flush.  In this case, the heat flush is used to compress the $^3$He into a ``sequestration volume'' (SV), from which it is dumped, and then subsequently refilled with pure $^4$He. The process of heat-flushing the $^3$He to the SV is performed in two stages.  In the first stage, the valve to the measurement cell is open; again, this process contributes to the overall deadtime of the measurement. Using a finite element simulation of the process, this step is estimated to take about 250 s.  Subsequently, the purification system is isolated from the measurement cells and the process of compressing the $^3$He to the SV can be completed. The initial step will remove 99.9\% of the $^3$He from the measurement cells; the second step will move all but 1\% of this $^3$He to the SV (in about 60~s), allowing this relatively small volume of helium highly concentrated in $^3$He to be pumped away. 

With the measurement cells isolated and during the time, $\sim 2000$~s, when UCNs are accumulated and the precession frequency measured, the purification system will be used to remove the remnant $^3$He from the system.  Superfluid tight valves must be used to clean both the injection and purification systems such that, overall, less than 1\% of the injected polarized $^3$He in a given cycle remains at the start of the next cycle.  These stages include, e.g., removing the remnant $^3$He from the injection volume in steps involving diffusion to IV1 and then heat flush from there to the SV.

The SV is a volume of about 100~cm$^3$ which is dumped each cycle.  By opening a valve, the mixture of $^3$He and $^4$He will be transferred (using gravity) to a dump volume from which it will be evaporated.  Evaporation will require heating the dumped liquid to a temperature of $\sim 1.2$~K where the vapor pressure is nearly 1 Torr, and then supplying the approximately 500 J required to provide the latent heat (in the evaporation phase, the dump essentially operates as a 1~K pot).  After dumping, the SV will then be refilled from a reservoir of pure $^4$He produced using an internal ``McClintock purifier''~\cite{McClintock87}.  The purifier, operating at $T \sim 1.2$~K where the heat flush process is significantly more efficient, consists of a long tube with a heater at one end, where the pure $^4$He is drawn off.  The purifier on which this design is based produces $^4$He with a $^3$He concentration $X_3 < 5 \times 10^{-13}$ at a rate of about 1~cm$^3$/s.  A significant amount of time is required to cool the purified $^4$He from about 1.2~K to the operating temperature of 0.45~K.  With an integrated heat capacity of about 35 mJ/cm$^3$, it will take of order 350 s for this cooling step.


\section{Cold Neutron Transport}
\label{sec:neutronics}
A high transport efficiency of neutrons from their production region to the experimental apparatus is essential for a precision EDM search. At the SNS, the neutrons are generated by a proton beam with a maximum power of 1.4~MW (1~GeV energy, up to 1.4~mA beam current). The protons are stopped in a liquid mercury spallation target, where an average neutron flux of $\sim 6.4 \times 10^{12}$~n/cm$^2$/s is available after the spallation process.  A liquid hydrogen moderator, kept at a temperature of 20~K, generates cold neutrons which are velocity-selected with beam choppers so that a flux of $\sim 1.92 \times 10^{8}$~n/cm$^2$/s/MW/\AA~ at a wavelength of 8.9~\AA~ is directed into the Fundamental Neutron Physics Beamline (FNPB, BL13). 

\begin{figure}[htbp]
\begin{center}

\includegraphics[scale=0.45]{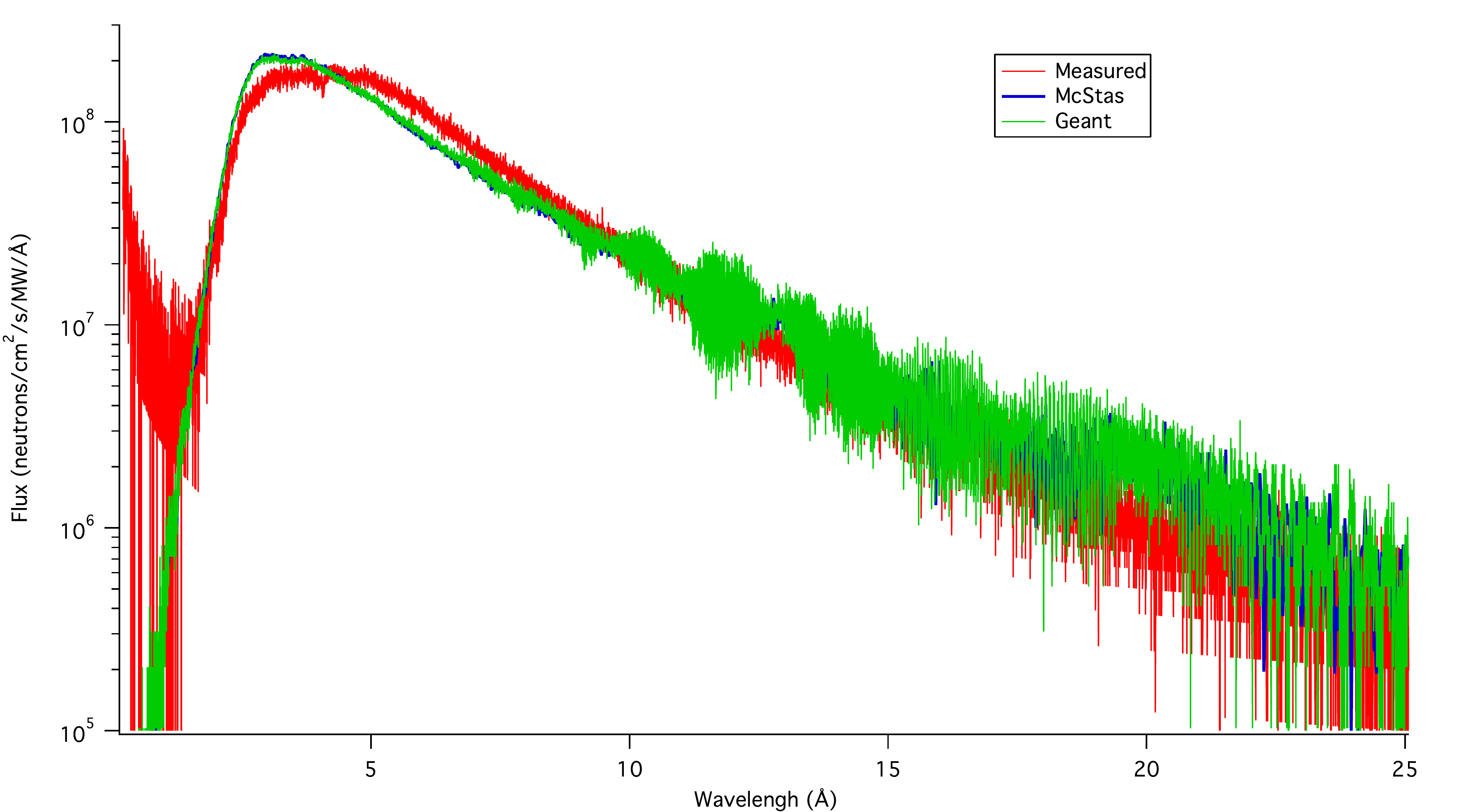}
\caption{Comparison of measured and calculated (McStas and GEANT4) neutron flux as a function of wavelength at the exit of the existing FNPB neutron guide.}
\label{fig:neutronflux}
\end{center}
\end{figure}

\begin{figure}[htbp]
\begin{center}
\includegraphics[scale=0.35]{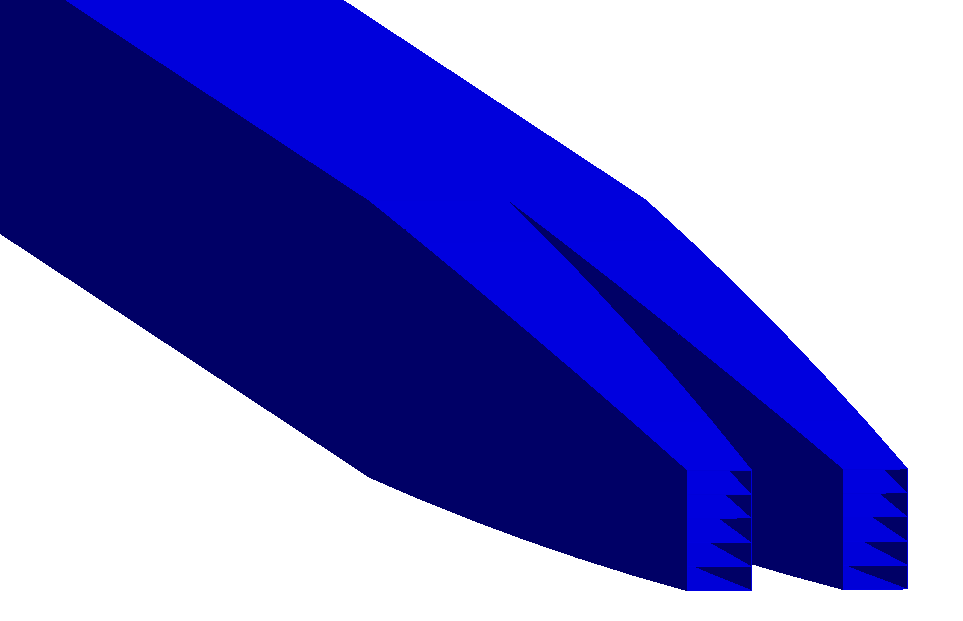}
\caption{Last section of the neutron guide. The beam is split into two separate beams for transport to the two measurement cells. Also shown, at the end of the guides, are horizontal panels to enhance focusing.}
\label{fig:splitter}
\end{center}
\end{figure}

The FNPB is used for precision measurements on the free neutron and starts 1~m downstream from the face of the cold moderator.  It consists of a 10.6~m 
long super-mirror guide with a constant cross section of $10 \times 12$~cm$^2$ (horizontal x vertical). Neutrons are deflected 2.42~degrees to the left (w.r.t. the beam direction) during the transport through this section. More details concerning the existing FNPB can be found in Ref.~\cite{fomin}.

The nEDM experiment will be mounted in a separate dedicated hall so that the existing neutron guide needs to be extended by about 40~m.  Two separate Monte Carlo simulation codes were developed to simulate and optimize the neutron transport to the nEDM measurement cells, one based on  GEANT4~\cite{geant4, atchison} and the other  based on McStas~\cite{mcstas}. Input data for the simulations were generated using MCNP by the SNS moderator group~\cite{iverson}. In order to validate the Monte Carlo calculations, the input distributions were tracked to the end of the existing guide where the results could be compared to measurements. The final comparisons are shown in Fig.~\ref{fig:neutronflux} and it can be concluded that both simulations agree within 5~\% and the measured data can be reproduced to better than 20~\%. Although detailed parameterizations of the experimentally determined supermirror reflectivities were applied, the source for the discrepancies for between the data and the simulations is unknown.

The existing polarizer of the FNPB~\cite{balascuta} will be replaced since it was optimized for neutron wavelengths around 4~\AA, and it exhibits a drop off in polarization at a wavelength of 8.9~\AA. In order to keep polarization values above 0.95 at longer wavelengths a gadolinium coated glass substrate is necessary. Otherwise, the general design of the existing SNS polarizer is well matched to the needs of nEDM assuming the neutrons are reflected to the right by 2.387 degrees. Following the polarizer is a ballistic section which expands the beam from a cross section of $10~\mbox{cm} \times 12~\mbox{cm}$ to  $26~\mbox{cm} \times 26~\mbox{cm}$. This section is 8~m long and the m-values decrease from m=3.5 to m=1.5 along its length.  Next is a 29~m long straight guide with constant cross section ($26~\mbox{cm} \times 26$~cm, m=1.5). At the end of this transition segment the neutron guide terminates with a 3~m long splitter design. The neutron beam gets divided symmetrically into two beams by two converging guide segments. Each splitter guide contains additional horizontal panes to improve focusing. The angles and m-values (m=3.5 for all surfaces) of the vertical and horizontal sides  were optimized for symmetric beam profiles inside the target cells. Figure~\ref{fig:splitter} shows the last few meters of the guide. The typical substrate for the supermirrors is borofloat glass. This glass, however, can cause significant gamma background due to the $n_{therm} +{^{10}}B \to {^{11}}B^{\star} \to {^{7}}Li + \alpha + \gamma$ reaction with a $\gamma$ energy of 480~keV. In order to minimize this background the substrate for some critical components like the polarizer and the splitter will most likely be boron-free soda-lime glass. Overall the total transmission of 8.9 \AA~ neutrons is about 3.5\% (total, i.e. both cells together) from the beginning of the FNPB to the center the measurement cells which were assumed to be positioned 1.17~m downstream of the end of the neutron guide exit window. Combining the input data with the transmission, a neutron flux of $d\phi/d\lambda = 7\cdot 10^{6}$~n/s/cm$^2$/MW/\AA~ for 8.9~\AA~ neutrons is expected. This yields a total UCN production rate, summed over both measurement cells, of $dN_{UCN}/dt$ = 1800 UCN/s for a cell geometry of 7.5~cm wide $\times$ 10~cm high $\times$ 40~cm long and a proton beam power of 2 MW~\cite{howell17}. While the transmission was evaluated without any windows in place, only a few $\%$ reduction is anticipated from these.


\section{Magnetically Shielded Enclosure}
\label{sec:shieldroom}

Traditionally, neutron electric dipole moment experiments have employed multi-layer, cylindrical-shaped magnetic shields made of $\mu$-metal \cite{Baker2014} or Permalloy \cite{Altarev1996}.  To set the scale, the radii and lengths of the outermost layers in these shielding configurations were typically of order $\sim 1$~m and $\sim 2$--3~m, respectively.  As these outermost layers were too large to have been annealed as a single contiguous piece, smaller subsections were instead individually annealed, and then later bolted together with overlapping magnetic material to form the cylindrical structure \cite{Baker2014}. Removable end-caps provided access to the interior.  Demagnetization was carried out by driving a $\sim$few~Hz~\cite{Baker2014} or $\sim 50$~Hz~\cite{Altarev1996} current through loops wound around each of the layers, with the amplitude of the current slowly decreasing from some maximum to zero over a time period of $\sim 20$--30 minutes \cite{Baker2014, Altarev1996}.  After demagnetization, typical residual fields were $< 2$~nT, with dynamic shielding factors in the axial and transverse directions ranging from $\sim 2 \times 10^3$ \cite{Altarev1996} to $\sim 2 \times 10^4$ \cite{Baker2014}.

In a revolution for the field, many next-generation neutron electric dipole moment experiments, including this one, are employing rectangular-shaped magnetically shielded enclosures (MSEs) \cite{Altarev2014, Altarev2015} in lieu of the traditional multi-layer, cylindrical-shaped configurations.  The rectangular geometry of such a MSE naturally permits the enclosure of a large experimental apparatus, with access to the interior generally provided via a ``door'' designed such that when it is closed, there is constant magnetic contact with the walls of the MSE.  Demagnetization of the walls of the MSE is achieved with sets of coils for each of the three spatial directions.  Despite their rather large interior volumes (e.g., $\sim 10$ m$^3$ in Ref.\ \cite{Altarev2014}), with the design of the magnetic contact at the door and the demagnetization procedure, residual fields of $< 1$ nT, field gradients $< 300$ pT/m over the central 1 m$^3$ interior volume, and dynamic shielding factors of $\sim 10^5$ at $\sim 10$ Hz were demonstrated in a MSE consisting of only two layers of $\mu$-metal \cite{Altarev2014}.

\begin{figure}[htbp]
\begin{center}
\includegraphics[width=\textwidth]{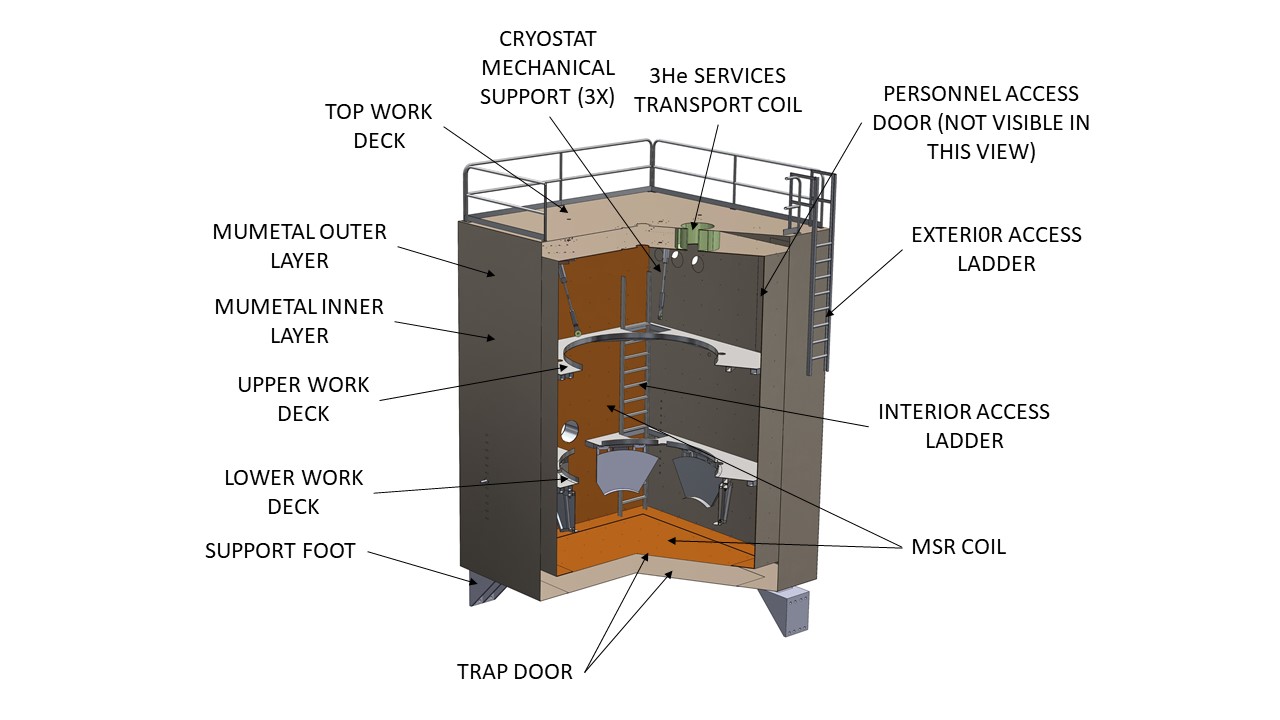}
\caption{Schematic diagram of the envisioned magnetically shielded enclosure (MSE).  Note the bottom ``trap door" providing access to the interior.}
\label{fig:magnetic_shield_room}
\end{center}
\end{figure}

The presently envisioned design for the MSE which will enclose the experiment is shown in Fig.\ \ref{fig:magnetic_shield_room}.  The MSE will consist of two (or possibly three) layers of $\mu$-metal, with a 40 cm spacing between the centers of the proposed 2-mm-thick inner layer and  4-mm-thick outer layer. The internal dimensions will be of order $\sim 4.1 \times 4.1 \times 6.1$ m$^3$.  Primary access to the interior will be provided via a bottom ``trap door'' (see Fig.\ \ref{fig:magnetic_shield_room}) whose dimensions will be $\sim 3.3 \times 3.3$ m$^2$. The MSE performance specifications include the following: field gradients $< 1$ nT/m over a 1 m$^3$ volume enclosing the measurement cells, dynamic shielding factors of at least 150 at 1 Hz, an internal structure capable of supporting (from above) the $\sim 12000$ kg experimental apparatus and the incorporation of several large penetrations through the MSE as required for the incoming 8.9~\AA~cold neutron guide, plumbing connections for the $^3$He services, a pumping line for the dilution refrigerator, emergency safety relief, etc.  The proposed dimensions for the layer spacing, wall thicknesses, etc.\ may change during the course of discussions with the candidate vendors.

The MSE will be surrounded by a system of tri-axial rectangular coils embedded into the infrastructure of the building housing the experiment's apparatus; the purpose of these coils will be to compensate for the Earth's and other (possibly time-varying) background fields, in order to reduce variations in the magnetic flux density within the MSE wall material. Active magnetic field compensation has been described, for example, in Ref.~\cite{Afach14}. As the dimensions of these coils will be on the scale of the building (e.g., $\sim 15$~m~$\times$~$\sim 15$~m), the power supply requirements for these coils will be significant.  As a simple example, a pair of $\sim 15$~m~$\times$~$\sim 15$~m coils configured as a rectangular Helmholtz pair requires a current of $\sim 500$ A-turns in order to generate a $0.5 \times 10^{-4}$~T field at the center of the coils.

\section{Cryogenics}
\label{sec:cryogenics}

Functionally and mechanically the cryogenic equipment of the nEDM experiment can be split into two main categories: one is LHe supply, to provide cooling at and above 4K, another is for cooling below 300 mK. Both components present a number of cryogenic challenges.

 The main challenge for LHe supply design is that the nEDM apparatus has three partially thermally-isolated cryostats: the magnet cryostat, the CDS volume and the 3HeS cryostat, each of which has specific features and operational requirements. Two of the cryostats, CDS and 3HeS, must have powerful dilution refrigerators (DR) , which require a vessel filled with LHe for operation. The third cryostat, the magnet system, has a large size superconducting magnetic shield made from lead that has $T_{c}=(7.193 \pm 0.005) K.$, which needs cold He cooling in continuous flow mode.

An additional challenge is given by the requirement that for all components inside the MSE (including common LHe plumbing and DRs) have to be fabricated from non-magnetic materials such as Titanium or composite materials. This is necessary since commonly used standard "non-magnetic" stainless steels (e.g. 316L) generally have too much residual magnetism. Even more stringent material requirements exist for the non-metal zone inside the CDS module. Therefore, a significant effort has gone into development of helium leak-tight demountable vacuum joints for the composite vessels as well as development of superfluid He leak-tight non-metal valves. 

A nEDM measurement period can last for many months and during that time the cryogenic part of the experiment has to be stable while also being able to adapt to changes of the operational modes. To meet these requirements, the experiment requires a reliable liquid helium production and transport facility that can be realized by a liquefaction/refrigeration plant that is an integral part of the nEDM apparatus.

\subsection{LHe production Requirements and design concept}

The nEDM apparatus will require substantial cryogenic services, necessitating a liquid helium liquefaction and refrigeration plant with associated liquid helium transmission pipes, recovery system, storage and LN service.

\begin{table}
\hspace*{-2cm}
\vspace*{-.5cm}
\center
\includegraphics[width=7in]{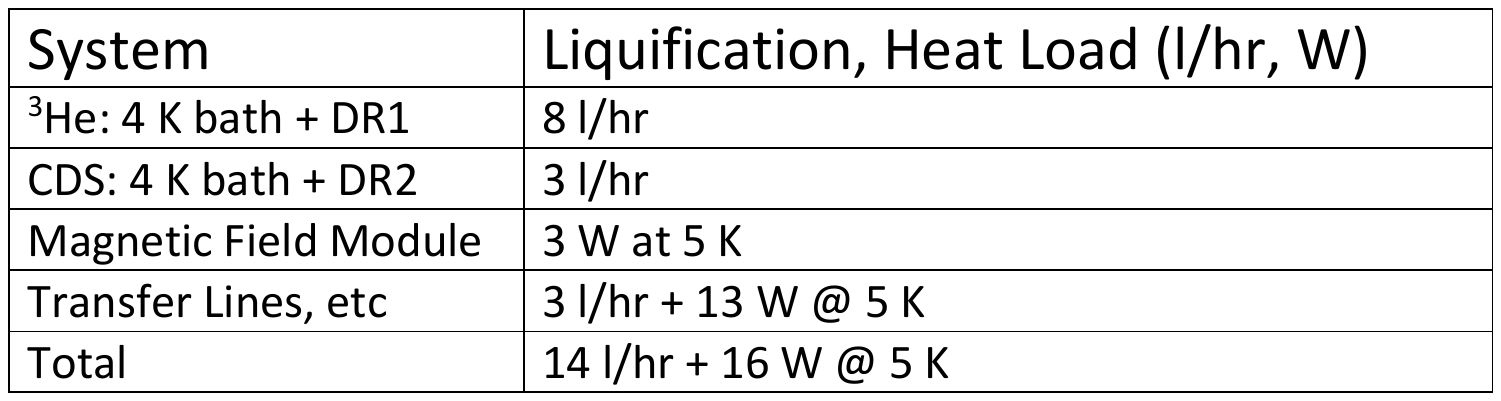}
\vskip -7in
\caption{Helium liquification rates and heat loads of the nEDM cryogenic system in steady-state operation.} 
\label{tab:cryo}
\end{table}


The running mode of the cryostats requires a mixed-mode liquefier operation and refrigeration capacity. To clarify,  the liquefaction load is, in general, the amount of LHe evaporated in cryostats and then returned as room temperature gas to the inlet of the liquefier compressor. The heat/refrigeration load is cold gas, less then 20~K, returned to the cold inlet of liquefier. 

 Table~\ref{tab:cryo} summarizes the heat and He liquefaction budgets for the helium cryogenic system, the transfer lines, etc when the system is operating in the steady state. The loads include projected heat sources from the nEDM cryostat modules as well as loads associated with the plant itself (transfer lines, cryogenic valves, bayonet connections, and dewar cooling).
 Note that shorter term requirements (initial cooling, for example) may require two to three times the values shown in Table~\ref{tab:cryo}.

The LHe requirements for the central detector and cryostats are dominated by the LHe usage of the 1-K pots in the dilution refrigerators. The high cooling power required of DRs implies high LHe consumption.

\begin{figure}[htbp]
\begin{center}
\includegraphics[width=\textwidth]{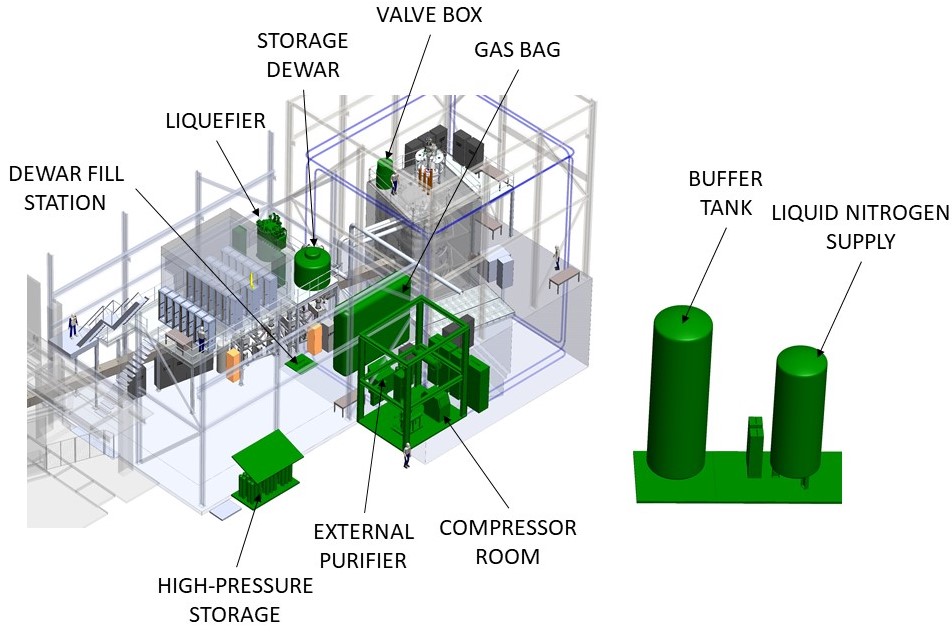}
\caption{Overview of the nEDM facility highlighting helium plant equipment.}
\label{fig:Cryoplant}
\end{center}
\end{figure}

As an example of possible LHe plants, a liquefier/refrigerator with liquid nitrogen pre-cooling could be used. 
In addition to the liquefier cold-box itself, the system would require at least a subset of the following equipment: liquid helium buffer storage (currently shown as a 3000 L dewar), low pressure and medium pressure storage, high-pressure storage, recovery compressor, vaporizors, condenser, valve box and liquid nitrogen storage.
A preliminary layout of suitable equipment has been performed in an effort to plan placement of components throughout the facility. This is shown in Fig.~\ref{fig:Cryoplant}. 

When laying out the equipment for such a plant, it is important to minimize the cold fluid transmission distances to minimize heat loads and thus increase the overall efficiency of operation.  Specifically, the distance from the liquefier to the storage dewar, from the storage dewar to the valve box, and from the valve box to the cryostat modules should all be minimized. If possible, LN cooled transfer lines or cold He return gas should be used to cool the transfer line. Because the high pressure gas is at room temperature, the distance to the compressor is not limited.  This allows the compressor to be located in a separate compressor room, which has the added benefits of isolating the vibration from the compressor through the floor to the nEDM apparatus and minimizing the noise impact to personnel in the facility. The collaboration is also actively investigating the use of cryocooler technology to simplify design and operation of the overall cryogenic system.

\begin{figure}[htbp]
\begin{center}
\includegraphics[width=4in]{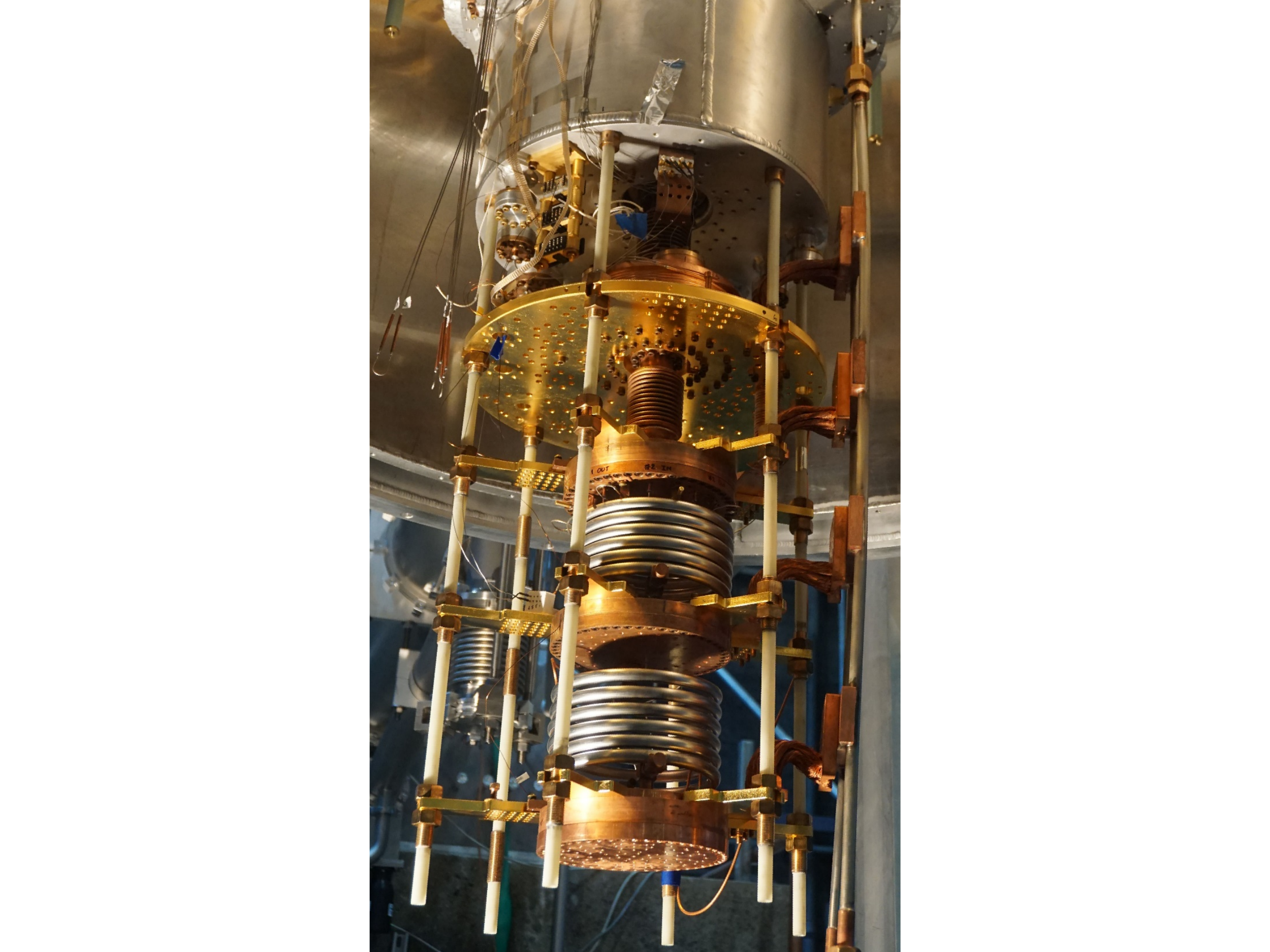}
\caption{Dilution Refrigerator for the $^3$He Services cryostat.}
\label{fig:DR}
\end{center}
\end{figure}

\subsection{Low Temperature ($< $1~K) Refrigeration}
To reach the operating temperatures necessary for the measurement, the system will make use of custom dilution refrigerators. The custom aspect results from the need to minimize magnetic interference in order to minimize magnetic noise, allow long polarization times for both the neutron and $^3$He and limit false EDM signals resulting from magnetic field gradients as discussed in Sec.~\ref{sec:falseEDM}. One DR will be used to cool the central volume, electrodes, and measurement cells, while the other will be used to cool the 3He Services injection cell, film burner, etc.

Both DRs are designed to provide 80 mW cooling power at 300 mK.  The mechanical design allows very  close integration of DR cooling stages with experimental  components to increase cooling efficiency of non-metal components. In addition, the design uses a modular approach  for easy modification if needed.  At present, the first DR, shown in Fig.~\ref{fig:DR} has been fabricated and is being commissioned.


\section{Conclusions}

This paper describes the design of an apparatus that will implement the neutron electric dipole moment measurement strategy laid out by Golub and Lamoreaux more than two decades ago~\cite{GL94}.  
Polarized ultracold neutrons ($E \leq 165$~neV) are produced {\it in situ} by superthermal scattering~\cite{UCNBOOK} of the incoming polarized cold neutrons ($\lambda \sim 8.9$~\AA) off phonons in isotopically purified ($^3\mbox{He}/^{4}\mbox{He} = R_3 < 1 \times 10^{-12}$) superfluid $^{4}$He. UCN upscattering is effectively eliminated by operating below 0.5~K. The UCN are simultaneously produced and stored in two measurement cells having opposite-sign $\vec E \cdot \vec B$ to cancel systematic uncertainties. Polarized $^{3}$He ($X_{3} \sim 1\times 10^{-10}$) is used as a co-magnetometer. The heat flush effect~\cite{Golub07} is used to move the $^{3}$He into the measurement cells at the start of each measurement cycle and remove it after it becomes depolarized. The $^{3}$He is also used as a spin analyzer. The strong spin-dependence of the n-$^{3}$He capture cross section allows the relative phase of the neutron and $^{3}$He spin vectors to be determined by the variation in the n-$^{3}$He capture rate. Capture events are identified through observation of 80~nm scintillation light produced by ionization of the decay products ($p+t$) in liquid helium~\cite{FLE59, THO59, SIM61, ADA95, BAN96, MCK03}. 

The statistical uncertainty per measurement cycle ($\sim 2,000$~s) is estimated to be $2-3 \times 10^{-26}$~e$\cdot$cm. Assuming 5,000 hr per year beam delivery and 50\% production running time gives an expected uncertainty of $2-3 \times 10^{-28}$~e$\cdot$cm in a three-year run. The uncertainty is proportional to $1/E\sqrt{N}$, where $E$ is the achievable electric field and $N$ is the incoming cold neutron flux. The uncertainty also depends on the UCN storage time, the rate of background events and the analysis technique. The latter dependencies are not describable with a simple formula because they can be optimized by tuning the $^{3}$He density and the fill and measurement cycle times. 

The experiment is designed to keep systematic uncertainties below the statistical uncertainty. Simultaneous measurements are made in storage cells with opposite sign $\vec E \cdot \vec B$. The electric field and magnetic field directions can both be reversed, as can the incoming neutron spin direction. Two different operational modes (free precession and spin-dressing) are available, each with different systematic uncertainties. The electric field dependence of the $n+^{3}$He-capture scintillation intensity~\cite{ITO12} provides an {\it in situ} monitor of electric field drifts. The largest systematic uncertainty is expected to be the geometric phase, an interaction between the $\vec v \times \vec E$ motional magnetic field and non-zero magnetic field gradients that results in a frequency shift proportional the electric field -- a false EDM~\cite{PEND04,COMMINS91,BS40,GEOPHASE}. This uncertainty is addressed first by minimizing magnetic field gradients and by using a co-magnetometer occupying the same volume as the stored UCN. The strong temperature dependence of the $^{3}$He geometric phase (due to the $1/T^{7}$-dependence of the $^{3}$He mean free path) can be exploited in two ways. At relatively warm temperatures ($T \sim 0.42$~K) the effect can be greatly reduced. At relatively cold temperatures the effect can be greatly magnified, providing an exquisitely sensitive mechanism to measure and minimize magnetic field gradients~\cite{SWANK16}.

The resulting apparatus is a large cryogenic system that must satisfy a wide variety of often conflicting restrictions on materials and material purity, magnetic field environment and electronic noise. The size is primarily driven by the high voltage electrode, which sets the radius of the ground return electrode, and therefore the inner radius of the magnetic coil package. The operating temperature is a compromise between considerations related to geometric phase, UCN upscattering and heat flush operation. The refrigeration system design is driven by the large heat load at 0.4~K ($\sim 80$~mW, dominated by heat flush operation) and the need to minimize the time needed to fill the central volume with $\sim 1,600$~liters of liquid helium and cool it to the operating temperature, thereby minimizing the duration of each cryogenic commissioning cycle. Additional requirements result from the need to control temperature differences between different volumes to enable heat flush operation. Magnetic materials are severely restricted; even so-called "non-magnetic" stainless steel is not allowed inside the cryovessel. Metals cannot be used inside the spin-dressing coils as eddy current heating would overwhelm the refrigeration system and Johnson noise would overwhelm the $^{3}$He spin precession signal on the SQUID magnetometers. This requirement drives the need for two non-metallic cryogenic neutron windows capable of surviving several atmospheres of differential pressure. Materials that have a superconducting transition are forbidden inside the superconducting lead shield, the outer-most layer of the magnet coil package. The neutron guide must be carefully shielded to avoid activation-induced backgrounds. Materials exposed to the beam must be carefully selected and screened for impurities. The $^{3}$He transport tubes must be selected (or treated) such that depolarization is $< 10^{-6}$/bounce. The measurement cells' depolarization requirement is even more severe ($< 10^{-7}$/bounce). The cells must also be UCN-compatible (in practice this means no hydrogen). And minute amounts of magnetic dust in the measurement cell will result in a false EDM due to geometric phase effects.

One final and crucial aspect of the experimental design is optimization for modularity and access. The three main subsystems (magnetic coil package, central detector system and $^{3}$He services) can each be operated independently, maximizing the ability for parallel development and debugging. All services enter the cryostat tops which are fixed to support frames. Access to the internal components is achieved by disconnecting room-temperature rubber vacuum seals and lowering the cryostat bottoms; no cryogenic seals need to be broken. Human entry to the room-temperature magnetic shield room that surrounds the magnet and central detector system is possible through a small door with a re-sealable joint. By improving access and minimizing the number of required vacuum/cryogenic seal disconnect operations these design features minimize the cooldown cycle length and improve the likelihood that a given cooldown will be successful.

In conclusion, an experimental design has been developed that will be capable of making a statistics-limited, world-leading measurement of the neutron electric dipole moment; a measurement that will have significant impact on our understanding of the origin of the baryon asymmetry of the universe, and the resulting existence of matter. All of the required technologies in at-or-near full-scale prototypes have been demonstrated. It is anticipated that several years will be required to fully commission the three large cryogenic subsystems in standalone cryostats and establish the necessary infrastructure at the FNPB. Full system installation is anticipated by 2023, followed by commissioning and physics data collection.

\acknowledgments
This work was supported in part by the U.S. Department of Energy under grants  
DE-AC02-06CH11357, 
DE-AC05-00OR22725, 
DE-AC52-06NA25396, 
DE-FG02-94ER40818, 
DE-FG02-97ER41042, 
DE-FG02-99ER41101, DE-SC0014622, DE-SC0008107, 
DE-SC0005367, 
and 2017LANLEEDM, 
and the National Science Foundation under grants 
1306547, 
1205977, 1506459, 1812340, 
1440011, 
1506416, 

\end{document}